\documentclass[final,nospthms,natbib]{svjour3}

\usepackage{url}
\usepackage{mathptmx}
\usepackage{amsmath}
\usepackage{amssymb}
\usepackage{bm}		
\usepackage{graphicx}

\journalname{General Relativity and Gravitation}

\def\program#1{{\sc #1}}

\def\Schw{\text{Schw}}
\def\rd{\text{rd}}		
\def\ss{\text{ss}}		

\def\rv{\texttt{rv}}
\def\ru{\texttt{ru}}
\def\rvu{\texttt{rvu}}

\def\Cplusplus{\hbox{C\raise.15ex\hbox{\footnotesize ++}}}

\def\tfrac#1#2{{\textstyle\frac{#1}{#2}}}
\def\thalf{\tfrac{1}{2}}

\def\O{\mathcal{O}}
\def\boxop{\Box}
\def\del{\nabla}
\def\eqdef{\equiv}
\def\Realpart{\mathop{\text{Re}}}
\def\Imagpart{\mathop{\text{Im}}}
\def\sinc{\mathop{\text{sinc}}}
\def\ltsim{\lesssim}
\def\gtsim{\gtrsim}

\newcounter{linenumber}
\newlength{\linenumberwidth}
\def\lnum{
  \stepcounter{linenumber}
  \hbox to\linenumberwidth{\hss\arabic{linenumber}}\hspace*{1.5em}}
\def\marker{$\bullet$}
\def\assign{\leftarrow}

\def\auxgrid#1{\mathsf{A}^{(#1)}}
\def\grid#1{\mathsf{G}^{(#1)}}
\def\kw#1{\textbf{#1}}

\def\var#1{\texttt{#1}}

\def\N{\mathsf{N}}
\def\S{\mathsf{S}}
\def\E{\mathsf{E}}
\def\W{\mathsf{W}}
\def\A{\mathsf{A}}
\def\B{\mathsf{B}}
\def\C{\mathsf{C}}

\begin{document}
\title{Adaptive Mesh Refinement for Characteristic Grids}
\author{Jonathan Thornburg}
\institute{Department of Astronomy,
	   Indiana University,
	   Bloomington, Indiana, USA,\\
	   \email{jthorn@astro.indiana.edu}}
\date{13 May 2010}

\maketitle


\begin{abstract}
I consider techniques for Berger-Oliger adaptive mesh refinement (AMR)
when numerically solving partial differential equations with wave-like
solutions, using characteristic (double-null) grids.  Such AMR algorithms
are naturally recursive, and the best-known past Berger-Oliger
characteristic AMR algorithm, that of Pretorius \& Lehner
(\textit{J.\ Comp.\ Phys.} \textbf{198} (2004), 10), recurses on
individual ``diamond'' characteristic grid cells.  This leads to the
use of fine-grained memory management, with individual grid cells
kept in 2-dimensional linked lists at each refinement level.  This
complicates the implementation and adds overhead in both space and
time.
Here I describe a Berger-Oliger characteristic AMR algorithm which
instead recurses on null \emph{slices}.  This algorithm is very similar
to the usual Cauchy Berger-Oliger algorithm, and uses relatively
coarse-grained memory management, allowing entire null slices to be
stored in contiguous arrays in memory.  The algorithm is very efficient
in both space and time.
I describe discretizations yielding both 2nd and 4th~order global
accuracy.  My code implementing the algorithm described here is
included in the electronic supplementary materials accompanying
this paper, and is freely available to other researchers under the
terms of the GNU general public license.
\end{abstract}




\begin{quote}
\textit{This paper is dedicated to the memory of Thomas Radke,
my late friend, colleague, and partner in many computational
adventures.}
\end{quote}


\section{Introduction}
\label{sect-introduction}

Adaptive mesh refinement (AMR) algorithms are now a vital part of
computational science and are particularly valuable in the numerical
solution of partial differential equations (PDEs) whose solutions have
a wide dynamic range across the problem domain.  Here I focus on
explicit finite difference methods and PDEs which have propagating-wave
solutions.  The most powerful and general AMR algorithms for problems
of this type are those based on the pioneering work of \citet{Berger-1984}
(see also \citet{Berger-1982,Berger86,Berger-1989}).  These algorithms
use locally uniform grids, refined in space and time as needed, with
fine grids (which generally cover only a small part of the problem
domain) overlaying coarse grids.  At each time step, coarse grids
are integrated first and spatial boundary conditions for fine-grid
integrations are obtained by time-interpolation from the coarse grids.
This whole process is applied recursively at each of the possibly-many
levels of mesh refinement.

Berger and Oliger's original work, as well as most following work,
used Cauchy-type coordinates and grids, where initial data is given
on a spacelike hypersurface and the solution is then computed one
spacelike slice at a time within a numerical problem domain with
(typically) timelike boundaries.  For problems where the propagating-wave
PDEs are naturally posed on an infinite domain, these finite-domain
timelike boundaries require radiation boundary conditions.  For many
problems of interest these boundary conditions can only be approximate,
and for the Einstein equations or similar constrained PDE systems
they may render the evolution system ill-posed, generate significant
boundary reflections, and/or generate significant constraint violations.
In practice it's often difficult and/or computationally expensive to
reduce these boundary-condition errors to an acceptably low level.
\footnote{
	 See \citet{Givoli-1991} for a general review of
	 numerical radiation boundary conditions, and
\citet{Kidder-2005:boundary-conditions,Rinne-2006,Buchman-Sarbach-2006,
Buchman-Sarbach-2007,Rinne-Lindblom-Scheel-2007,Ruiz-Rinne-Sarbach-2007,
Seiler-etal-2008,Rinne-etal-2009}
	 for recent progress towards non-reflecting and
	 constraint-preserving radiation boundary conditions
	 for the Einstein equations.
	 }

As an alternative to Cauchy formulations, here I consider
\emph{characteristic} formulations, where the numerical problem domain's
boundaries are null geodesics.  This makes it very easy to impose
boundary conditions on the continuum PDEs in a well-posed and
constraint-preserving manner, and to approximate these boundary
conditions very accurately in the finite differencing scheme.  While
Cauchy-type AMR is now widely used in numerical relativity, and
characteristic formulations are also not uncommon, there has been
much less study of Berger-Oliger AMR using characteristic formulations.
This is the topic of this paper.

The best-known work on Berger-Oliger characteristic AMR is that
of \citet{Pretorius:2003wc}, who describe an algorithm which treats
the two null coordinates symmetrically, and whose fundamental unit
of recursion is the ``diamond'' double-null characteristic grid cell.
This leads to their code using very fine-grained memory management,
with each individual grid point at each refinement level containing
linked-list pointers to its neighboring grid points in each null
direction.  This makes the programming more complicated and adds
some space and time overhead.  Their algorithm has $\O(\Delta^2)$~global
accuracy, where $\Delta$ is the grid resolution.

In contrast, the AMR algorithm I describe here is much closer to the
earlier work of \citet{Hamade-Stewart-1996}, treating the two null
coordinates asymmetrically and only recursing on null \emph{slices}.
(In Cauchy-evolution terms, the slice-recursion algorithm treats one
null coordinate as a ``time'' coordinate labelling null slices and
the other as a ``space'' coordinate labelling events on a null slice.)
My algorithm uses relatively coarse-grained memory management, with
all the grid points in a single null slice level stored in a single
set of arrays which can easily be stored contiguously in memory.
This leads to relatively simple programming with only a small loss
of efficiency from the coarser-grained adaptivity.  I describe
finite differencing schemes and interpolation operators which yield
$\O(\Delta^4)$~global accuracy, as well as the usual $\O(\Delta^2)$.
By using \Cplusplus{} templates, my code is able to support both
cases with no run-time overhead.

To demonstrate the slice-recursion AMR algorithm I use a simple
model problem, the spherically symmetric real or complex scalar wave
equation on a Schwarzschild-spacetime background, with a time-dependent
Dirac $\delta$-function source term.  This problem is generally
representative of a wide range of black-hole perturbation problems
and, more generally, of PDEs where characteristic AMR algorithms may
be appropriate.

The remainder of this paper is organized as follows:
the remainder of this section outlines the notation used in this paper.
Section~\ref{sect-model-problem} describes the model problem.
Section~\ref{sect-unigrid-FD} gives a brief outline of the unigrid
finite differencing schemes I use for globally 2nd and 4th~order
accuracy.
Section~\ref{sect-LTE-estimation} describes how the local
truncation error of the finite differencing scheme can be estimated.
Section~\ref{sect-AMR} describes the slice-recursion AMR algorithm
and compares it to other Cauchy and characteristic Berger-Oliger
algorithms.  Section~\ref{sect-numerical-tests} presents tests of
the AMR algorithm to demonstrate that it is accurate and efficient.
Section~\ref{sect-conclusions} draws general conclusions.
Appendix~\ref{app-FD-details} gives a detailed description of the
unigrid finite differencing schemes I use.
Appendix~\ref{app-implementation} discusses some implementation
aspects of the AMR algorithm.


\subsection{Notation}

I generally follow the sign and notation conventions of \citet{Wald84},
with a $(-,+,+,+)$ metric signature.  I use the Penrose abstract-index
notation, with Latin indices $abc$ running over spacetime coordinates.
$\del_a$ is the covariant derivative operator associated with the 4-metric.

I use upper-case sans-serif letters $\A$, $\B$, $\C$, \dots to label
grid points and (in section~\ref{sect-AMR} and appendix~\ref{app-FD-details})
finite difference grids.  I describe my notation for finite difference
grids in detail in section~\ref{sect-AMR/Berger-Oliger-algorithm}.
I use \program{Small Capitals} for the names of software packages
and (in appendix~\ref{app-implementation/data-structures}) major
data structures in my AMR code.
$\lceil x \rceil$ denotes the smallest integer $\ge x$.

I use a pseudocode notation to describe algorithms:
Lines are numbered for reference,
but the line numbers are not used in the algorithm itself.
\# marks comment lines, while
keywords are typeset in \kw{bold font}
and most variable names in \var{typewriter font}
(a few variable names are mathematical symbols, such as ``$\ell_{\max}$'').
\hbox{``$\var{X} \assign Y$''} means that the variable $\var{X}$
is assigned the value of the expression $Y$.
Variables are always declared before use.
The declaration of a variable explicitly states the variable's type
and may also be combined with the assignment of an initial value, as in
\hbox{``\kw{integer} $\var{j} \assign 0$''}.
The looping construct
``\kw{for} \kw{integer} \var{X} \kw{from} $A$ \kw{to} $B$ \kw{by} $C$''
is inspired by BASIC but also includes a declaration of the loop variable
(with scope limited to the loop body, as in \Cplusplus{} and Perl).
The looping semantics are the same as Fortran's
``\texttt{do X = $A$, $B$, $C$}'', with the increment~$C$ defaulting
to~$1$ if omitted.
Conditional statements use PL/I-inspired syntax (\kw{if}-\kw{then}-\kw{else}).
$\{$ and $\}$ delimit the scope of procedures, loop bodies, and either
of the branches of conditional statements.
Procedures (subroutines) are marked with the keyword \kw{procedure},
and are explicitly invoked with a \kw{call} statement.  Procedure names
are typeset in \var{typewriter font}.  When referring to a procedure as
a noun in a figure caption or in the main text of this paper, the procedure
name is suffixed with ``\var{()}'', as in ``\var{foo()}''.


\section{Model Problem}
\label{sect-model-problem}

The basic AMR algorithm presented here is quite general, but for ease
of exposition I present it in the context of a simple model problem.
This model problem derives from the calculation of the radiation-reaction
``self-force'' on a scalar particle orbiting a Schwarzschild black
hole, but for purposes of this paper the model problem may be considered
by itself, divorced from its physical context.

Thus, consider Schwarzschild spacetime of mass~$M$ and introduce ingoing
and outgoing null coordinates $u$ and $v$ respectively, so the line element
is
\begin{equation}
ds^2 = - f(r) \, du \, dv + r^2 d\Omega^2
								      \,\text{,}
\end{equation}
where $r$ is (thus defined to be) the usual areal radial coordinate,
$f(r) \eqdef 1 - 2M/r$, and $d\Omega^2$ is the line element on a 2-sphere
of constant~$r$.  It's also useful to define the Schwarzschild time
coordinate $t_\Schw = \thalf(v+u)$ and the ``tortise'' radial coordinate
\begin{equation}
r_* = \thalf(v-u) = r + 2M \log \left| \frac{r}{2M} - 1 \right|
							  \label{eqn-rstar-defn}
								      \,\text{.}
\end{equation}
In this paper I only consider the region outside the event horizon,
$r > 2M$, so the coordinates $t_\Schw$, $r$, and $r_*$ are always
nonsingular, $t_\Schw$ is always timelike, and both $r$ and $r_*$
are always spacelike.  My computational scheme requires numerically
inverting~\eqref{eqn-rstar-defn} to obtain $r(r_*)$; I discuss this
inversion in appendix~\ref{app-implementation/computing-r(r_*)}.

The model problem is the spherically symmetric scalar wave equation
on this background spacetime, with a time-dependent Dirac $\delta$-function
source term (stationary in space),
\begin{equation}
\boxop \phi + V(r) \phi \equiv
\frac{\partial^2 \phi}{\partial u \, \partial v}
+ V(r) \phi
	= S(t_\Schw) \delta(r - r_p)
								      \,\text{,}
								\label{eqn-wave}
\end{equation}
where
$\boxop = \del^a \del_a$ is the usual curved-space wave operator,
$\phi$ is a real or complex scalar field, $r_p > 2M$ is a specified
``particle'' radius giving the spatial position of the source-term
worldline, $V(r)$ is a specified (smooth) position-dependent potential
which varies on a typical spatial scale $\gtsim M$ and which vanishes
at spatial infinity, and $S(t_\Schw)$ is a specified time-dependent
(typically highly-oscillatory) source term defined along the source
worldline $r = r_p$.

I define $\| \cdot \|$ to be a pointwise norm on the scalar field $\phi$.
For reasons discussed in section~\ref{sect-LTE-estimation/smoothing},
in the complex-scalar-field case $\| \cdot \|$ should be the complex
magnitude rather than (say) the $L_1$ norm
$\|\phi\|_1 = \bigl| \Realpart[\phi] \bigr| + \bigl| \Imagpart[\phi] \bigr|$,
even though the latter is slightly cheaper to compute.

\begin{figure}[!bp]
\begin{center}
\includegraphics[scale=0.50]{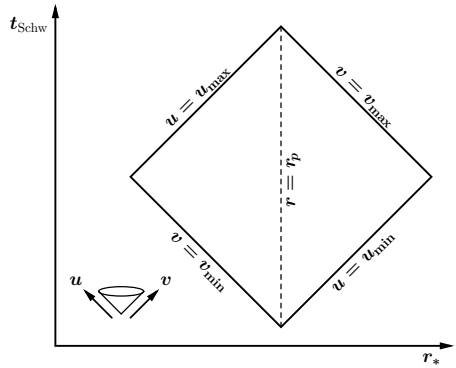}
\end{center}
\caption[Overall Problem Domain]
	{
	This figure shows the overall problem domain,
	and the $(u,v)$ and $(t_\Schw,r_*)$ coordinates.  The
	vertical dashed line marks the particle worldline.
	}
\label{fig-problem-domain}
\end{figure}

As shown in figure~\ref{fig-problem-domain},
the problem domain is a square in $(v,u)$ space,
$(v,u) \in [v_{\min}, v_{\max}] \times [u_{\min}, u_{\max}]$.
I take the particle worldline $r = r_p$ to symmetrically bisect the
domain.  For a given domain, it's convenient to introduce
``relative~$v$'' ($\rv$) and ``relative~$u$'' ($\ru$) coordinates
$\rv = v - v_{\min}$ and $\ru = u - u_{\min}$ respectively.

Initial data must be specified along the ``southwest'' and ``southeast''
faces of the domain, $v = v_{\min}$ and $u = u_{\min}$ respectively.
The ``northwest'' and ``northeast'' faces of the problem domain,
$u = u_{\max}$ and $v = v_{\max}$ respectively, are ingoing null
characteristics with respect to the problem domain, so no boundary
conditions need be posed there.  This is a key advantage of a
characteristic evolution scheme.  In contrast, using a Cauchy
evolution scheme an outgoing-radiation boundary condition is
generally required on each timelike boundary.

Assuming smooth initial data on the southwest and southeast grid
faces, $\phi$ is $C^\infty$ everywhere in the problem domain except
at the particle worldline.  In practice, $\phi$ and its spacetime
gradients display high dynamic ranges across the problem domain,
varying rapidly near the particle worldline but only slowly far
from the worldline.  This makes a unigrid scheme quite inefficient
and is the primary motivation for using AMR.

Because of the $\delta$-function source term, at the particle worldline
$\phi$ is generically only $C^1$, i.e., $\phi$ is continuous but its
gradient generically has a jump discontinuity across the particle
worldline.  Any finite differencing or interpolation operators
which include grid points on both sides of the particle worldline
must take the discontinuity into account.


\section{Unigrid Finite Differencing}
\label{sect-unigrid-FD}

The foundation of any mesh-refinement scheme is a stable and locally
consistent unigrid discretization.  To describe this, I introduce a
uniform finite-difference grid over the problem domain, with equal
grid spacing $\Delta$ in $v$ and $u$.  I use $j$ and $i$ as the integer
grid-point coordinates corresponding to $v$ and $u$ respectively,
\begin{subequations}
\begin{eqnarray}
v_j	& \eqdef &	v_{\min} + j\Delta				\\
u_i	& \eqdef &	u_{\min} + i\Delta
								      \,\text{,}
\end{eqnarray}
\end{subequations}
where $j$ and $i$ range from $0$ to $N \eqdef D/\Delta$ inclusive,
where $D = v_{\max} - v_{\min} = u_{\max} - u_{\min}$ is the
problem-domain size.  As is common in finite-differencing computations,
I use the notation that subscripting a grid function denotes its value
at the specified grid point, for example
\begin{equation}
\phi_{j,i} \eqdef \phi(v{=}v_j, u{=}u_i)
								      \,\text{.}
\end{equation}

To simplify the finite differencing near the particle worldline, I
require that the grid be placed such that if the particle worldline
passes through a grid cell, it does so symmetrically, bisecting through
the center of the cell.
\footnote{
\label{footnote-particle-symmetric-in-cell}
	 This symmetric passage is only possible because $r_p$ is
	 time-independent.  For the more generic case where $r_p$
	 is time-dependent, then in general the particle worldline
	 would pass obliquely through the cell.  As discussed by,
	 for example,
	 \protect\citet{Martel-Poisson-2002,Lousto-2005,Haas-2007},
	 this would considerably complicate the finite differencing
	 of cells intersecting the particle worldline.  However,
	 it wouldn't alter the overall character of the AMR
	 algorithm.
	 }
{}  This assumption considerably simplifies the finite differencing
near the particle worldline, and makes it easy to represent the
effects of the $\delta$-function source term accurately.  (See
\citet{Tornberg-Enquist-2004} for a general discussion of the numerical
treatment of $\delta$-function terms in PDEs.)

\begin{figure}[!bp]
\begin{center}
\begin{picture}(100,40)
%
\put(10,15.7){\includegraphics[scale=0.50]{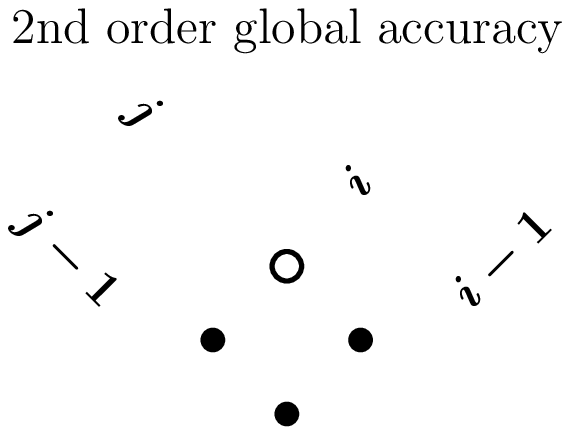}}
\put(50,12){\includegraphics[scale=0.50]{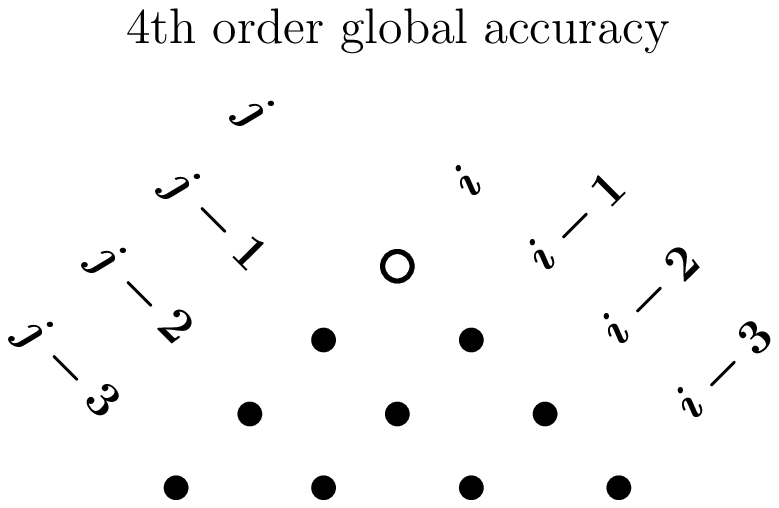}}
\put(19.6,2){\includegraphics[scale=0.50]{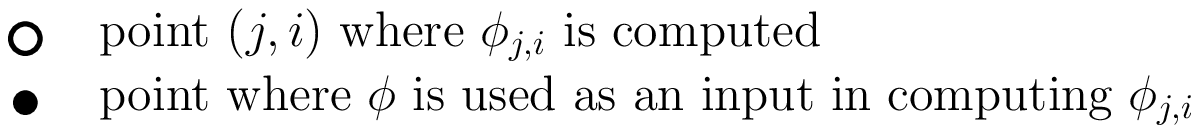}}
\end{picture}
\end{center}
\caption[Unigrid Finite Differencing Schemes]
	{
	This figure summarizes the unigrid finite differencing molecules.
	The left subfigure shows the molecule used for 2nd~order
	global accuracy.  The right subfigure shows the molecule
	used for 4th~order global accuracy far from the particle
	worldline; near the worldline the molecules are more complicated,
	and are shown in figure~\ref{fig-FD-molecules-details}.
	Open circles show the point where $\phi$ is being computed;
	solid circles show points where $\phi$ is used as an input
	in the computation of $\phi_{j,i}$.
	}
\label{fig-FD-molecules-summary}
\end{figure}

The finite difference schemes I consider here are all explicit, with
molecules summarized in figure~\ref{fig-FD-molecules-summary}.  Briefly,
for 2nd~order global accuracy I use a standard diamond-cell integration
scheme (\citet{Gomez-Winicour-1992:sssf-2+2-evolution-and-asymptotics,
Gomez-Winicour-Isaacson-1992:sssf-2+2-numerical-methods,
Gundlach-Price-Pullin-1994a,Burko-Ori-1997,Lousto-Price-1997,
Lousto-2005,Winicour-2009:living-review}),
while for 4th~order global accuracy I use a modified version of
the scheme described by \citet{Lousto-2005,Haas-2007}.
I describe the finite differencing schemes in detail in
appendix~\ref{app-FD-details}.  With one exception discussed
in appendix~\ref{app-FD-details/coarse-grid-instability},
these finite differencing schemes are stable.

It's important to know the domain of dependence of the finite
difference computation of $\phi_{j,i}$, i.e., the set of grid points
$(j{+}\beta, i{+}\alpha)$ where $\phi_{j{+}\beta, i{+}\alpha}$ is used
as an input in computing $\phi_{j,i}$.  (I define the ``radius'' of
such a finite difference computation as the maximum of all such
$|\beta|$ and $|\alpha|$, the radius in the $j$~direction as the
maximum of all such $|\beta|$, and the radius in the $i$~direction
as the maximum of all such $|\alpha|$.)  For 2nd~order global accuracy,
the domain of dependence is precisely that shown in
figure~\ref{fig-FD-molecules-summary}, i.e., it comprises the
3~grid points $\{(j,i{-}1), (j{-}1,i), (j{-}1,i{-}1)\}$.
For 4th~order global accuracy, this set depends on the position of
$(j,i)$ relative to the particle worldline, but it never includes
points outside the 4~slices $\{j, j{-}1, j{-}2, j{-}3\}$ or outside
the 4~$u$~positions $\{i, i{-}1, i{-}2, i{-}3\}$.

Given this domain of dependence, there are many possible orders in
which the $\phi_{j,i}$ may be computed in a unigrid integration.
However, the algorithms I consider here all integrate the grid points
in ``raster-scan'' order, i.e., using an outer loop over~$j$ (so that
each iteration of the outer loop integrates a single $v = \text{constant}$
null slice) and an inner loop over~$i$.

\subsection{Local versus Global Truncation Errors for Characteristic Schemes}
\label{sect-unigrid-FD/LTE-vs-GTE}

Recall the (standard) definitions of the local and global truncation
error of a finite differencing scheme
(\citet{Kreiss73,Choptuik-1991:FD-consistency,
Richtmyer-Morton-2nd-edition,LeVeque-2007:finite-diff-methods}):
The local truncation error (LTE) is a pointwise norm of the discrepancy
that results when the exact solution of the PDE is substituted into
the finite difference equations at a grid point.  The global truncation
error (GTE) is a pointwise norm of the difference between the exact
solution of the PDE and the result of solving the finite difference
equations using exact arithmetic (i.e., without floating-point roundoff
errors).  The LTE and GTE are both grid functions.

For a stable and consistent Cauchy evolution scheme, the GTE and LTE
are normally of the same order in the grid spacing $\Delta$
(\citet{Kreiss73,Choptuik-1991:FD-consistency,
Richtmyer-Morton-2nd-edition,LeVeque-2007:finite-diff-methods}).
However, in a characteristic evolution, errors can build up cumulatively
over many grid points, so the GTE is generically worse than the LTE.
For a $1{+}1$ ($\text{space} + \text{time}$) dimensional problem such
as that considered here, errors can accumulate over $\O(N^2)$~grid
points where $N = \O(1/\Delta)$, so generically an $\O(\Delta^{n{+}2})$~LTE
is required to guarantee an $\O(\Delta^n)$ GTE.  Corresponding to this,
the finite differencing schemes I present here for 2nd and 4th~order
global accuracy (GTE) actually have 4th and 6th~order LTE respectively,
except near the particle worldline where one order lower GTE is acceptable
because there are only $\O(N)$ near-worldline grid cells.


\subsection{Initial Data}
\label{sect-unigrid-FD/initial-data}

The globally--4th-order finite differencing scheme summarized in
figure~\ref{fig-FD-molecules-summary} and described in detail in
appendix~\ref{app-FD-details/4th-order} is a 4-level scheme:  the
finite-difference molecule for computing $\phi_{j,i}$ includes points
on 3~previous slices and at 3~previous $u$~positions.  Physical boundary
conditions are (only) given on the southeast and southwest grid boundaries,
so starting the numerical integration requires that ``extended initial
data'' somehow be computed, comprising the physical boundary data,
the next 2~slices after the southwest grid boundary, and the next
2~grid points on each succeeding slice after the southeast grid boundary.
I describe several different schemes for constructing the extended
initial data in appendix~\ref{app-FD-details/4th-order/initial-data}.


\section{Estimation of the Local Truncation Error}
\label{sect-LTE-estimation}

As is common in AMR algorithms, I implement the ``adaptive''
part of AMR using an estimate of the numerical solution's local
truncation error (LTE).  I compute this via the standard technique
(\citet[section~A.6]{LeVeque-2007:finite-diff-methods})
of comparing the main numerical integration with that from a
coarser-resolution integration.  In the context of Berger-Oliger
style AMR, where a hierarchy of grids are being integrated concurrently
at varying resolutions, there are several possible choices for the
coarser-resolution comparison solution.  It can come from a separate
``shadow'' AMR hierarchy (each level of which is coarser than the
corresponding level of the main AMR hierarchy), or it can come from
the next-coarsest level of the (single) AMR hierarchy (the
``self-shadow hierarchy'' technique of
\citet{Pretorius-PhD,Pretorius:2003wc}).

Here I use a different technique, also used by \citet{Hamade-Stewart-1996},
where the coarser-resolution comparison is obtained locally within
the (unigrid) evolution at each level of the grid hierarchy, by simply
subsampling every 2nd~grid point of every 2nd~slice.  That is, suppose
we have a finite differencing scheme with $\O(\Delta^n)$ LTE, whose
domain of dependence for computing $\phi_{j,i}$ is the set of $K$~grid
points $\{(j{+}\beta_k, i{+}\alpha_k) \big| 1 \le k \le K\}$ for some
constants $\{\beta_k\}$ and $\{\alpha_k\}$.  For any $(j,i)$ and
$\Delta$, let $\phi^{(\Delta)}_{j,i}$ be the value of $\phi_{j,i}$
obtained from the usual numerical integration with grid spacing~$\Delta$.
Let $\phi^{(\Delta \to 2\Delta)}_{j,i}$ be the value of $\phi_{j,i}$
obtained by taking a single step of size~$2\Delta$ using as inputs
the $\phi^{(\Delta)}_{j,i}$ values at the set of $K$~grid points
$\{(j{+}2\beta_k, i{+}2\alpha_k) \big| 1 \le k \le K\}$.  Then I
estimate the LTE in computing $\phi^{(\Delta)}_{j,i}$ as
\begin{equation}
\text{LTE} \approx
	\frac{1}{2^n-4}
	\left\|
	\phi^{(\Delta)}_{j,i} - \phi^{(\Delta \to 2\Delta)}_{j,i}
	\right\|
							\label{eqn-LTE-estimate}
\end{equation}
where the normalization factor is obtained by comparing the LTE of
a single $2\Delta$-sized step with that accumulated in 4~separate
$\Delta$-sized steps covering the same region of spacetime.

This scheme is easy to implement and works well, although it does
limit the locations where the LTE can be estimated to those where
(i)~the data for a $2\Delta$-sized step is available, and where
(ii)~both the $\Delta$- and $2\Delta$-sized steps use the same finite
differencing scheme.  Constraint~(i) implies, for example, that when a
new refinement level is created, LTE estimates aren't available for
it until a few $v=\text{constant}$ slices have been integrated on
that level.  For the finite differencing schemes described here,
constraint~(ii) implies that the LTE estimate isn't available for
cells within a few grid points of the particle worldline.  I haven't
found either of these constraints to be a problem in practice.


\subsection{Cost of Computing the LTE Estimate}

The cost in space (memory usage) associated with this LTE-estimation
scheme is the requirement that sufficiently many adjacent slices be
kept in memory simultaneously for the $2\Delta$-sized steps.  For the
globally-2nd-order finite differencing scheme described in
section~\ref{sect-unigrid-FD}, the LTE estimator requires data from
3~adjacent slices.  As discussed in
section~\ref{sect-AMR/slice-recursion-algorithm}, for 2nd~order
global accuracy my AMR algorithm uses interpolation operators which
use data from up to 4~adjacent slices, so the LTE estimator is ``free''
in the sense that it doesn't increase the number of slices needing
to be kept in memory beyond what the rest of the computation already
requires.  For the globally--4th-order finite differencing scheme
described in section~\ref{sect-unigrid-FD}, the LTE estimator
requires 7~adjacent slices, while the interpolation operators only
need 6~adjacent slices, so the LTE estimator adds a
$\tfrac{1}{6} \approx 17\%$ fractional overhead in memory usage.

I discuss the CPU-time cost of computing the LTE estimate in
footnote~\ref{footnote-cost-of-LTE-estimate} in
section~\ref{sect-AMR/slice-recursion-algorithm}.


\subsection{Smoothing the LTE Estimates}
\label{sect-LTE-estimation/smoothing}

In an AMR scheme, mesh-refinement boundaries and regridding operations
tend to introduce small amounts of interpolation noise into the numerical
solution, which tends to be amplified in the LTE estimate.  In particular,
for the scheme described here, there are often isolated points with
anomolously high LTE estimates.  To avoid having these falsely trigger
(unwanted) mesh refinements, I smooth the LTE estimates on each slice
with a moving median-of-3 filter before comparing them to the error
threshold.

\citet{Pretorius:2003wc} describe the use of a moving-average smoothing
of the LTE estimate to address a slightly different problem in their
characteristic AMR algorithm:
	\begin{quote}
	The point-wise TE [truncation error] computed using
	solutions to wave-like finite-difference equations is
	in general oscillatory in nature, and will tend to go
	to zero at certain points within the computational
	domain \dots, even in regions of relatively high
	truncation error. We do not want such isolated points
	of (anomalously) small TE to cause temporary
	unrefinement, \dots
	\end{quote}

For the model problem I consider here, ``temporary unrefinement''
doesn't seem to be a problem in practice so long as the norm
$\| \cdot \|$ is chosen to be the complex magnitude of $\phi$.  This
appears to be because while the complex phase of $\phi$ oscillates
rapidly along the particle worldline, $\phi$'s complex magnitude tends
to remain relatively constant.  (In an early version of my AMR code
where I used the $L_1$ norm
$\|\phi\|_1 = \bigl| \Realpart[\phi] \bigr| + \bigl| \Imagpart[\phi] \bigr|$,
I found that temporary unrefinement did indeed tend to occur, as the
rapidly changing complex phase of $\phi$ translated into corresponding
changes in $\|\phi\|_1$.)

For other physical systems, further smoothing of the estimated LTE
might be necessary.


\section{Adaptive Mesh Refinement}
\label{sect-AMR}


\subsection{The Berger-Oliger Algorithm}
\label{sect-AMR/Berger-Oliger-algorithm}

The Berger-Oliger AMR algorithm for Cauchy evolutions of
hyperbolic or hyperbolic-like PDEs (\citet{Berger-1984}; see also
\citet{Berger-1982,Berger86,Berger-1989}) was first used in numerical
relativity by \citet{Choptuik86,Choptuik89,
Choptuik-in-d'Inverno:self-similarity-and-AMR,
Choptuik-1993:self-similarity}, and is now widely used
for a variety of problems.
\citet{Schnetter-etal-03b} give a nice summary of some of the
considerations involved in using the Berger-Oliger algorithm with
evolution systems which contain 2nd~spatial derivatives but only
1st~time derivatives.
\citet{Lehner-Liebling-Reula-2006,Bruegmann-etal-2008:BAM-calibration,
Husa-etal-2008:6th-order-FD-for-2BH-evolution} discuss adjustments
to the algorithm (particularly interpolation and prolongation operators)
necessary to obtain higher-than-2nd-order global finite differencing
accuracy in a Berger-Oliger scheme.
\citet{Pretorius:2005amr} discuss refinements to the standard
Berger-Oliger algorithm to accomodate coupled elliptic-hyperbolic
systems of PDEs.
\citet{MacNeice00,Burgarelli-etal-2006:Quadtree-AMR} discuss
quadtree/octtree grid structures and their use with Berger-Oliger
mesh refinement.
Many individual codes also have published descriptions,
including (among many others),
\program{AD}
  (\citet{Choptuik89,Choptuik-in-d'Inverno:self-similarity-and-AMR,
Choptuik-1994:AD-code}),
\program{AMRD}/\program{PAMR}
  (\citet{Pretorius-PhD,Pretorius-2002:AMRD-code,Pretorius-2002:PAMR-code}),
\program{BAM}
  (\citet{Bruegmann-1996:3D-AMR-evolution-of-Schw,
Bruegmann-Tichy-Jansen-2004,Bruegmann-etal-2008:BAM-calibration}),
\program{Carpet}/\program{Cactus}
  (\citet{Schnetter-etal-03b,Schnetter-2001:Carpet-code,
Goodale02a,Goodale-etal-1999:Cactus-code}),
\program{Chombo}/\program{AmrLib}/\program{BoxLib}
  (\citet{Colella-etal-2009:Chombo-design,
Su-Wen-Yelick-2006:reimplementing-Chombo-subset-in-Titanium,
Colella-etal-2009:Chombo-code,Rendleman-etal-2000:BoxLib,
Lijewki-Beckner-Rendleman-2006:Boxlib-and-AmrLib-codes}),
the Choptuik~\etal{} axisymmetric code
  (\citet{Choptuik-etal-2003:axisym-grav-collapse-code,
Choptuik-etal-2003b:axisym-critical-collapse-of-massless-scalar-field,
Pretorius:2005amr}),
\program{HAD}
  (\citet{Liebling-2002,Liebling-2004,Anderson-etal-2006}),
\program{GrACE}/\program{HDDA}/\program{AMROC}/\program{DAGH}
  (\citet{Parashar-Li-2009:GrACE,Parashar00a,
Deiterding-2006:simulations-using-AMROC,
Deiterding-2005:AMROC-parallelization,Deiterding-2005:AMROC-code,
Mitra-Parashar-Browne-1995:DAGH-code}),
\program{Overture}
  (\citet{Brown-etal-1997:Overture-software,
Brown-Henshaw-Quinlan-1999:Overture-software,
Henshaw-etal-2002:Overture-code}),
\program{PARAMESH}
  (\citet{MacNeice00,Olson-MacNeice-2005:PARAMESH,Olson-2006:PARAMESH,
Olson-MacNeice-1999:PARAMESH-code}),
and
\program{SAMRAI}
  (\citet{Hornung-Kohn-2002:SAMRAI-design,
Hornung-Kohn-1999:SAMRI-object-oriented-design,
Hornung-Wissink-Kohn-2006:SAMRAI-design,
Hornung-Kohn-2002:SAMRAI-code}).

Although the focus of this paper is on characteristic Berger-Oliger
AMR, it's useful to begin with a brief review of the Cauchy
Berger-Oliger algorithm.  I will only present a few of the algorithm's
properties that are particularly relevant here; see the references
cited in the previous paragraph for more extensive discussions of the
algorithm, its rationale (i.e., why the algorithm is constructed in
the way that it is), and how it may be modified to meet various
situations.

To describe the Berger-Oliger algorithm it's convenient to consider
a generic PDE with propagating-wave solutions in $1{+}1$
($\text{space}{+}\text{time}$) dimensions, and define global
timelike and spacelike coordinates $t$ and~$x$ respectively.
\footnote{
	 For the model problem of section~\ref{sect-model-problem}
	 these coordinates may be taken to be $t = t_\Schw$ and
	 $x = r_*$.
	 }
{}  I consider uniform finite difference grids in these coordinates,
with indices~$j$ and~$i$ indexing the~$t$ and~$x$ dimensions
respectively.  In contrast to the characteristic-grid case discussed
in section~\ref{sect-unigrid-FD}, I don't assume that the grid cells
are square, i.e., I don't assume anything about the Courant number
(the ratio of the time step to the spatial resolution).

The basic data structure of the Berger-Oliger algorithm is that of
a hierarchy of such uniform grids, each having a different resolution.
The grids are indexed by an integer ``refinement level'' $\ell$ in
the range $0 \le \ell \le \ell_{\max}$.  (In general $\ell_{\max}$
is time-dependent, but for convenience I don't explicitly show this
in the notation.)  I refer to the grid at refinement level~$\ell$ as
$\grid{\ell}$, to the time level (``slice'') $j$ (i.e., the slice
$t = t_j$) of $\grid{\ell}$ as $\grid{\ell}_j$, to the grid point
$(t{=}t_j, x{=}x_i)$ as $(j,i)$, and to the grid function value(s)
at this grid point as $\grid{\ell}_{j,i}$.  In general any
$\grid{\ell}_j$ may consist of several connected components, but
here I focus on the simpler case where each $\grid{\ell}_j$ contains
only a single connected component
\footnote{
\label{footnote-each-grid-has-only-a-single-component}
	 For the model problem of section~\ref{sect-model-problem}
	 this isn't a significant restriction, since apart from
	 the (ignorable) spurious radiation discussed in
	 section~\ref{sect-AMR/slice-recursion-algorithm/avoiding-undesired-mesh-refinement},
	 the resolution required to adequately represent $\phi$
	 tends to decrease monotonically with distance from the
	 particle worldline, so that for a given LTE threshold,
	 the region of a slice needing a given resolution is just
	 a single closed interval.  For other problems this might
	 not be the case, requiring some or all of the $\grid{\ell}_j$
	 to have multiple connected components for good efficiency.
	 This would somewhat complicate the data structures, but
	 it wouldn't alter the overall character of the AMR
	 algorithm.
	 }
{} covering some closed interval
$x \in [\grid{\ell}_j.x_{\min}, \grid{\ell}_j.x_{\max}]$,
with corresponding grid-point indices
$i \in [\grid{\ell}_j.\var{i}_{\min}, \grid{\ell}_j.\var{i}_{\max}]$.
As suggested by the notation, in general the domain of $\grid{\ell}$
isn't rectangular, i.e., in general
$\grid{\ell}_j.x_{\min}$, $\grid{\ell}_j.x_{\max}$,
$\grid{\ell}_j.\var{i}_{\min}$, and $\grid{\ell}_j.\var{i}_{\max}$
all vary with $j$.

The coarsest or ``base'' grid $\grid{0}$ covers the entire problem domain.
For each integer $\ell$ with $0 < \ell \le \ell_{\max}$, $\grid{\ell}$
has a resolution $2^\ell$~times finer than that of $\grid{0}$,
\footnote{
\label{footnote-choice-of-refinement-ratio}
	 This can easily be generalized to any other integer
	 refinement ratio $> 1$ (including having the refinement
	 ratio vary from one level to another).  Larger refinement
	 ratios reduce the $\ell_{\max}$ needed for a given
	 total dynamic range of resolution in the refinement
	 hierarchy and thus reduce some of the ``bookkeeping''
	 overheads in the computation.
	 However, smaller refinement ratios give a smoother
	 variation of the grid resolution (i.e., a variation
	 with smaller jumps) across the problem domain,
	 allowing the resolution to be better matched to
	 that needed to just obtain the desired LTE at
	 each event, which improves the overall efficiency
	 of the computation.  For this latter reason I use
	 a refinement ratio of $2\!:\!1$ in my AMR algorithm
	 and code.
	 }
{} and typically covers only some proper subset of the problem domain.
Here I consider only ``vertex-centered'' grids, where every~2nd
$\grid{\ell{+}1}$~point coincides with a $\grid{\ell}$~point.
\footnote{
	 An alternative approach uses ``cell-centered'' grids,
	 where grid points are viewed as being at the center
	 of grid cells and it is these grid \emph{cells} which
	 are refined (so that the $\grid{\ell{+}1}$~grid points
	 are located $\tfrac{1}{4}$ and $\tfrac{3}{4}$ of
	 the way between adjacent $\grid{\ell}$~grid points).
	 This approach is particularly useful with finite volume
	 discretizations (\citet{LeVeque-2002:finite-volume-methods})
	 and is used by, for example, the \program{PARAMESH}
	 mesh-refinement framework
  (\citet{MacNeice00,Olson-MacNeice-2005:PARAMESH,Olson-2006:PARAMESH,
Olson-MacNeice-1999:PARAMESH-code})
	 and the \program{BAM} code
  (\citet{Bruegmann-Tichy-Jansen-2004,Bruegmann-etal-2008:BAM-calibration}).
	 }
{}  The Berger-Oliger algorithm requires that the grids always be
maintained such that for any $\ell$, on any slice common to both
$\grid{\ell}$ and $\grid{\ell{+}1}$, the region of the problem domain
covered by $\grid{\ell{+}1}$ is a (usually proper) subset of that
covered by $\grid{\ell}$.  I refer to this property as the
``proper nesting'' of the grids.
\footnote{
	 The proper-nesting requirement is actually slightly
	 stronger:  each $\grid{\ell{+}1}$ grid point must be
	 far enough inside the region covered by $\grid{\ell}$
	 to allow interpolating data from $\grid{\ell}$.
	 In practice, in the $v$~direction this requirement is
	 enforced implicitly by the Berger-Oliger algorithm,
	 while in the $u$~direction this requirement must be
	 enforced explicitly in the regridding process
	 (procedure \var{shrink\_to\_ensure\_proper\_nesting()}
	 in figure~\ref{fig-Berger-Oliger-algorithm/regrid}).
	 }

Each grid $\grid{\ell}$ maintains its own current slice for the
integration, denoted $\grid{\ell}_\var{current\_j}$; the $j$~coordinate
of this slice is denoted $\grid{\ell}.\var{current\_j}$.  Each slice is
integrated with the same finite differencing scheme and Courant number.
\footnote{
	 Using the same Courant number at each refinement
	 level (i.e., scaling the time step on each grid
	 proportional to the spatial resolution) is known
	 as ``subcyling in time'' and is widely, though not
	 universally, used in Berger-Oliger AMR codes.
	 \citet{Dursi-Zingale-2003} disuss some of the tradeoffs
	 determining whether or not subcycling is worthwhile.
	 }
{}  Although I describe each $\grid{\ell}$ here as containing the
entire time history of its integration, in practice only the most
recent few time slices need to be stored in memory.  The precise
number of time slices needed is set by the larger of the number
needed by the unigrid finite differencing scheme, the LTE estimation,
and by the interpolations used in the Berger-Oliger algorithm.  This
is discussed further in section~\ref{sect-AMR/slice-recursion-algorithm}.

\begin{figure}[bp!]
\begin{flushleft}
\includegraphics[scale=0.75]{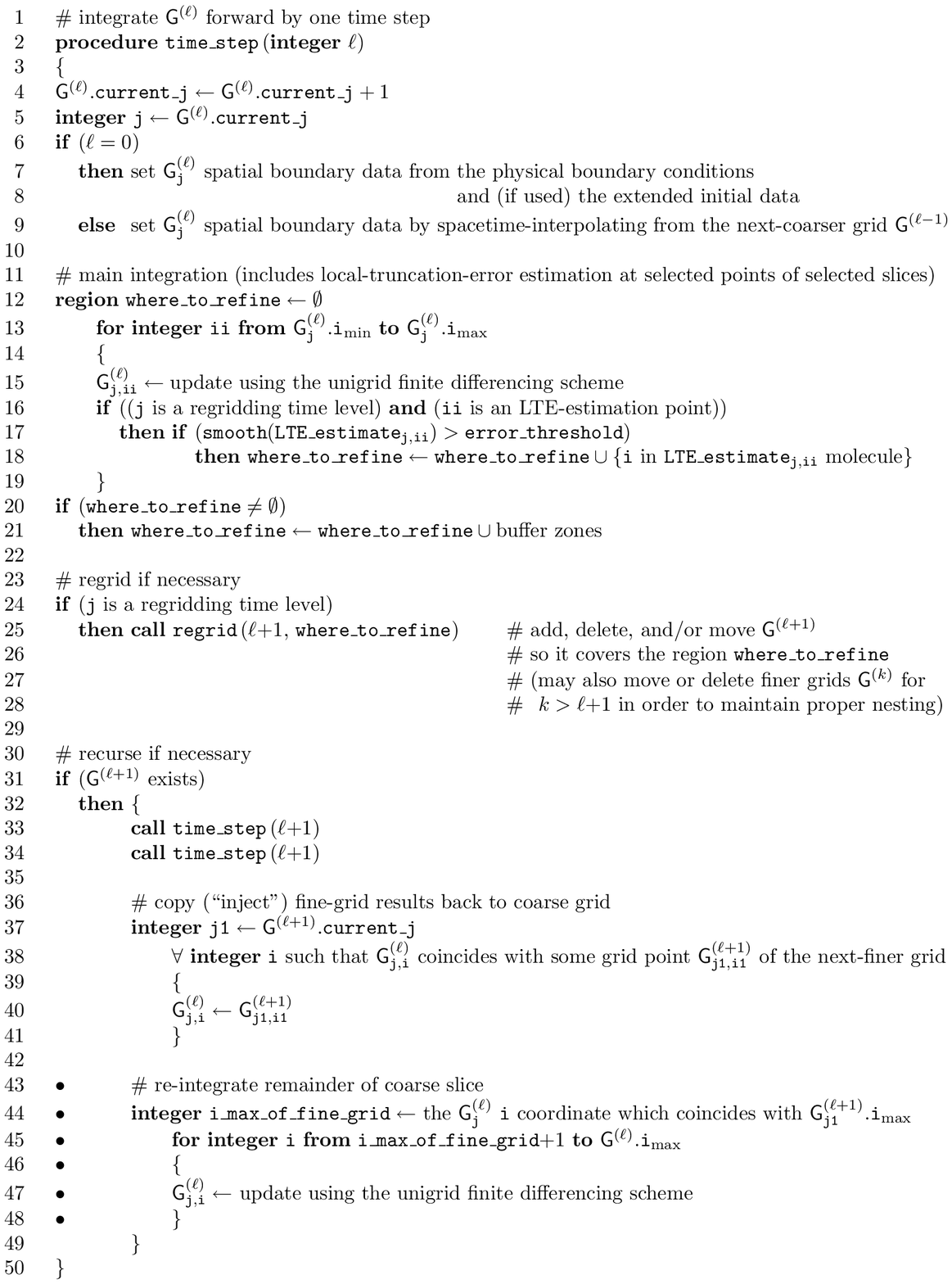}
\end{flushleft}
\caption[Outline of the Berger-Oliger Algorithm]
	{
	This figure gives an outline of the standard Berger-Oliger
	AMR algorithm (if the lines marked with~\marker{} are omitted),
	and of the slice-recursion algorithm
	(if the lines marked with~\marker{} are included).
	See figure~\ref{fig-Berger-Oliger-algorithm/regrid}
	for an outline of the procedure \var{regrid()}
	which is called at line~25.
	}
\label{fig-Berger-Oliger-algorithm/main}
\end{figure}

\begin{figure}[bp!]
\begin{flushleft}
\includegraphics[scale=0.75]{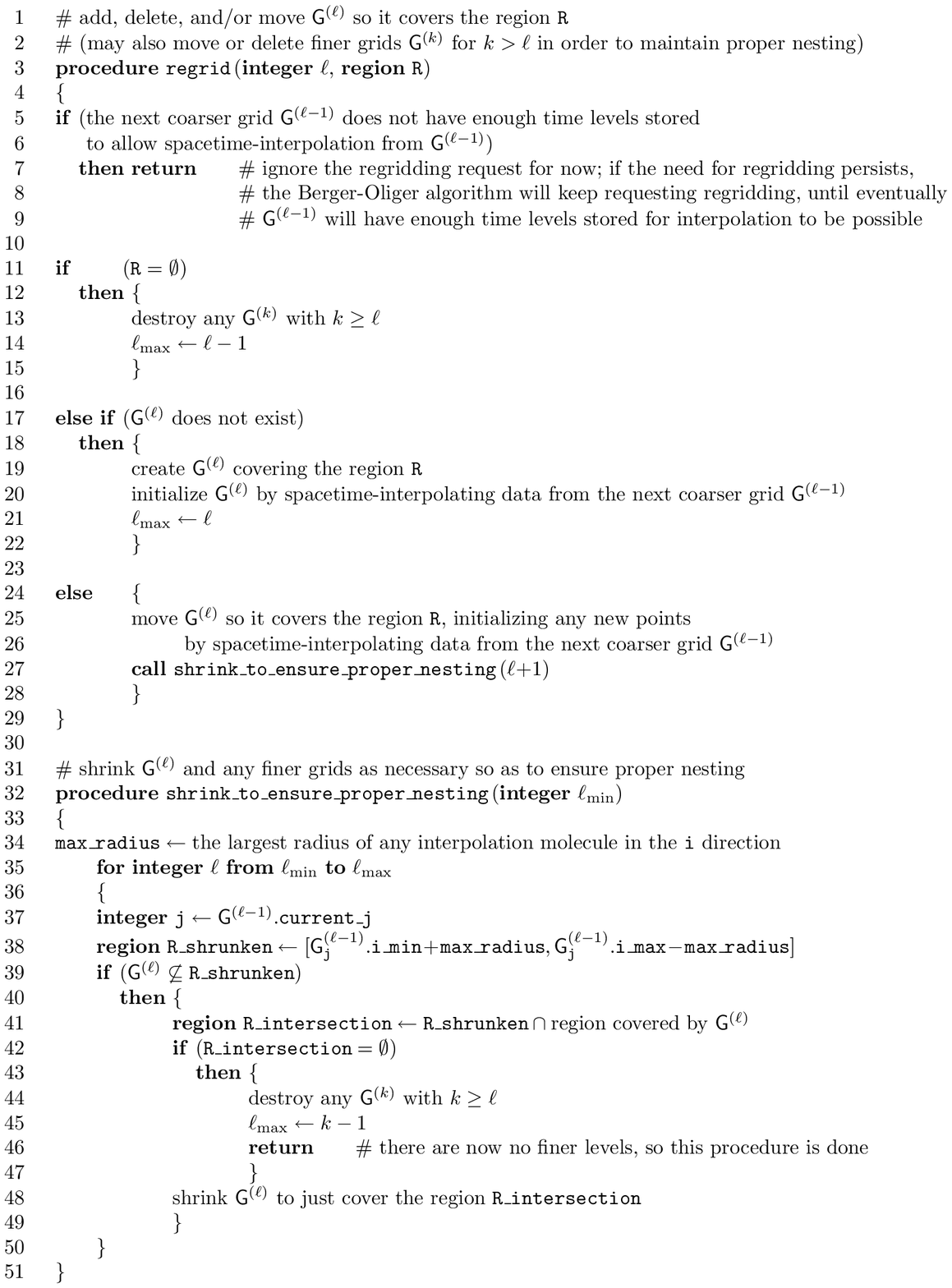}
\end{flushleft}
\caption[Outline of the Regridding Algorithm]
	{
	This figure gives an outline of the regridding procedure
	\var{regrid()} which is called by the main Berger-Oliger
	algorithm (figure~\ref{fig-Berger-Oliger-algorithm/main}),
	and of the auxiliary procedure
	\var{shrink\_to\_ensure\_proper\_nesting()} which is
	called by \var{regrid()}.
	}
\label{fig-Berger-Oliger-algorithm/regrid}
\end{figure}

Figure~\ref{fig-Berger-Oliger-algorithm/main}
(ignoring the lines marked with~\marker{}, whose purpose will be
discussed in section~\ref{sect-AMR/slice-recursion-algorithm})
gives a pseudocode outline of the Berger-Oliger algorithm.
Notice that the algorithm is recursive, and that this recursion
is at the granularity of an entire slice.  That is, the algorithm
integrates an entire slice
(lines~11--19 in figure~\ref{fig-Berger-Oliger-algorithm/main})
before recursing to integrate the finer grids (if any).

\begin{figure}[!bp]
\begin{center}
\begin{picture}(120,180)
\put(0,117){\includegraphics[scale=0.380]{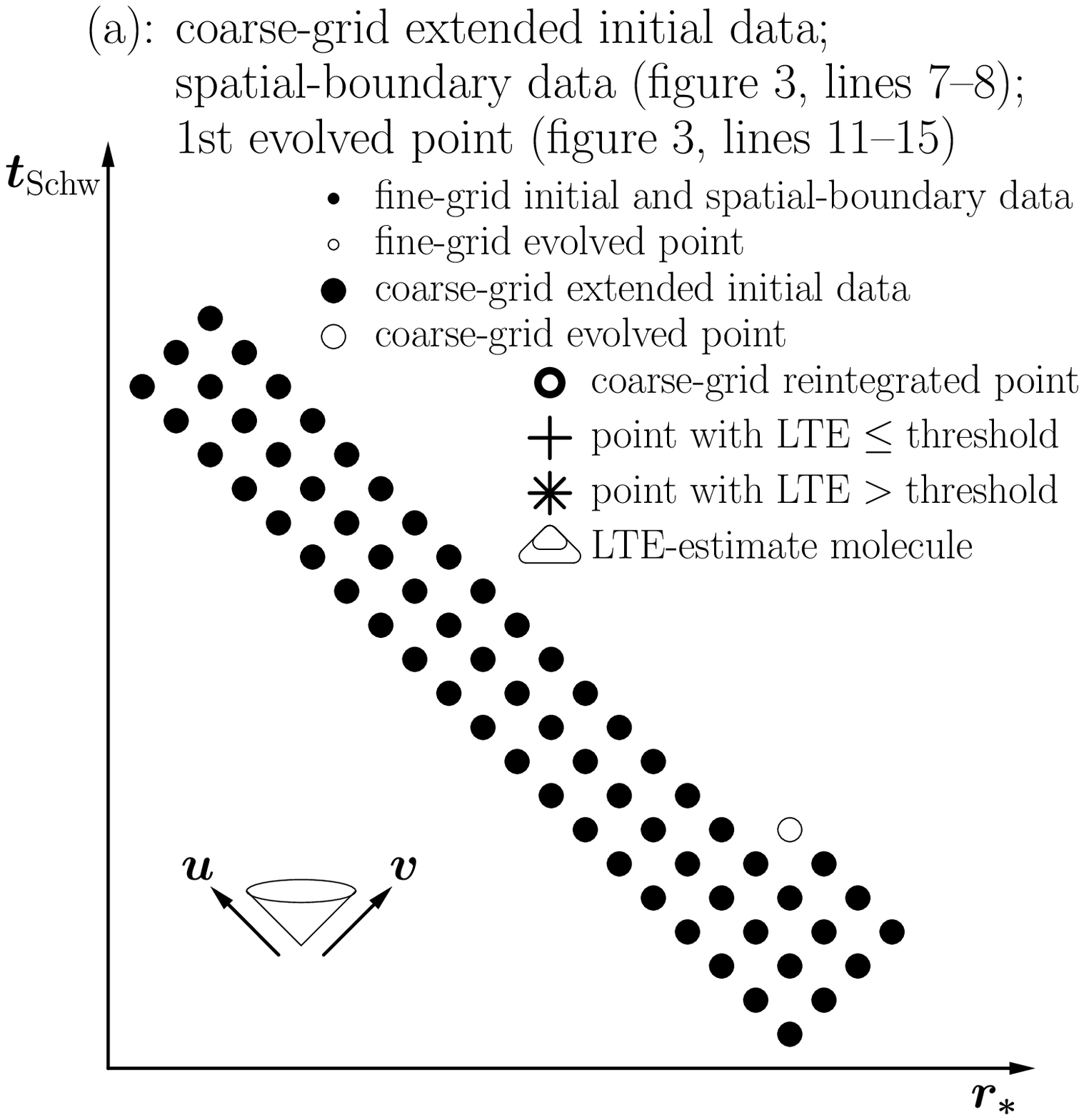}}
\put(62,117){\includegraphics[scale=0.380]{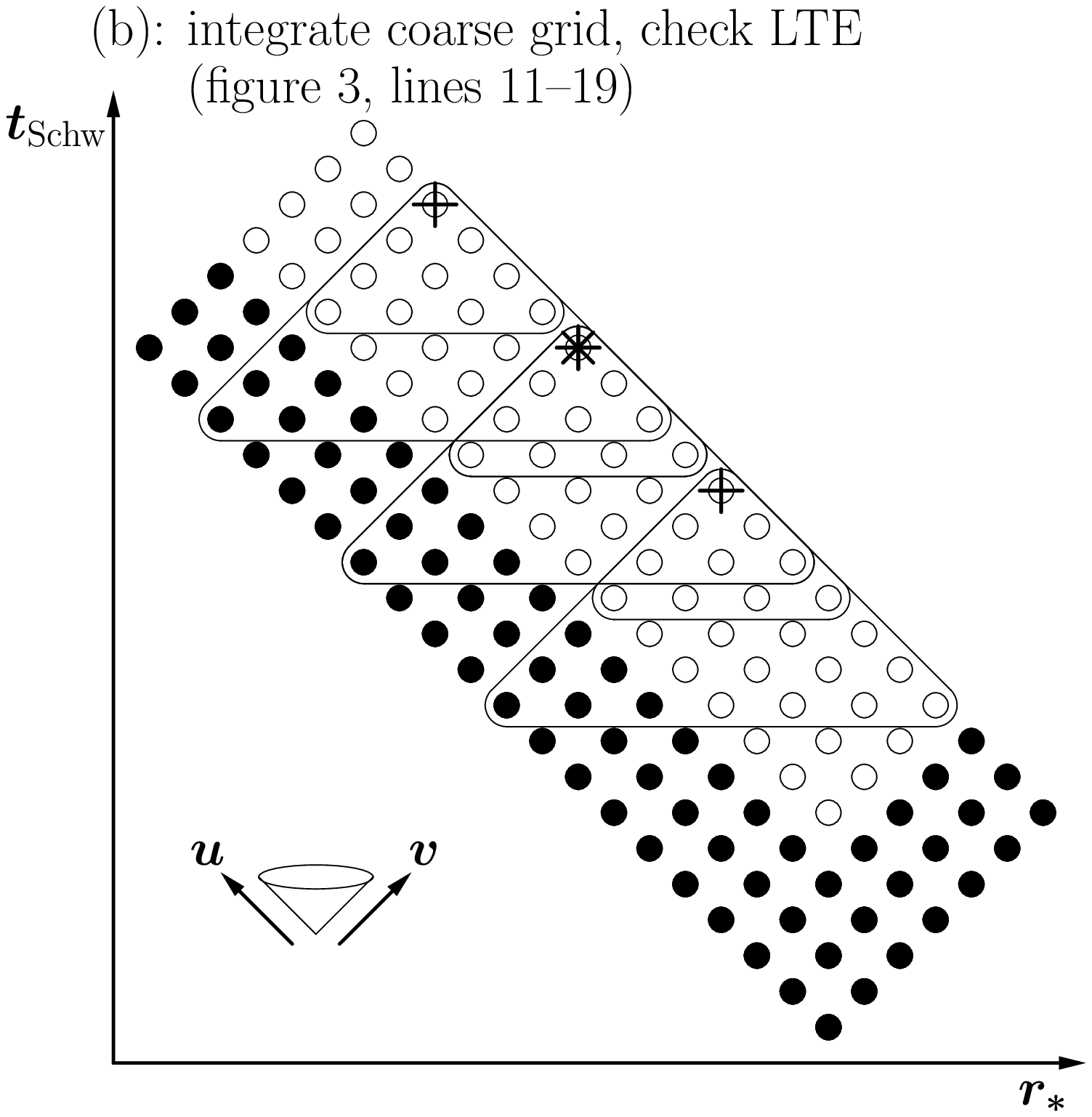}}
\put(0,57){\includegraphics[scale=0.380]{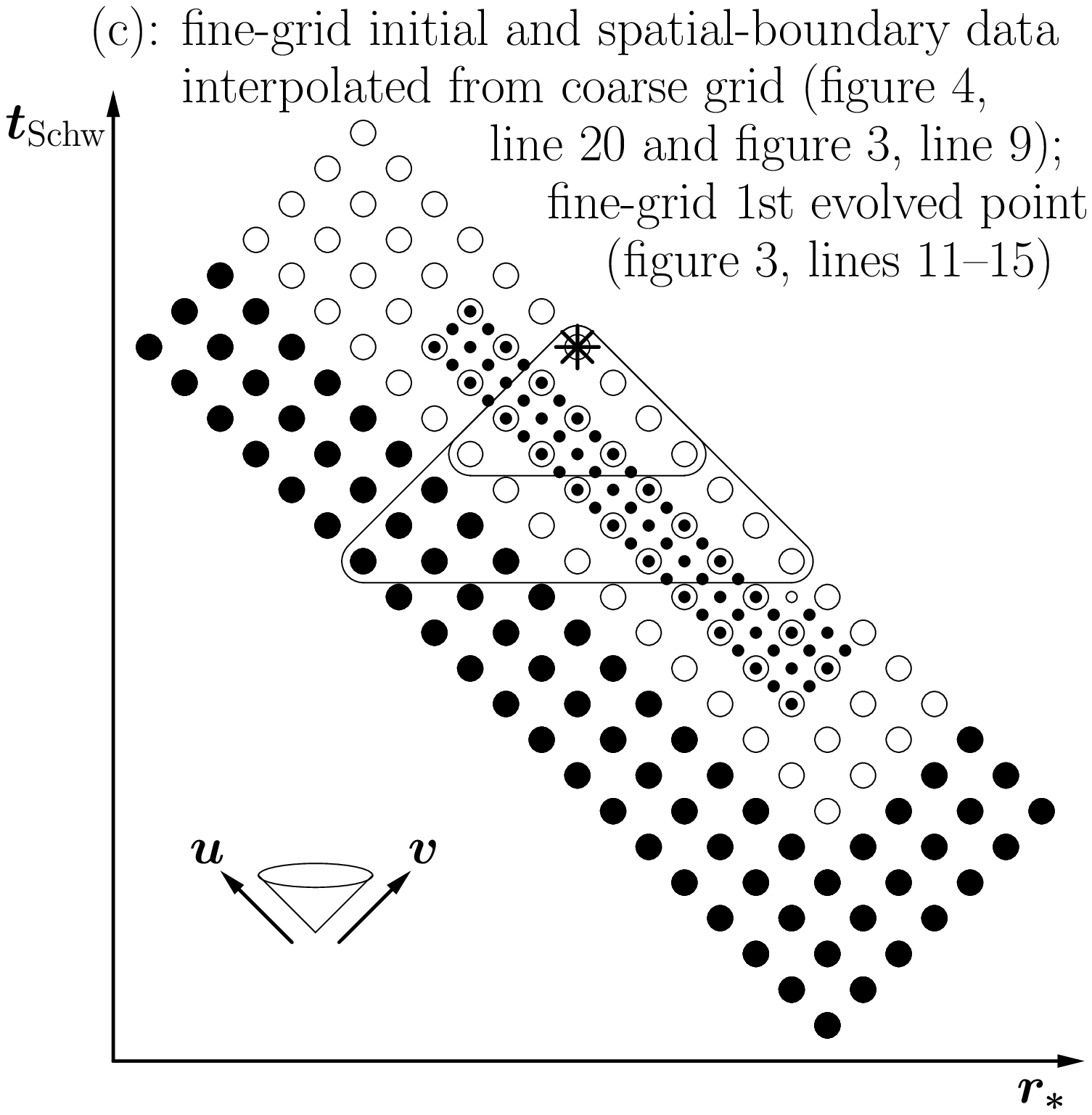}}
\put(62,57){\includegraphics[scale=0.380]{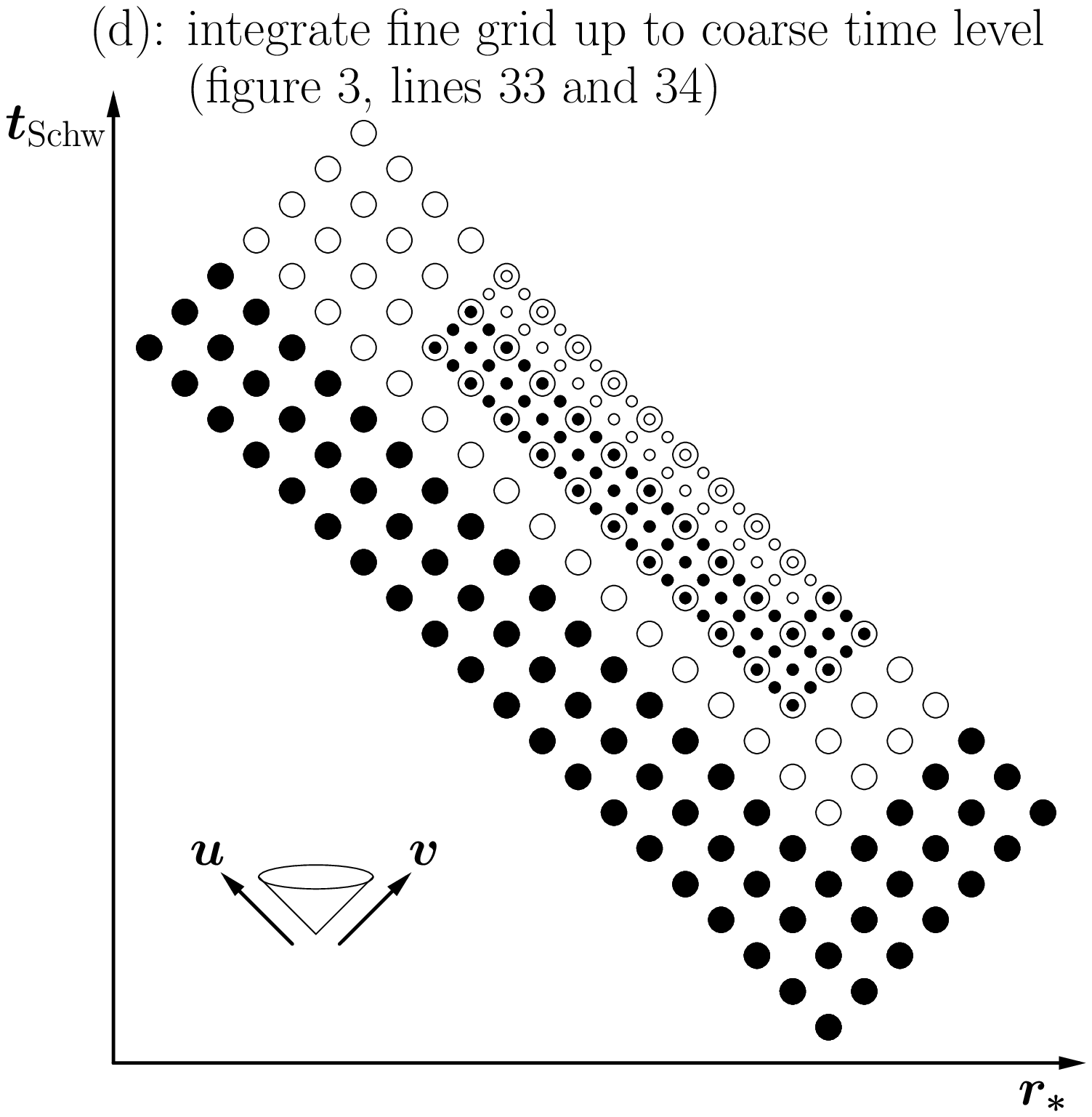}}
\put(0,-3){\includegraphics[scale=0.380]{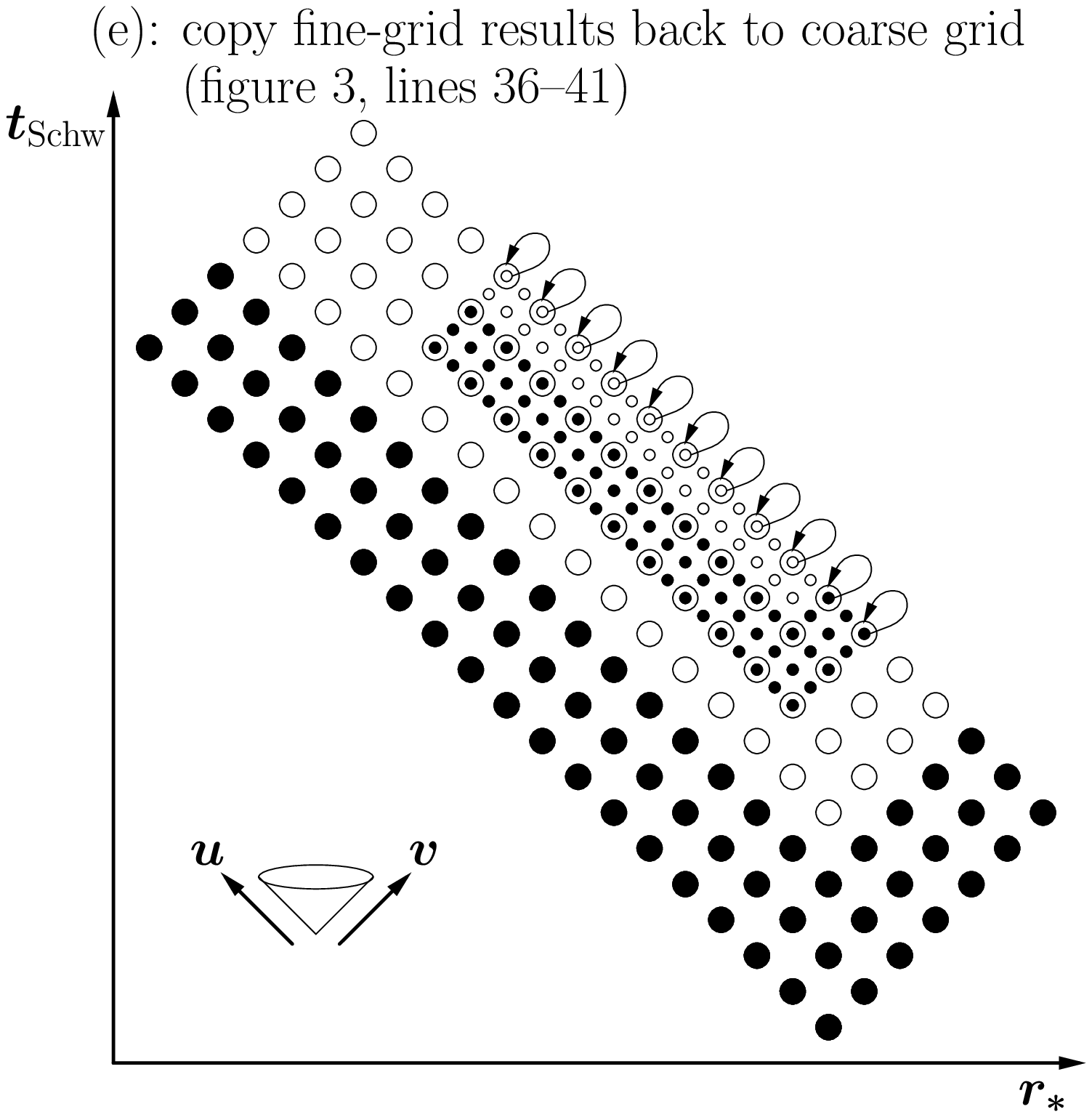}}
\put(62,-3){\includegraphics[scale=0.380]{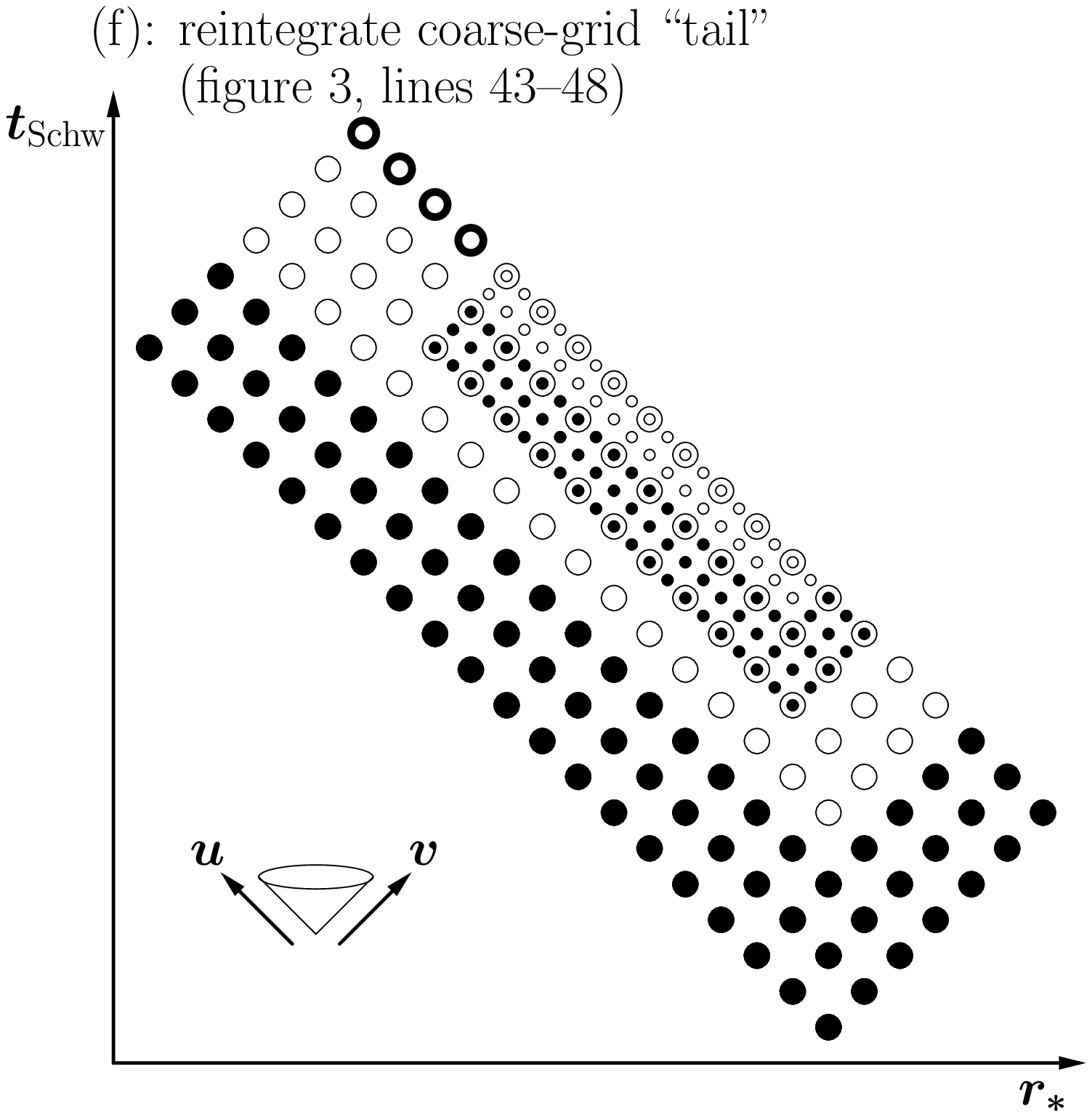}}
\end{picture}
\end{center}
\caption[Example of the Slice-Recursion Algorithm]
	{
	This figure shows an example of the operation of the
	slice-recursion algorithm, using the globally--4th-order
	finite differencing scheme described in
	section~\ref{sect-unigrid-FD}.  Parts~(a)--(e) can also
	be interpreted as an example of the operation of the
	(Cauchy) Berger-Oliger algorithm, with $u$ as the spatial
	coordinate and $v$ as the time coordinate.
	See the main text for further discussion.
	}
\label{fig-slice-recursion-example}
\end{figure}

A key part of the Berger-Oliger algorithm is regridding
(lines~23--28 in figure~\ref{fig-Berger-Oliger-algorithm/main}),
where the grid hierarchy is updated so that each $\grid{\ell}$
covers the desired spatial region for the current time.  As shown
in more detail in figure~\ref{fig-Berger-Oliger-algorithm/regrid},
if this requires adding a new $\grid{\ell}$ to the hierarchy, or
moving an existing $\grid{\ell}$ to cover a different set of
spatial positions than it previously covered, then data must be
interpolated from coarser refinement levels to initialize the new
fine-grid points.  Finer grids may also need to be updated to maintain
proper nesting.  Because of this, and because of the data copying
discussed below, regridding is moderately expensive, typically costing
$\O(1)$~times as much (at each level of the refinement hierarchy) as
integrating a single time step.  To prevent this cost from dominating
the overall computation, regridding is only done on ``selected'' slices;
in practice a common choice is to regrid on every $k$th~slice at each
level of the refinement hierarchy, for some (constant) parameter
$k \sim 4$.  For similar reasons, the LTE estimate is often only
computed at every $k$th~grid point.
\footnote{
\label{footnote-cost-of-LTE-estimate}
	 The LTE estimate discussed in section~\ref{sect-LTE-estimation}
	 roughly doubles the cost of integrating a single grid point,
	 so estimating the LTE at every $k$th grid point of every
	 $k$th slice adds a fractional overhead of roughly $1/k^2$
	 to the computation.  For $k=4$ (the example shown in
	 figure~\ref{fig-slice-recursion-example}b) this is only
	 about~$6\%$.
	 }

Figure~\ref{fig-slice-recursion-example} shows an example
of the operation of the slice-recursion algorithm discussed in
section~\ref{sect-AMR/slice-recursion-algorithm}.  However,
parts~(a)--(e) of this figure can also be interpreted as an example
of the operation of the (Cauchy) Berger-Oliger algorithm discussed
in this section, with $u$ as the spatial coordinate and $v$ as the
time coordinate.  [The example shown is unrealistic in one way: to
allow the figure to show a relatively small number of grid points
(and thus be at a larger and more legible scale), the figure ignores
the limits on regridding discussed in
section~\ref{sect-AMR/slice-recursion-algorithm/avoiding-undesired-mesh-refinement},
which my code actually enforces.]

Because the globally--4th-order finite differencing scheme illustrated
in figure~\ref{fig-slice-recursion-example} is a 4-level scheme, the
extended initial data for the base grid comprises 3~slices and 3~points
on each succeeding slice.  Figure~\ref{fig-slice-recursion-example}a
shows the base grid just after the computation of the first evolved
point of its first evolved slice (i.e., the first slice which isn't
entirely part of the extended initial data).

Figure~\ref{fig-slice-recursion-example}b shows an example of the LTE
being checked at several points, and (after smoothing) exceeding the
error threshold at one of these.  The regridding procedure thus creates a
new fine grid (lines 18-22 in figure~\ref{fig-Berger-Oliger-algorithm/regrid}).
Figure~\ref{fig-slice-recursion-example}c shows this new fine grid
just after the computation of the first evolved point of its first
evolved slice.  Notice that the actual spatial extent of the newly-created
fine grid is larger than just the set of points where the (smoothed)
LTE exceeds the error threshold, for two reasons:
\begin{itemize}
\item	If the (smoothed) LTE estimate exceeds the error threshold
	at some location on the current slice, then logically we don't
	know which point(s) in the LTE-estimate molecule have inaccurate
	data.  The algorithm thus includes \emph{all} the LTE-estimate
	molecule's points in the region-to-be-refined
	(line~18 of figure~\ref{fig-Berger-Oliger-algorithm/main}).
\item	The algorithm also uses ``buffer zones'' to further enlarge
	the region-to-be-refined in the spatial direction beyond the
	set of points just described (lines~20--21 in
	figure~\ref{fig-Berger-Oliger-algorithm/main}).
	The buffer zones are 2~grid points on each side of this set for
	the example shown in figure~\ref{fig-slice-recursion-example}c.
	The buffer zones are used for two reasons:
	\begin{itemize}
	\item	The buffer zones ensure that moving solution features
		and their finite-difference domains of dependence will
		remain within the refined region -- and thus be
		well-resolved -- throughout the time interval before
		the next regridding operation.
	\item	The buffer zones also help to ensure that if there are
		any finer grids in the grid hierarchy, the finite-difference
		domains of dependence for interpolating the next-finer
		grid's spatial boundary data from the current grid
		(line~9 in figure~\ref{fig-Berger-Oliger-algorithm/main};
		figure~\ref{fig-slice-recursion-example}c,d) will
		avoid regions where the solution is not well-resolved
		by the current grid (and hence the interpolation would
		be inaccurate).
	\end{itemize}
\end{itemize}

When a new fine grid $\grid{\ell{+}1}$ is created, at what time level
should it be placed relative to the next-coarser grid $\grid{\ell}$?
There are a number of possible design choices here, ranging from the
time level of the most recent all-points-below-the-error-threshold
$\grid{\ell}$ LTE-estimate check up to the time level of the
some-points-above-the-error-threshold $\grid{\ell}$ LTE-estimate
check which triggered the creation of the new fine grid.  [Placing
the fine grid at an earlier time level makes the total integration
slightly more expensive, but lessens the use of insufficiently-accurate
coarse-grid data (i.e., $\grid{\ell}$ data whose LTE estimate exceeds
the error threshold) in interpolating the fine grid's initial data.]
As shown in the example of figure~\ref{fig-slice-recursion-example}c,
I have (somewhat arbitrarily) chosen to place the newly-created fine
grid $\grid{\ell{+}1}$ with the last (most-future) of its initial-data
slices (those which are entirely interpolated from the next-coarser grid
$\grid{\ell}$; line~20 of figure~\ref{fig-Berger-Oliger-algorithm/regrid})
at the time level $\grid{\ell}_{\var{current\_j} \!-\! 1}$, one
coarse-grid time step before the time level on which the over-threshold
LTE estimates were computed.

Each slice $\grid{\ell}_j$ is a standard 1-dimensional grid function
or set of grid functions, and so may be stored as a contiguous array
or set of arrays in memory.  This is easy to program, and allows the
basic time integration
(lines~11--19 in figure~\ref{fig-Berger-Oliger-algorithm/main})
to be highly efficient.
\footnote{
	 Because the algorithm primarily sweeps sequentially
	 through contiguously-stored grid functions, it should
	 have fairly low cache miss rates.  Moreover, many modern
	 computer systems have special (compiler and/or hardware)
	 support for automatically prefetching soon-to-be-used
	 memory locations in code of this type, further reducing
	 the average memory-access time and thus increasing
	 performance.
	 }
$^,$
\footnote{
	 In fact, with appropriate software design the basic
	 integration routine can often be reused intact, or
	 almost intact, from an existing unigrid code.  For
	 example, this is common when using the \program{Carpet}
	   (\citet{Schnetter-etal-03b,Schnetter-2001:Carpet-code}),
	 \program{PARAMESH}
	   (\citet{MacNeice00,Olson-MacNeice-2005:PARAMESH,
Olson-2006:PARAMESH,Olson-MacNeice-1999:PARAMESH-code}),
	 and \program{SAMRAI}
	   (\citet{Hornung-Kohn-2002:SAMRAI-design,
Hornung-Wissink-Kohn-2006:SAMRAI-design,
Hornung-Kohn-2002:SAMRAI-code})
	 mesh-refinement infrastructures.
	 }
{}  It also means that the amount of additional ``bookkeeping''
information required to organize the computation is very small,
proportional only to the maximum number of distinct grids at any
time.  The one significant disadvantage of contiguous storage is
that when regridding requires expanding $\grid{\ell}_j$ then the
existing data must be copied to a new (larger) set of arrays.
However, the cost of doing this is still relatively small (less
than the cost of a single time step for all the grid points involved).

After the recursive calls to integrate finer grids
(lines~33 and~34 in figure~\ref{fig-Berger-Oliger-algorithm/main}),
$\grid{\ell{+}1}$ has been integrated to the same time level as
$\grid{\ell}$.  Figure~\ref{fig-slice-recursion-example}d shows an
example of this.  Since $\grid{\ell{+}1}$ has twice as fine a resolution
as $\grid{\ell}$, it presumably represents the solution more accurately
at those events common to both grids.  To prevent the coarser grids
from gradually becoming more and more inaccurate as the integration
proceeds (which would eventually contaminate the finer grids via
coarse-to-fine interpolations in regridding), the algorithm copies
(``injects'') $\grid{\ell{+}1}$ back to $\grid{\ell}$ at those
events common to both grids.  This is done at
lines~36--41 in figure~\ref{fig-Berger-Oliger-algorithm/main};
figure~\ref{fig-slice-recursion-example}e shows an example of this.

\subsection{The Pretorius-Lehner Algorithm}
\label{sect-AMR/Pretorius-Lehner-algorithm}

\citet{Pretorius:2003wc} discuss modifications to the standard
Berger-Oliger algorithm to accomodate characteristic evolution.
Their algorithm treats the two null directions symmetrically,
and instead of using a separate regridding step, interleaves the
integration, injection, LTE estimation, and updating of the mesh-refinement
hierarchy at a very fine granularity (essentially that of individual
diamond cells).  This gives an elegant algorithm where the integration
can proceed simultaneously in both null directions, ``flowing''
across the problem domain in a way that's generally not known in
advance.

Because of this unpredictable flow, Pretorius and Lehner don't
use contiguous arrays to store the grid functions.  Rather, they
use a fine-grained linked-list data structure, where each grid point
at each refinement level~$\ell$ stores explicit pointers to its
4~(null) neighboring points at that refinement level, and also to
the grid points at that same event at refinement levels $\ell{\pm}1$.
The Pretorius-Lehner algorithm explicitly walks these pointer chains
to locate neighboring points for the (unigrid) integration at each
refinement level, to create finer grids, and to inject fine-grid
results back into coarser grids at each level of the refinement
hierarchy.

In comparison to the contiguous storage possible with the standard
(Cauchy) Berger-Oliger algorithm, this linked-list storage allocation
avoids data copying when grids must be grown.  However, the per--grid-point
pointers require extra storage, and following the pointer chains adds
some programming complexity and extra execution time.
\footnote{
	 The largest execution-time cost is probably that due
	 to (nearby) grid points which are accessed in succession
	 not being in contiguous, or even nearby, memory locations.
	 This leads to increased cache miss rates and thus poorer
	 performance.
	 }
$^,$
\footnote{
	 The execution time of the dynamic storage allocation
	 routines themselves (C's \texttt{malloc()} and \texttt{free()},
	 or other languages' equivalents) may also be substantial.
	 However, this can be greatly reduced by using customized
	 storage-allocation routines that allocate grid points
	 in large batches.  These and other optimization techniques
	 for dynamic storage allocation are discussed by, for example,
\citet{Wilson-etal-1995:dynamic-storage-allocation-review,Lea-malloc-web-page}.
	 }


\subsection{The Slice-Recursion Algorithm}
\label{sect-AMR/slice-recursion-algorithm}

Here I describe a different variant of the Cauchy Berger-Oliger
algorithm for characteristic evolution.  The basic concept of this
algorithm is to treat one null direction ($v$) as a ``time'' and the
other ($u$) as a ``space'', then apply the standard Berger-Oliger
algorithm as discussed in section~\ref{sect-AMR/Berger-Oliger-algorithm}
(with one significant modification discussed below).
Figure~\ref{fig-Berger-Oliger-algorithm/main} (now including the
lines marked with~\marker{}) gives a pseudocode outline of this
``slice-recursion'' algorithm, and figure~\ref{fig-slice-recursion-example}
shows an example of the algorithm's operation for the globally--4th-order
finite differencing scheme described in section~\ref{sect-unigrid-FD}.

For a Cauchy evolution, the future light cone of a grid point contains
only $\O(1)$~grid points on the next $t=\text{constant}$ slice.  In
contrast, for a characteristic evolution, the future light cone of a
grid point $(j_*,i_*)$ contains all points $(j,i)$ with $j \ge j_*$
and $i \ge i_*$.  To see the impact of this difference on the
Berger-Oliger algorithm, suppose that on some $v = \text{constant}$
slice we have a coarser grid $\grid{\ell}$ overlaid by a finer grid
$\grid{\ell{+}1}$ covering the coarse-grid coordinate region
$i \in [i_1, i_2]$.  Then the injection of the fine-grid results
back to the coarse grid
(lines~36--41 in figure~\ref{fig-Berger-Oliger-algorithm/main};
figure~\ref{fig-slice-recursion-example}e)
restores the coarse-grid solution to the fine-grid accuracy for
$i \in [i_1, i_2]$.  However, unlike in the Cauchy case, the
coarse-grid solution for the slice ``tail'' $i > i_2$ remains
inaccurate, because its computation was affected by the (inaccurate)
pre-injection coarse-grid region $i \in [i_1, i_2]$.  The solution
to this problem is to re-integrate the ``tail'' $i > i_2$ of the
slice after the injection
(lines~43--48 of figure~\ref{fig-Berger-Oliger-algorithm/main};
figure~\ref{fig-slice-recursion-example}f).

Depending on the placement of the fine grid relative to the coarse
grid, the cost of the re-integration may vary from negligible up to
roughly the cost of a single time step for the coarse grid (i.e., a
factor of~2 increase in the cost of the coarse-grid part of the computation).
This overhead only affects grids with $0 \le \ell < \ell_{\max}$; the
finest grid ($\ell = \ell_{\max}$) never needs to be re-integrated.
In practice the re-integration overhad is generally modest; I present
numerical test results quantifying this in section~\ref{sect-numerical-tests}.

The slice-recursion algorithm I present here is quite similar to
that outlined by \citet{Hamade-Stewart-1996}.  Their algorithm
shares the basic Berger-Oliger mesh-hierarchy structure,
uses the same LTE estimator (section~\ref{sect-LTE-estimation}),
imposes the same nesting requirements on the grid hierarchy,
and does the same ``tail'' re-integration
(lines~43--48 of figure~\ref{fig-Berger-Oliger-algorithm/main};
figure~\ref{fig-slice-recursion-example}f).
However, their algorithm uses a $4\!:\!1$~refinement ratio between
adjacent levels in the mesh-refinement hierarchy, whereas I use a
$2\!:\!1$ ratio in the slice-recursion algorithm for the reasons
outlined in footnote~\ref{footnote-choice-of-refinement-ratio}.
They discuss only globally--2nd-order finite differencing.

When the grid hierarchy contains 3~or more refinement levels, the
evolution and regridding schemes described by \citet{Hamade-Stewart-1996}
are somewhat different than those presented here:  Their algorithm
integrates child grids at \emph{all} levels of the grid hierarchy up
to the same time level before doing any fine-to-coarse-grid injections
(lines~36--41 of figure~\ref{fig-Berger-Oliger-algorithm/main};
figure~\ref{fig-slice-recursion-example}e) or regridding
(figure~\ref{fig-Berger-Oliger-algorithm/regrid}), whereas the
algorithms presented here follow the standard Berger-Oliger pattern
where injections and regridding are interleaved with the evolution
of different refinement levels in an order corresponding to the
depth-first traversal of a complete binary tree.

Unlike the algorithm of \citet{Hamade-Stewart-1996}, the slice-recursion
algorithm presented here is purely recursive, treating all levels of
the refinement hierarchy in exactly the same way except for the setup
of the base grid's extended initial data and the details of how the
spatial boundary data are determined at the start of each slice's
integration (lines~6--9 of figure~\ref{fig-Berger-Oliger-algorithm/main}).
An important consequence of the algorithm being organized this way
is that for any integer $k \ge 1$, the algorithm's computations on
a grid hierarchy with $k$~refinement levels are identical (apart from
the initial-data setup just noted) to those in each of the recursive
calls (lines~33 and~34 of figure~\ref{fig-Berger-Oliger-algorithm/main})
for a problem with $k{+}1$~refinement levels.  More generally, the
treatment of the $k$~finest refinement levels in the grid hierarchy
is independent of the presence of any coarser level(s).  I find that
this simplifies debugging, by making the algorithm's behavior on
small test problems with only a few refinement levels very similar
to its behavior on large ``physics'' problems with many refinement
levels.

Like any Berger-Oliger algorithm, the slice-recursion algorithm
needs to interpolate data from coarse to fine grids
(line~9 of figure~\ref{fig-Berger-Oliger-algorithm/main} and
lines~20 and~25--26 of figure~\ref{fig-Berger-Oliger-algorithm/regrid},
figure~\ref{fig-slice-recursion-example}c).
I use a mixture of 1-dimensional and 2-dimensional Lagrange polynomial
interpolation for this, in all cases chosen so as to avoid crossing
the particle worldline.  Appendix~\ref{app-implementation/interp-ops}
describes the interpolation operators in detail.  These operators
may use data from up to 4~[6]~adjacent slices for the 2nd~[4th]~order
GTE schemes, which sets a lower bound on the number of slices of each
$\grid{\ell}$ which must be kept in memory.  For 2nd~order GTE, my
code keeps only the minimum~(4) number of slices in memory; for
4th~order GTE, it keeps one extra slice in memory (for a total of~7)
to accomodate the LTE estimator (section~\ref{sect-LTE-estimation};
figure~\ref{fig-slice-recursion-example}b,c).


\subsubsection{Avoiding Undesired Mesh Refinement}
\label{sect-AMR/slice-recursion-algorithm/avoiding-undesired-mesh-refinement}

When adding a new refinement level to the mesh-refinement hierarchy,
the interpolation of initial data
(line~20 of figure~\ref{fig-Berger-Oliger-algorithm/regrid})
tends to introduce low-level noise into the field variables.  The
same is true when an existing fine grid is moved to a new position.
In either case, this noise can cause the LTE estimate to be inaccurate.
It's thus useful to allow this noise to decay (i.e., be damped out
by the inherent dissipation in the finite differencing scheme) before
using the LTE estimate to determine the placement of another new
refinement level.  To this end, my code suppresses regridding
operations for the first 8~[16]~time steps of a new or newly-moved
grid, for the 2nd~[4th]~order GTE finite differencing scheme respectively.

My code also suppresses creating a new fine grid if insufficient data
is available for the interpolation of all 4~[7]~slices kept in memory
for the 2nd~[4th]~order GTE finite differencing scheme respectively.

When using the slice-recursion algorithm for the self-force computation,
the arbitrary initial data on the southwest and southeast grid faces
induces spurious radiation near these grid faces.  This radiation is
of no physical interest, so there's no need for the mesh-refinement
algorithm to resolve it.  Moreover, as discussed in
footnote~\ref{footnote-each-grid-has-only-a-single-component}, not
resolving the spurious radiation also allows a significant
simplification of the code's data structures.  Thus my code suppresses
mesh refinement for the first $\sim 100M$ of the integration, and for
the first $\sim 100M$ of each slice thereafter.

To reduce the effects of the interpolation noise when adding new
refinement levels, my code also turns on the mesh refinement gradually,
adding new refinement levels only at the rate of one each $10M$ of
evolution.  More precisely, the code limits the maximum refinement
level to
\begin{equation}
\ell_{\max}
	\le \left\{
	    \begin{array}{ll}
	    0				& \text{if $\rvu < 100M$}	\\
	    \displaystyle
	    \left\lceil \frac{\rvu - 100M}{10M} \right\rceil
					& \text{if $\rvu \ge 100M$}	
	    \end{array}
	    \right.
								      \,\text{,}
\end{equation}
where $\rvu = \min(\rv,\ru)$ is the distance from the closest point
on the southeast or southwest grid faces.


\section{Numerical Tests of the AMR Algorithm}
\label{sect-numerical-tests}

As a test case for the slice-recursion AMR algorithm, I consider a
particular example of the model problem of section~\ref{sect-model-problem},
which arises in the course of calculating the radiation-reaction
``self-force'' on a scalar particle orbiting a Schwarzschild black
hole (\citet{Barack-Ori-2002}, see \citet{Barack-2009:self-force-review}
for a general review).  I take $\phi$ to be a complex scalar field,
with the potential $V(r)$ and source term $S(t_\Schw)$ given by
\begin{eqnarray}
V_\ell(r)
	& = &	\frac{f(r)}{4}
		\left( \frac{2M}{r^3} + \frac{\ell(\ell+1)}{r^2} \right)
							    \label{eqn-V_ell(r)}
									\\
S_{\ell m}(t_\Schw)
	& = &	\frac{\pi q f^2(r_p) a_{\ell m}}{r_p E(r_p)}
		\exp \bigl(-i m \omega(r_p) t_\Schw \bigr)
								      \,\text{,}
\end{eqnarray}
where
\begin{equation}
\omega(r) = \sqrt{\frac{M}{r^3}}
\end{equation}
and
\begin{equation}
E(r) = f(r) \left(1 - \frac{3M}{r}\right)^{-1/2}
								      \,\text{.}
\end{equation}
are respectively the orbital frequency and energy per unit mass of
a particle in circular orbit at areal radius~$r$ in Schwarzschild
spacetime.  The coefficients $a_{\ell m}$ are defined such that the
spherical harmonic
$Y_{\ell m}(\theta{=}\tfrac{\pi}{2}, \varphi) = a_{\ell m} e^{im\varphi}$,
i.e.,
\begin{equation}
a_{\ell m} = \left\{
	     \begin{array}{l@{\quad}l}
	     \displaystyle
	     (-1)^{(\ell{+}m)/2}
	     \sqrt{\frac{2\ell+1}{4\pi}}
	     \sqrt{\frac{(\ell+m-1)!! \, (\ell-m-1)!!}
			{(\ell+m)!! \, (\ell-m)!!}}
				& \text{if $\ell{-}m$ is even}	\\
	     \displaystyle
	     0			& \text{if $\ell{-}m$ is odd}	
	     \end{array}
	     \right.
								      \,\text{,}
\end{equation}
where the ``double factorial'' function is defined by
\begin{equation}
n!! = \left\{
      \begin{array}{ll}
      n \cdot (n-2)!!	& \text{if $n \ge 2$}		\\
      1			& \text{if $n \le 1$}		
      \end{array}
      \right.
								      \,\text{.}
\end{equation}

For this test, I take $\ell = m = 10$, $r_p = 10M$, and use a problem
domain size of $D = 200M$ on a side.  Here $v_{\min} = 12.773M$,
$u_{\min} = -12.773M$, and the grid extends from
$\rv = 0M$ to $\rv=200M$ and from $\ru = 0M$ to $ru = 200M$.

The physical boundary data is zero along the southwest and southeast
grid faces, and the extended initial data is computed using the 2-level
subsampling scheme described in
appendix~\ref{app-FD-details/4th-order/initial-data/two-level-subsampling}.
The base grid has a resolution of $0.25M$, the finite differencing
is the globally-4th-order scheme, and the error tolerance
for the LTE estimate is $10^{-16}$.  As discussed in
section~\ref{sect-AMR/slice-recursion-algorithm/avoiding-undesired-mesh-refinement},
mesh refinement is suppressed for the first $100M$ of the evolution
and the first $100M$ of each slice thereafter, and then turned on
gradually at the rate of one refinement level for each further $10M$
of evolution.

\begin{figure}[!bp]
\begin{center}
\includegraphics[width=125mm]{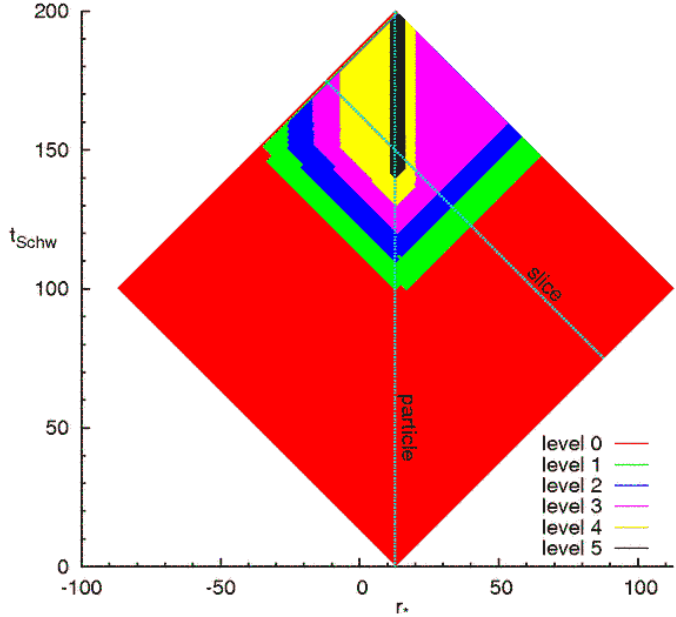}
\end{center}
\caption[Mesh-Refinement Map]
	{
	This figure shows a ``map'' of the mesh refinement,
	giving the highest refinement level at each event in the
	problem domain.  The vertical dashed line labelled
	``particle'' shows the particle worldline.  The
	diagonal dashed line labelled ``slice'' shows the
	$\rv=150M$ slice; figure~\ref{fig-convergence}
	shows convergence tests on that slice.
	}
\label{fig-reflevel-map}
\end{figure}

Figure~\ref{fig-reflevel-map} shows a ``map'' of the mesh refinement,
giving the highest refinement level at each event in the problem domain.
Notice that the highest-refinement grids only cover small regions close
to the particle worldline.

\begin{figure}[!bp]
\begin{center}
\begin{picture}(120,160)
%
\put(-2,110){\includegraphics[scale=0.50]{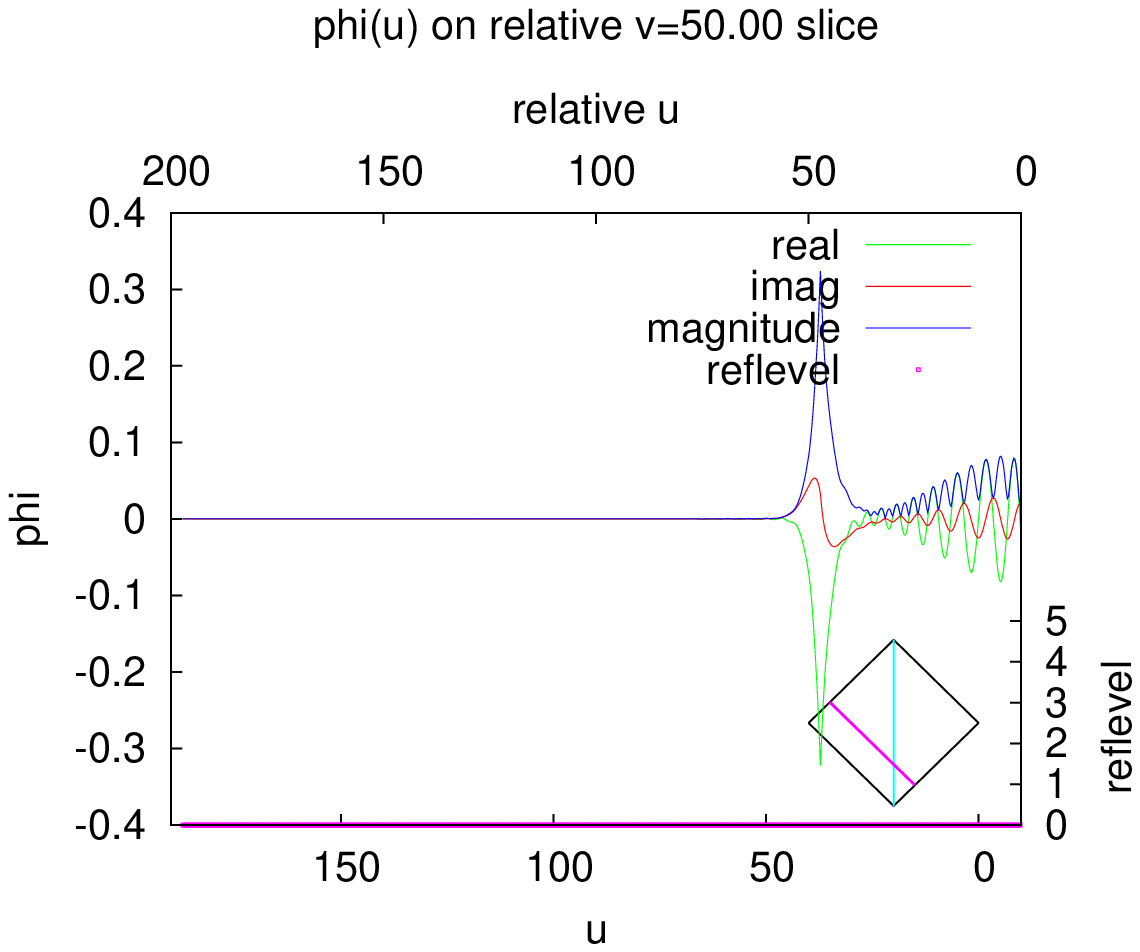}}
\put(58,110){\includegraphics[scale=0.50]{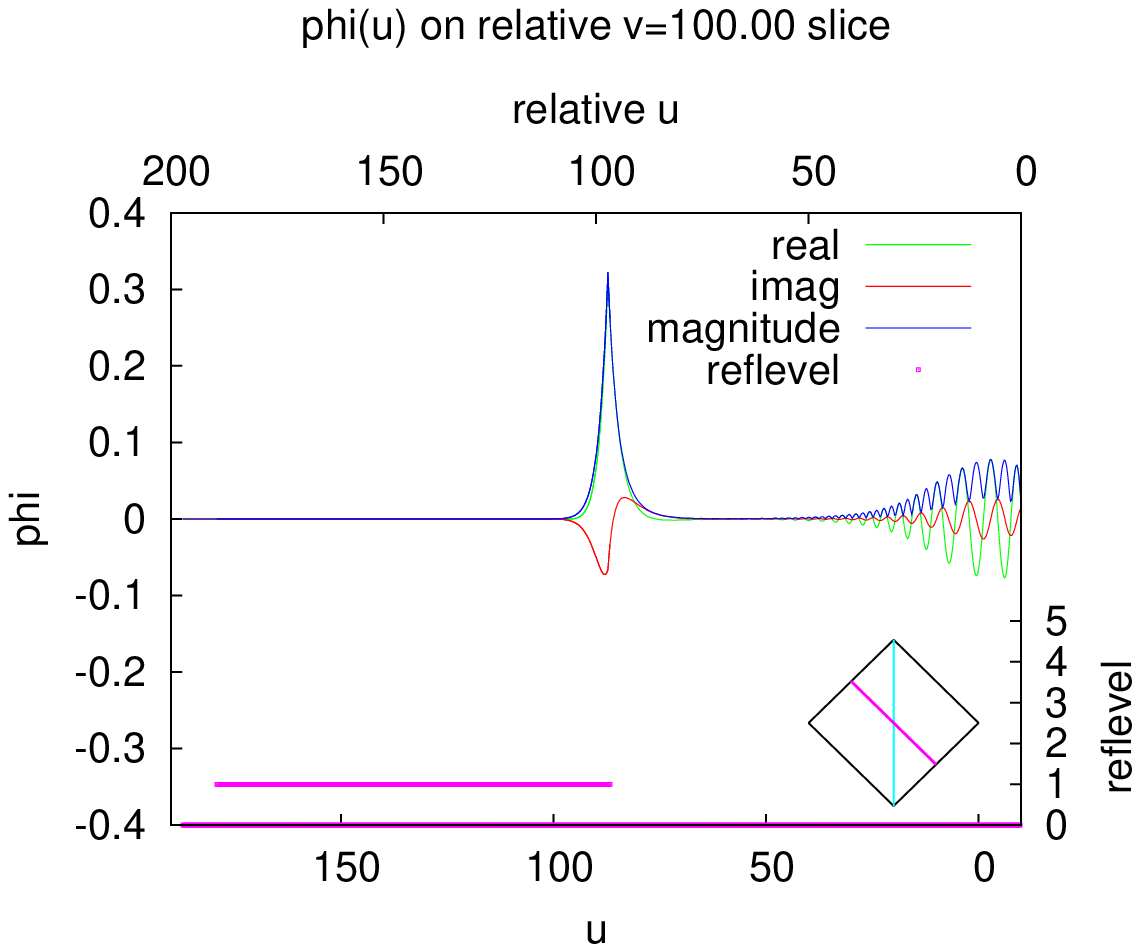}}
\put(-2,55){\includegraphics[scale=0.50]{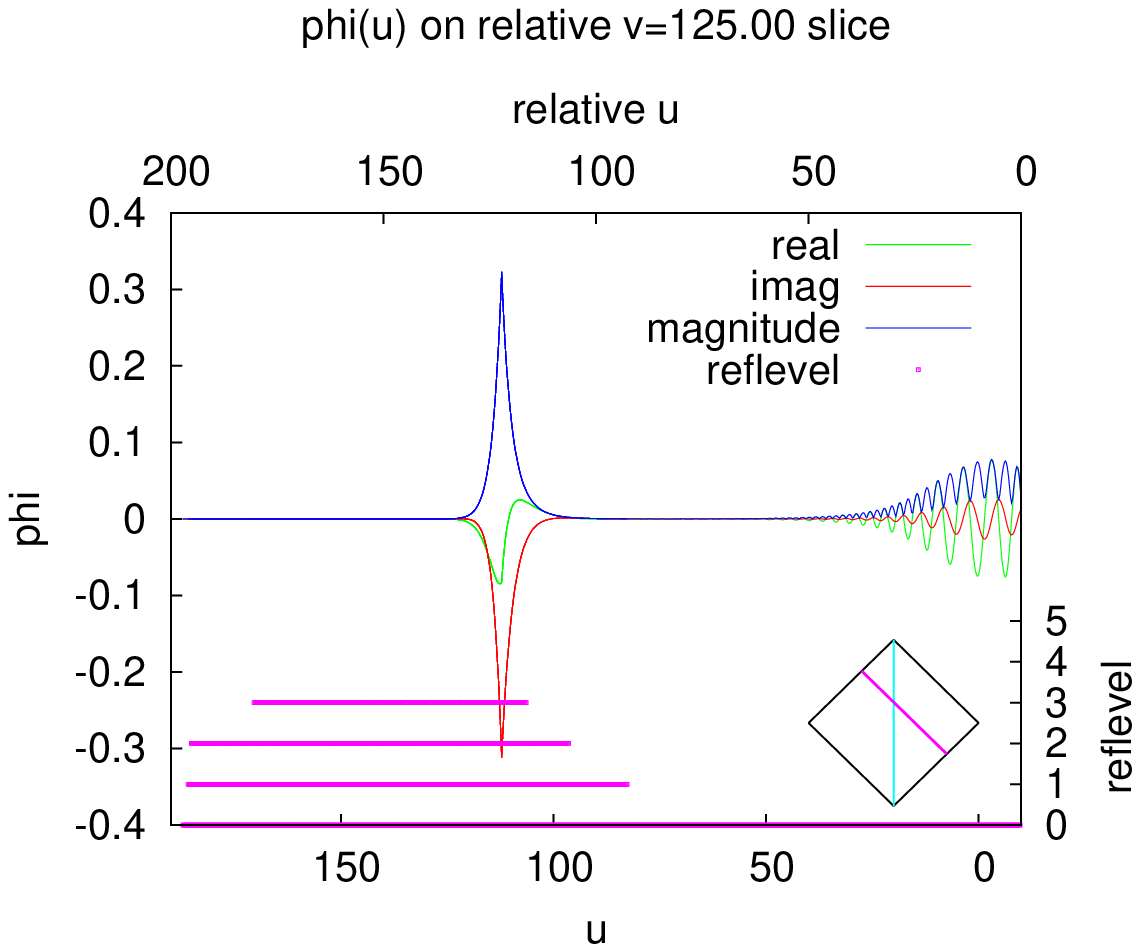}}
\put(58,55){\includegraphics[scale=0.50]{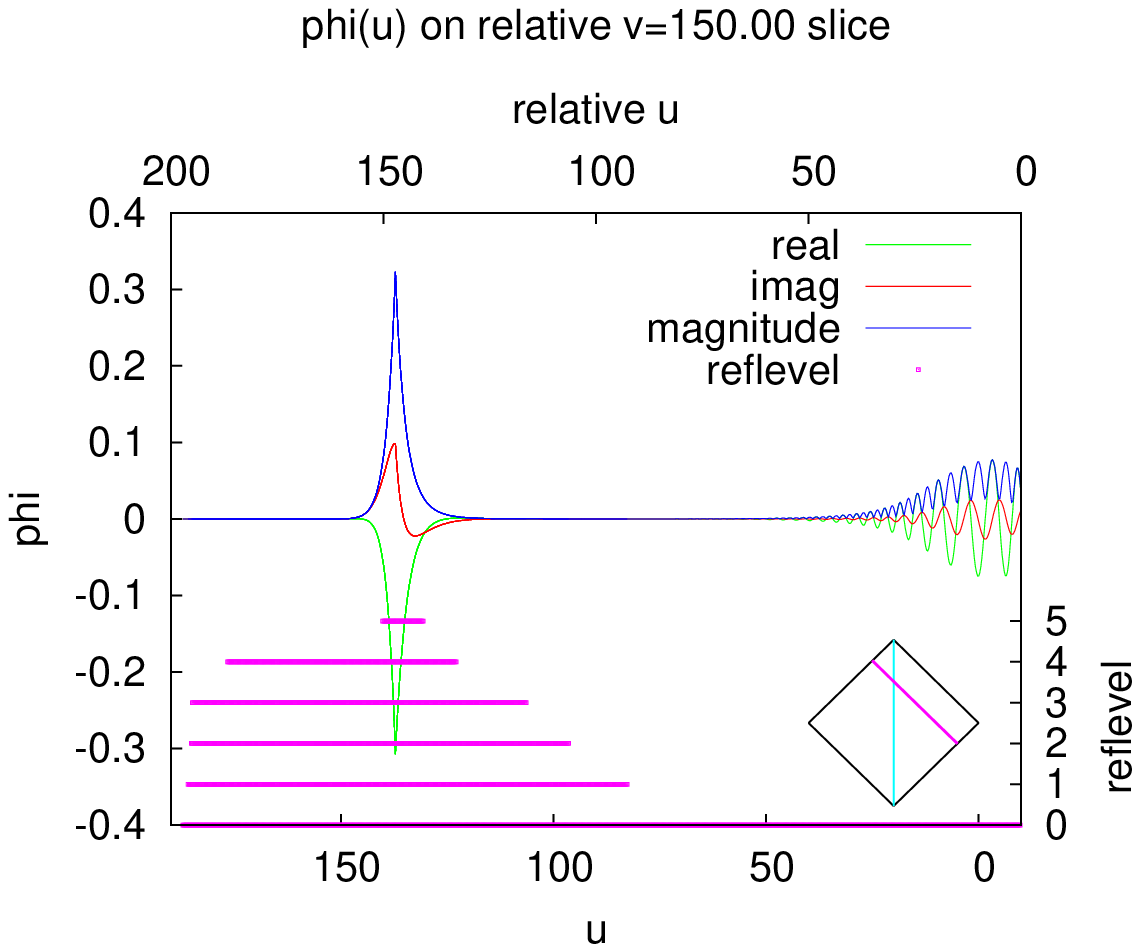}}
\put(-2,0){\includegraphics[scale=0.50]{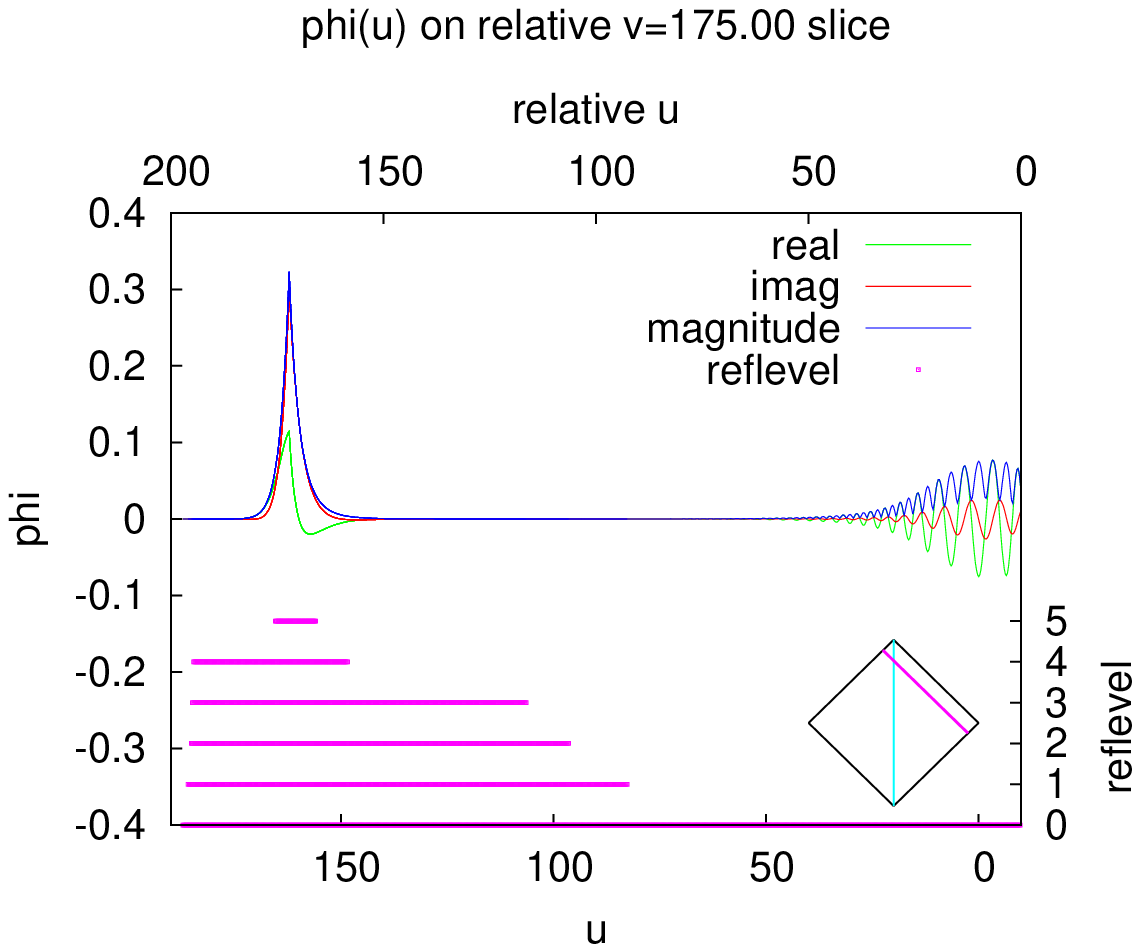}}
\put(58,0){\includegraphics[scale=0.50]{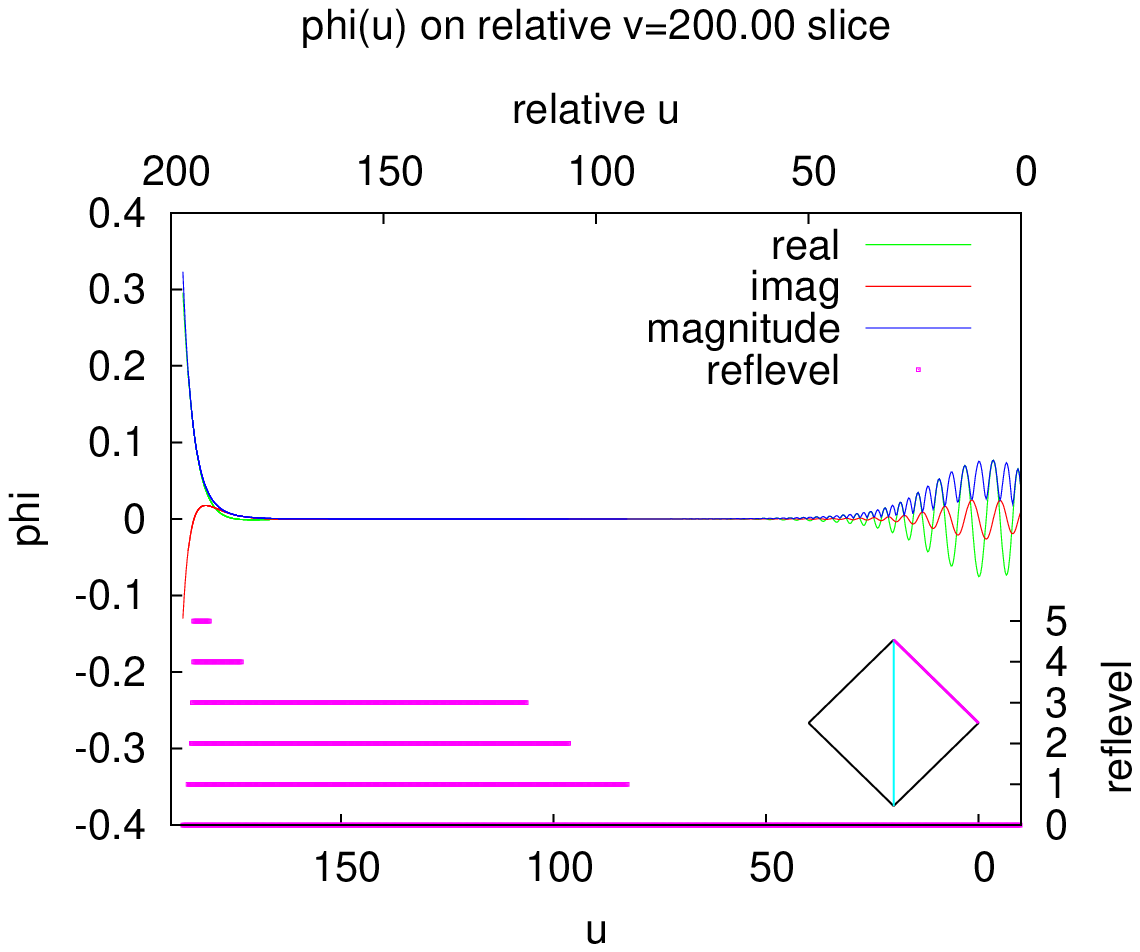}}
\end{picture}
\end{center}
\caption[Sample Frames from Movie]
	{
	This figure shows sample frames from the movie accompanying
	this paper (online resource~1).  Each frame of the movie shows
	a single $\rv = \text{constant}$ slice, and plots
	$\Realpart[\phi]$, $\Imagpart[\phi]$, and $\|\phi\|$ on the
	left scale, and the spatial extent of each refinement level
	(on the right scale) as the horizontal lines.  The $u$, $\ru$,
	and $\rv$~coordinates are all shown in units of the Schwarzschild
	spacetime mass~$M$.  Each frame of the movie also includes a
	subplot (in the lower right corner) showing the location of that
	plot's $\rv = \text{constant}$ slice within the entire problem
	domain.  The vertical line in the subplot shows the particle
	worldline.
	}
\label{fig-movie-sample-frames}
\end{figure}

The online supplemental materials which accompany this paper include
a movie (online resource~1) showing the spacetime-dependence of $\phi$
and the placement of the refined grids; figure~\ref{fig-movie-sample-frames}
shows several sample frames and explains them in more detail.

Because of the adaptive placement of refined grids, it's difficult to
do a standard convergence test (\citet{Choptuik-1991:FD-consistency}).
Instead, I have followed \citet{Choptuik-in-d'Inverno:self-similarity-and-AMR}
in adding an option to my code to write out a ``script'' of the position
and grid spacing of each grid generated by the AMR algorithm, and a
related option to ``play back'' such a script.  For a convergence
test, I first run a ``record'' evolution, then generate a
``$\text{playback}{\times}N$'' script by refining each grid in
the script by a chosen (small integer) factor $N$, and finally
``play  back'' the refined script.

\begin{figure}[!bp]
\begin{center}
\includegraphics[width=125mm]{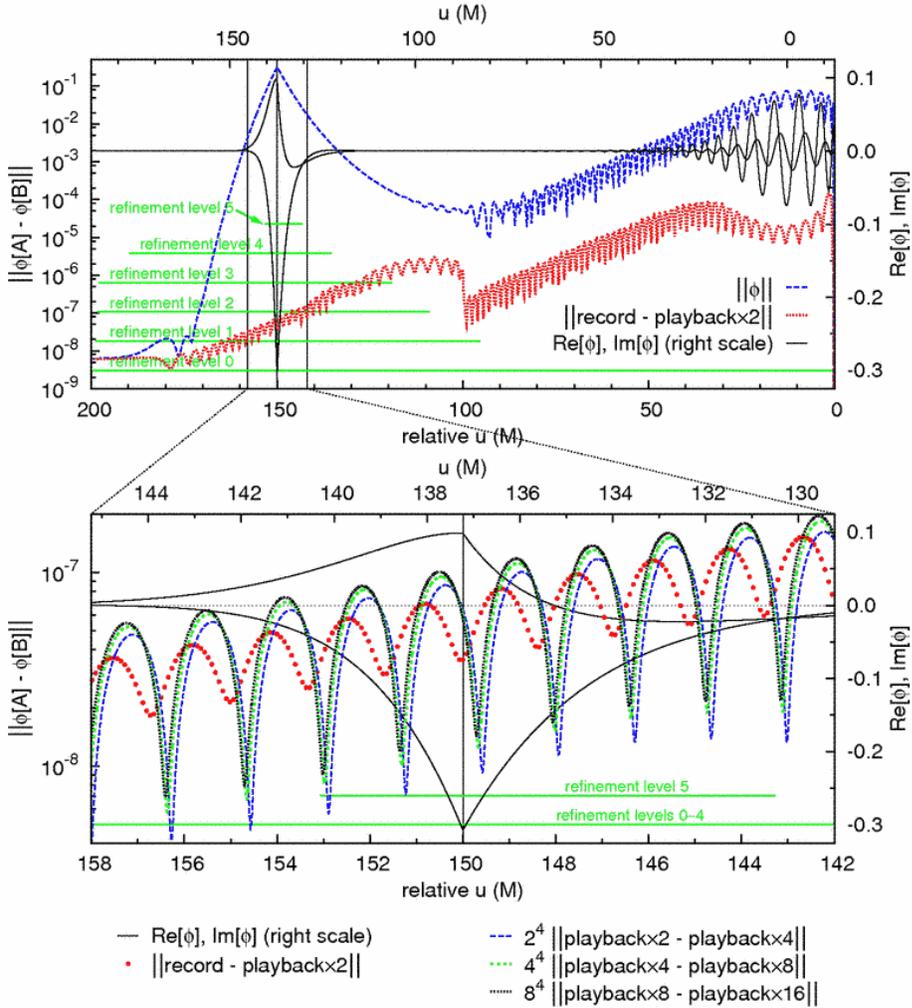}
\end{center}
\caption[Convergence at $\rv=150M$]
	{
	This figure shows $\phi$ and its convergence on the
	$\rv=150M$ slice.  In each subfigure, the solid vertical
	line at $\ru=150M$ shows the particle position.  The upper
	subfigure (which shows the entire slice) plots the real and
	imaginary parts of $\phi$ on the right (linear) scale, and
	the complex norm $\|\phi\|$ on the left (logarithmic) scale.
	It also plots one of the convergence differences
	$\|\text{record} - \text{playback}{\times}2\|$
	on the left (logarithmic) scale.  The horizontal lines show
	the portion of this slice covered by each refinement level.
	The lower subfigure (which shows an expanded view of
	a $\pm 8M$ region centered on the particle position)
	plots the real and imaginary parts of $\phi$ on the right
	(linear) scale, and the four convergence differences
	(scaled by the 4th~power of the resolution) on the left
	(logarithmic) scale.  Notice that the three higher-resolution
	scaled-convergence-difference curves are almost superimposed,
	indicating excellent 4th-order convergence, and that this
	convergence is not degraded either across the mesh-refinement
	boundaries at~$\rv = 143.3M$ and~$\rv = 153.1M$, or near
	the particle at $\rv = 150M$.
	}
\label{fig-convergence}
\end{figure}

Figure~\ref{fig-convergence} shows an example of such a convergence
test for the field $\phi$ on the $\rv = 150M$ slice.  The code shows
excellent 4th~order convergence in the interior of each grid, across
mesh-refinement boundaries, and near to the particle.

Profiling the code shows that almost all of the CPU time is consumed
in the basic diamond-cell integration code, and the code's overall
running time is closely proportional to the number of diamond cells
integrated.  In other words, the AMR bookkeeping consumes only a
negligible fraction of the CPU time.

For this evolution, the AMR evolution integrates a total of
$35.6 \,{\times}\, 10^6$~diamond cells.  Of these, $16\%$ are accounted
for by the re-integration of coarse slices after fine-grid recursion
(lines~43--48 of figure~\ref{fig-Berger-Oliger-algorithm/main}).
A hypothetical unigrid evolution covering the entire problem domain
at the resolution of the finest AMR grid (level~5) would require
integrating $655 \,{\times} 10^6$~diamond cells, $18.4$~times as many
as the AMR evolution.  That is, for this problem the AMR evolution
was approximately a factor of $18$~faster than an equivalent-resolution
(and thus equivalent-accuracy) unigrid evolution.

\begin{figure}[bp!]
\begin{center}
\begin{picture}(120,45)
\put(-3,0){\includegraphics[scale=0.50]{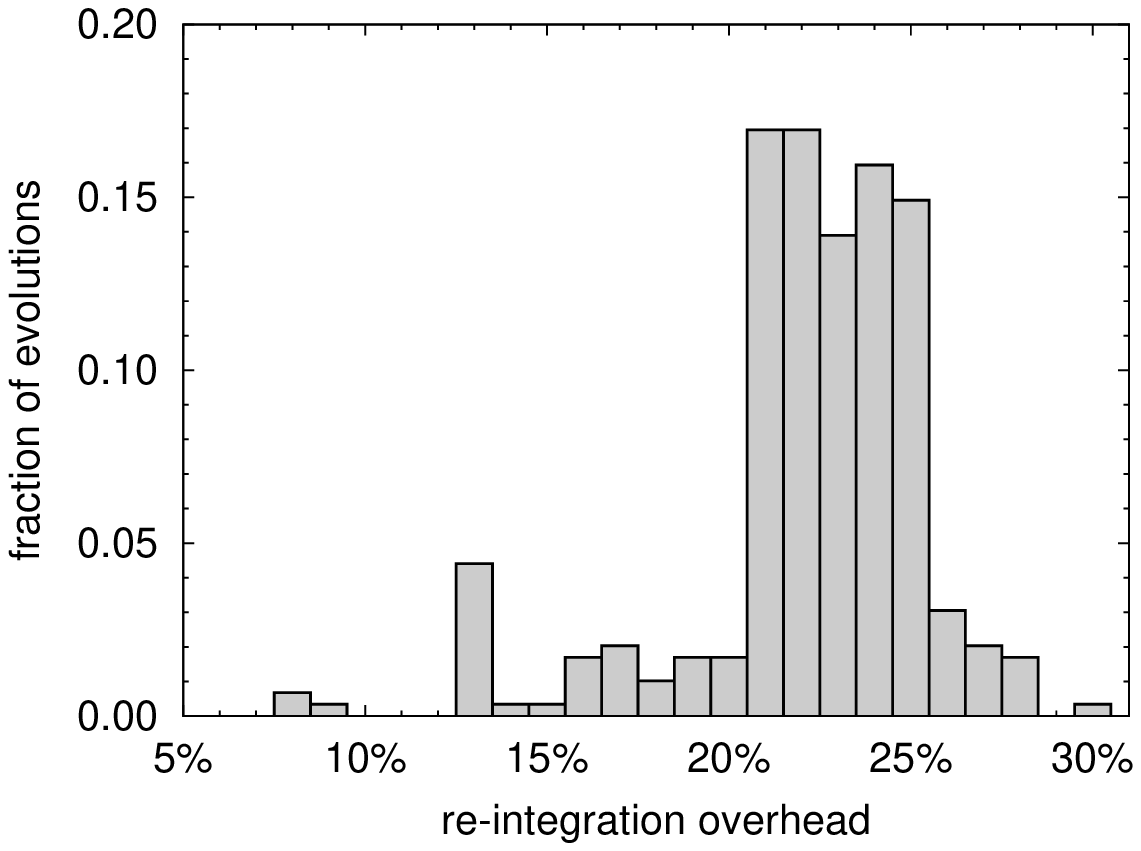}}
\put(58,0){\includegraphics[scale=0.50]{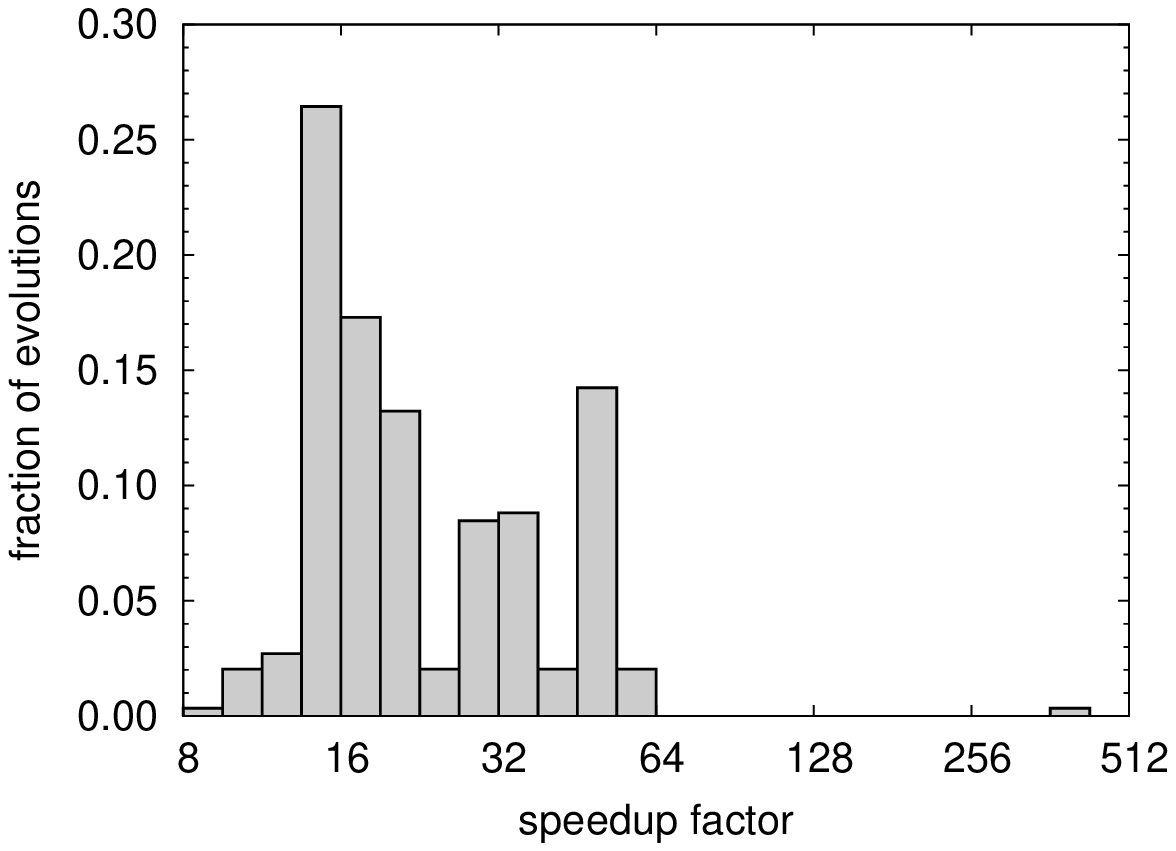}}
\end{picture}
\end{center}
\caption[Histograms of Performance Statistics]
	{
	This figure shows histograms of the re-integration overhead
	(the fraction of all diamond-cell integrations which occur as
	part of coarse-grid re-integrations, i.e., those in
	lines~43--48 of figure~\ref{fig-Berger-Oliger-algorithm/main})
	and of the AMR speedup factor (the ratio of the number of cells
	that would have been integrated in hypothetical unigrid evolution
	covering the entire problem domain at the resolution of the
	finest AMR grid, to the number of cells actually integrated
	in the AMR evolution) for a sample of 295~evolutions.
	The re-integration overhead is usually between $20\%$ and $25\%$,
	and never exceeds $30\%$, while the speedup factor varies
	much more widely.
	}
\label{fig-performance-stats}
\end{figure}

To further characterize the performance of the slice-recursion
algorithm, I consider a sample of 295~separate evolutions of the
same model problem, with each evolution having a different $(\ell,m)$
in the range $0 \le \ell \le 40$ and $0 \le m \le \ell$, and a
problem-domain size between $400M$ and $30\,000M$.
\footnote{
	 This set of evolutions arises naturally as part of
	 a calculation of the radiation-reaction ``self-force''
	 on a scalar particle orbiting a Schwarzschild black
	 hole.  Details of this calculation will be reported
	 elsewhere; for present purposes these evolutions
	 provide a useful set of test cases for the AMR algorithm.
	 }
{}  The globally-4th-order scheme is used for all of these evolutions,
with an error tolerance for the LTE estimate of $10^{-16}$.
Figure~\ref{fig-performance-stats} shows histograms of the
re-integration overhead (the fraction of all diamond-cells
integrations which occur as part of coarse-grid re-integrations)
and the AMR speedup factor for this sample of evolutions.
The re-integration overhead is usually between $20\%$ and $25\%$,
and never exceeds $30\%$.

For this sample the AMR speedup factor varies over a much wider
range, from as low as~$8.9$ to as high as~$400$.  The median speedup
factor is~$19$, and $95\%$ of the speedup factors are between $12$
and~$51$.

\section{Conclusions}
\label{sect-conclusions}

The main result of this paper is that only a small modification
(the tail re-integration, i.e., the lines marked with \marker{} in
figure~\ref{fig-Berger-Oliger-algorithm/main}) is needed to adapt
the standard Cauchy Berger-Oliger AMR algorithm to characteristic
coordinates and grids.  The resulting ``slice-recursion'' algorithm
uses relatively coarse-grained control for the recursion and the
adaptivity part of AMR, which greatly simplifies the memory management,
allowing entire null slices to be stored in contiguous arrays in memory.
The algorithm can readily accomodate any order of finite differencing
scheme; I present schemes for both 2nd and 4th~order global truncation
error.

The numerical tests results presented here demonstrate that the
slice-recursion algorithm is highly efficient and displays excellent
4th~order convergence.  This algorithm readily accomodates a moving
``particle'' Dirac $\delta$-function source term (which leads to a
solution which is generically only $C^1$ at the particle position);
there is no loss of convergence there.

The main potential disadvantage of the slice-recursion algorithm over
the standard Cauchy Berger-Oliger algorithm is the cost of the tail
re-integration.  This cost depends on the placement of fine grids
relative to their coarser parent grids; the worst possible case is
a factor of~2 overhead in the coarse-grid part of the integration.
There is no overhead for the finest level of the mesh-refinement
hierarchy.  In practice I find the re-integration cost to be quite
modest, typically about~$20$--$25\%$ of the total computation, and
(in my tests) never exceeding $30\%$.  Compared to an equivalent-resolution
unigrid evolution, I find that the slice-recursion algorithm is
always much faster, with the exact speedup factor varying widely
from one problem to another: for a sample of 295~test cases, $95\%$
of the speedup factors are between $12$ and $51$.

The largest practical obstacle to the use of Berger-Oliger mesh
refinement algorithms, including the slice-recursion algorithm presented
here, is probably their implementation (programming) complexity.
This is much less of an obstacle if codes can be shared across projects
and researchers.  To this end, my code implementing the slice-recursion
algorithm is included in the electronic supplementary materials
accompanying this paper (online resource~2), and is freely available
to other researchers under the terms of the GNU general public license.
This code uses \Cplusplus{} templates to support both 2nd and 4th~order
global truncation error, and both real and complex scalar fields, with
no run-time overhead.  It would be fairly easy to adapt the code to
other finite differencing schemes and/or PDEs.

\citet[sections~4.2 and 4.3]{Pretorius:2003wc} describe how their
characteristic Berger-Oliger algorithm can be extended to problems
of higher dimensionality than $1{+}1$ (i.e., problems whose domains
have multiple spacelike dimensions), and how the algorithm may be
parallelized.  The techniques they describe should all apply equally
to the slice-recursion algorithm presented here.


\begin{acknowledgements}
I thank Leor Barack, Darren Golbourn, and Norichika Sago for
introducing me to the self-force problem, and both they and Ian Hawke
for many valuable conversations.  I thank Virginia J.\ Vitzthum for
useful comments on this manuscript.  I thank the University of Southampton
and the Max-Planck-Institut f\"{u}r Gravitationsphysik for their
generous support during various stages of the research described
in this manuscript.
\end{acknowledgements}


\appendix

\section{Details of the Unigrid Finite Differencing Scheme}
\label{app-FD-details}

In this appendix, I describe the finite differencing schemes in detail
for the model problem~\eqref{eqn-wave}, for both 2nd and 4th~order GTE.

In this appendix only, when describing the computation of $\phi$ at a
particular grid point $(j,i)$, I sometimes use an abbreviated notation
for grid-point indexing, denoting the grid point $(j-n, i-m)$ by a
subscript $nm$.  Such subscripts can be distinguished from the usual
grid-point indices by the absence of a comma between the two indices.
These abbreviated subscripts may be either integral or half-integral,
and in this latter context (only) I also use the abbreviation
$h \eqdef \thalf$.  For example, $\phi_{12}$ denotes $\phi_{j-1,i-2}$,
while $\phi_{0h}$ denotes $\phi_{j,i-1/2}$.


\subsection{Second Order Global Accuracy}
\label{app-FD-details/2nd-order}

To discretize the wave equation~\eqref{eqn-wave} to 2nd~order global
accuracy in the grid spacing~$\Delta$, I use a standard diamond-cell
integration scheme
(\citet{Gomez-Winicour-1992:sssf-2+2-evolution-and-asymptotics,
Gomez-Winicour-Isaacson-1992:sssf-2+2-numerical-methods,
Gundlach-Price-Pullin-1994a,Burko-Ori-1997,Lousto-Price-1997,
Lousto-2005,Winicour-2009:living-review}):
Consider a double-null ``diamond'' grid cell of side $\Delta$, with
``north'', ``west'', ``east'', and ``south'' vertices (grid points)
$\N$, $\W$, $\E$, and $\S$ respectively, and central point $\C$, as
shown in figure~\ref{fig-FD-molecules-details}.

Consider first the vacuum case, where the particle worldline $r = r_p$
doesn't intersect the cell and hence the right hand side of the wave
equation~\eqref{eqn-wave} vanishes everywhere in the cell.  Integrating
this equation over the cell then gives
\begin{equation}
\Bigl( \phi_\N + \phi_\S - \phi_\E - \phi_\W \Bigr)
+ \left( \Delta^2 V_\C \frac{\phi_\E + \phi_\W}{2} + \O(\Delta^4) \right)
	= 0
								      \,\text{,}
\end{equation}
or equivalently
\begin{subequations}
							      \label{eqn-FD-2nd}
\begin{equation}
\phi_\N =
	\phi_\E + \phi_\W - \phi_\S
	- \Delta^2 V_\C \frac{\phi_\E + \phi_\W}{2}
	+ \O(\Delta^4)
								      \,\text{,}
						       \label{eqn-FD-2nd/vacuum}
\end{equation}
where subscripts denote the value of the field at the corresponding
grid point.

If the particle worldline $r = r_p$ intersects the cell, then as
discussed in section~\ref{sect-unigrid-FD}, I assume it does so
symmetrically.  Integrating the right-hand-side source term
in~\eqref{eqn-wave} over the grid cell then adds an extra term
\begin{equation}
\int_\text{cell} S(t_\Schw) \delta(r-r_p) \, dv \,du
	  = \frac{2}{f(r_p)}
	    \int_{t_\C-\Delta/2}^{t_\C+\Delta/2} S(t_\Schw) \, dt_\Schw
						  \label{eqn-FD-2nd/source-term}
\end{equation}
\end{subequations}
to the right hand side of~\eqref{eqn-FD-2nd/vacuum},
where $t_\C \eqdef {(t_\Schw)}_\C$.  For the test case considered
in section~\ref{sect-numerical-tests}, this source-term integral
becomes
\begin{equation}
\frac{2}{f(r_p)}
\int_{t_\C-\Delta/2}^{t_\C+\Delta/2} S_{\ell m}(t_\Schw) \, dt_\Schw
	= \frac{2 \pi q f(r_p) a_{\ell m}}{r_p E(r_p)}
	  \exp \bigl(-i m \omega(r_p) t_\C \bigr)
	  \Delta
	  \sinc \bigl( \thalf m \omega(r_p) \Delta \bigr)
	  +
	  \O(\Delta^3)
								      \,\text{.}
\end{equation}


\subsection{Fourth Order Global Accuracy}
\label{app-FD-details/4th-order}

To discretize the wave equation~\eqref{eqn-wave} to 4th~order global
accuracy in the grid spacing~$\Delta$, I use a scheme based on that of
\citet{Haas-2007} (see also \citet{Lousto-2005}), but modified in its
treatment of cells near the particle.  The modification makes the scheme
fully explicit, removing the need for an iterative computation at each
cell intersecting the particle.  However, the scheme is only valid with
the assumption noted above, that if the particle intersects a cell it
does so symmetrically.  In practice
(cf.~footnote~\ref{footnote-particle-symmetric-in-cell}), this means
that the scheme is only valid for the case where $r_p$ is constant,
i.e., the particle is in a circular orbit around the central black hole.

Consider the computation of $\phi_{j,i}$, and suppose the particle
worldline intersects the $v = v_j$ slice at the grid point $i = i_p$.
There are several different cases for the finite differencing scheme,
depending on the sign and magnitude of $i - i_p$.
Figure~\ref{fig-FD-molecules-details} shows all of these cases.


\subsubsection{$|i - i_p| \ge 3$ (Cell far from the particle)}

If $|i - i_p| \ge 3$ (so that the particle worldline doesn't intersect
any part of the finite difference molecule), then I use the finite
differencing scheme of Haas's equations~(2.7), (2.10), and~(4.10).
That is (following \citet{Lousto-2005} and \citet{Haas-2007}) I first
define the new grid function $G \eqdef V \phi$.  I then interpolate
$G_{hh}$ via Haas's equation~(2.7), which in my notation reads
\begin{equation}
G_{hh} = \frac{1}{16}
	 \Bigl[
	 (8 G_{10} + 8 G_{01})
	 + (- 4 G_{20} + 8 G_{11} - 4 G_{02})
	 + (G_{30} - G_{21} - G_{12} + G_{03})
	 \Bigr]
	 + \O(\Delta^3)
								      \,\text{.}
						  \label{eqn-FD-4th/G_hh-vacuum}
\end{equation}

I then compute $G_\Sigma \eqdef G_{h0} + G_{0h} + G_{1h} + G_{h 1}$ via
Haas's equation~(2.10), which in my notation reads
\begin{eqnarray}
G_\Sigma
	& = &	  2 \Bigl( 1 - \thalf \Delta_2^2 V_{hh} \Bigr) G_{hh}
								\nonumber\\
	&   &
		+ \Bigl( 1 - \thalf \Delta_2^2 V_{h0} \Bigr) V_{h0} \phi_{10}
		+ \Bigl( 1 - \thalf \Delta_2^2 V_{0h} \Bigr) V_{0h} \phi_{01}
								\nonumber\\
	&   &
		+ \thalf
		  \Bigl( V_{h0} - 2 V_{hh} + V_{0h} \Bigr)
		  (\phi_{10} + \phi_{01})
								      \,\text{,}
						      \label{eqn-FD-4th/G_Sigma}
\end{eqnarray}
where $\Delta_2 \eqdef \Delta/2$.

Finally, I compute $\phi_{00}$ via Haas's equation~(4.7), which in my
notation reads
\begin{eqnarray}
\phi_{00}
	& = &	  - \phi_{11}
								\nonumber\\
	&   &
		  + \left[
		    1 - \Bigl(
			\tfrac{1}{4}\Delta_3^2 - \tfrac{1}{16} \Delta_3^4 V_{hh}
			\Bigr) (V_{hh} + V_{10})
		    \right]
		    \phi_{10}
								\nonumber\\
	&   &
		  + \left[
		    1 - \Bigl(
			\tfrac{1}{4}\Delta_3^2 - \tfrac{1}{16} \Delta_3^4 V_{hh}
			\Bigr) (V_{hh} + V_{01})
		    \right]
		    \phi_{01}
								\nonumber\\
	&   &
		  - \left[
		    \Bigl( 1 - \tfrac{1}{4}  \Delta_3^2 V_{hh} \Bigr) \Delta_3^2
		    \right]
		    (G_\Sigma + 4G_{hh})
								      \,\text{,}
						       \label{eqn-FD-4th/phi_00}
\end{eqnarray}
where $\Delta_3 \eqdef \Delta/3$.


\subsubsection{$i = i_p \pm 1$ or $i = i_p \pm 2$ (Cell near the particle)}

If $i - i_p = -1$ or $-2$ (so we are computing $\phi$ at a grid point
near to, but to the right of, the particle), then the
interpolation~\eqref{eqn-FD-4th/G_hh-vacuum} would use data from both
sides of the particle's worldline (hereinafter I refer to this as
``crossing'' the worldline), violating the smoothness assumptions
used in the interpolation's derivation.  Instead, I use the ``right''
interpolation
\begin{equation}
G_{hh} =   \tfrac{3}{8} G_{10} + \tfrac{5}{8} G_{01}
	 + \tfrac{1}{4} G_{11} - \tfrac{1}{4} G_{02}
	 - \tfrac{1}{8} G_{21} + \tfrac{1}{8} G_{12}
	 + \O(\Delta^3)
								      \,\text{,}
						   \label{eqn-FD-4th/G_hh-right}
\end{equation}
which (since $i < i_p$) doesn't cross the worldline.

If $i - i_p = +1$ or $+2$ (so we are computing $\phi$ at a grid point
near to, but to the left of, the particle), then again the
interpolation~\eqref{eqn-FD-4th/G_hh-vacuum} would cross the particle's
worldline, so I instead use the ``left'' interpolation
\begin{equation}
G_{hh} =   \tfrac{3}{8} G_{01} + \tfrac{5}{8} G_{10}
	 + \tfrac{1}{4} G_{11} - \tfrac{1}{4} G_{20}
	 - \tfrac{1}{8} G_{12} + \tfrac{1}{8} G_{21}
	 + \O(\Delta^3)
								      \,\text{,}
						    \label{eqn-FD-4th/G_hh-left}
\end{equation}
which (since $i > i_p$) doesn't cross the worldline.

[Each of the interpolations~\eqref{eqn-FD-4th/G_hh-vacuum},
\eqref{eqn-FD-4th/G_hh-right}, and \eqref{eqn-FD-4th/G_hh-left} is
actually valid for any grid function which is smooth (has a convergent
Taylor series in $v$ and $u$) throughout the region spanned by the
interpolation molecule.]

Once $G_{hh}$ has been computed, $G_\Sigma$ and then $G_{00}$ can be
computed in the same manner as before, i.e., via~\eqref{eqn-FD-4th/G_Sigma}
and~\eqref{eqn-FD-4th/phi_00} respectively.
\footnote{
	 Notice that once $G_{hh}$ is known,
	 \eqref{eqn-FD-4th/G_Sigma} and~\eqref{eqn-FD-4th/phi_00}
	 use only the grid-function values $\phi_{10}$, $\phi_{01}$,
	 and $\phi_{11}$, and thus require only that the particle
	 worldline doesn't pass between these points; this condition
	 is satisfied whenever $i \ne i_p$.
	 }


\subsubsection{$i = i_p$ (Cell symmetrically bisected by the particle)}

If $i = i_p$, then for each $k \in \{1,2,3\}$ I first compute an
estimate $\phi_{j,i}^{(k)}$ for $\phi_{j,i}$ using the globally--2nd-order
scheme~\eqref{eqn-FD-2nd} applied to the diamond cell of size $k\Delta$
defined by the four grid points $\N^{(k)} = (j,i)$, $\W^{(k)} = (j-k,i)$,
$\E^{(k)} = (j,i-k)$, and $\S^{(k)} = (j-k,i-k)$.  I then assume
a Richardson expansion for $\phi^{(k)}(v,u)$ in the effective grid
spacing~$k\Delta$ (\citet{Choptuik-1991:FD-consistency}),
\begin{equation}
\phi^{(k)}(v,u) = \phi(v,u)
		  + (k\Delta)^3 P(v,u) + (k\Delta)^4 Q(v,u)
		  + \O(\Delta^5)
								      \,\text{,}
					    \label{eqn-FD-4th/FD-2nd-Richardson}
\end{equation}
where $P$ and $Q$ are smooth functions which do not depend on the
grid spacing, and where the $\O(k\Delta^3)$ leading-order error term
is that of~\eqref{eqn-FD-2nd/source-term}.  Finally, I
Richardson-extrapolate the value of $\phi_{j,i} \equiv \phi(v,u)$
from the three estimates $\phi_{j,i}^{(k)}$ to obtain
\begin{equation}
\phi_{j,i} =   \frac{108}{85} \phi_{j,i}^{(1)}
	     - \frac{ 27}{85} \phi_{j,i}^{(2)}
	     + \frac{  4}{85} \phi_{j,i}^{(3)}
	     + \O(\Delta^5)
								      \,\text{.}
					    \label{eqn-FD-4th/phi_ji-Richardson}
\end{equation}

There are only $\O(N)$ $i = i_p$ cells, so this $\O(\Delta^5)$ LTE
suffices to keep these cells' collective contribution to the global
error at $\O(\Delta^4)$.

\begin{figure}[!bp]
\begin{center}
\begin{picture}(106,125)
\put(12.7,91){\includegraphics[scale=0.50]{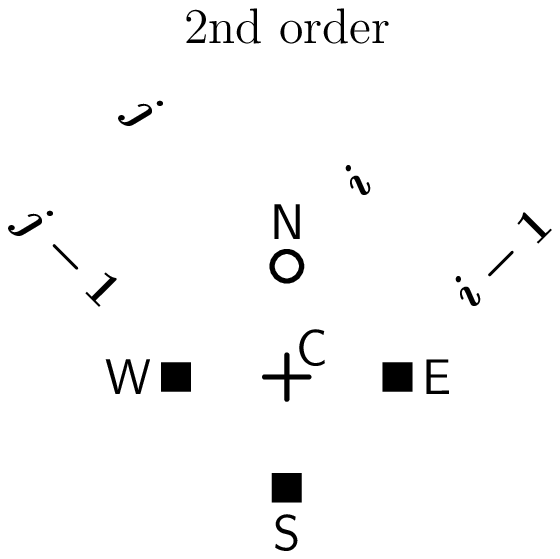}}
\put(66,93.25){\includegraphics[scale=0.50]{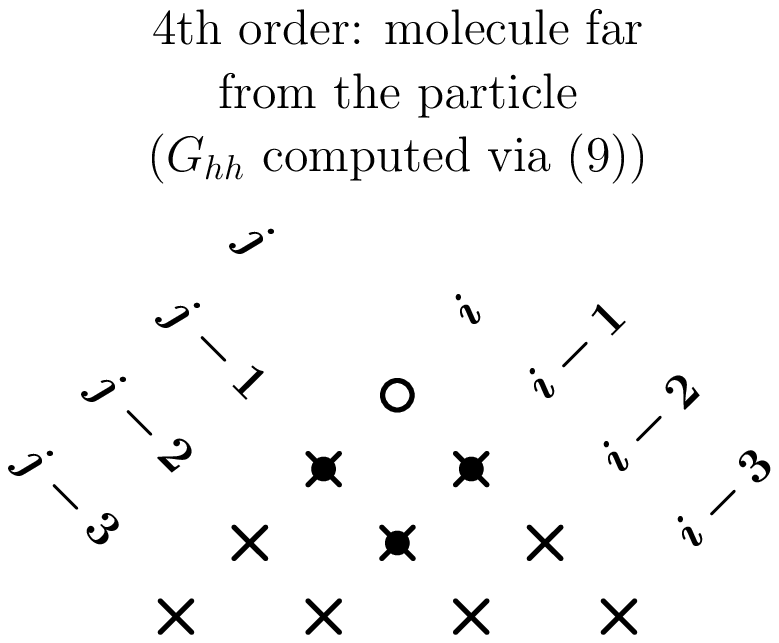}}
\put(7,50){\includegraphics[scale=0.50]{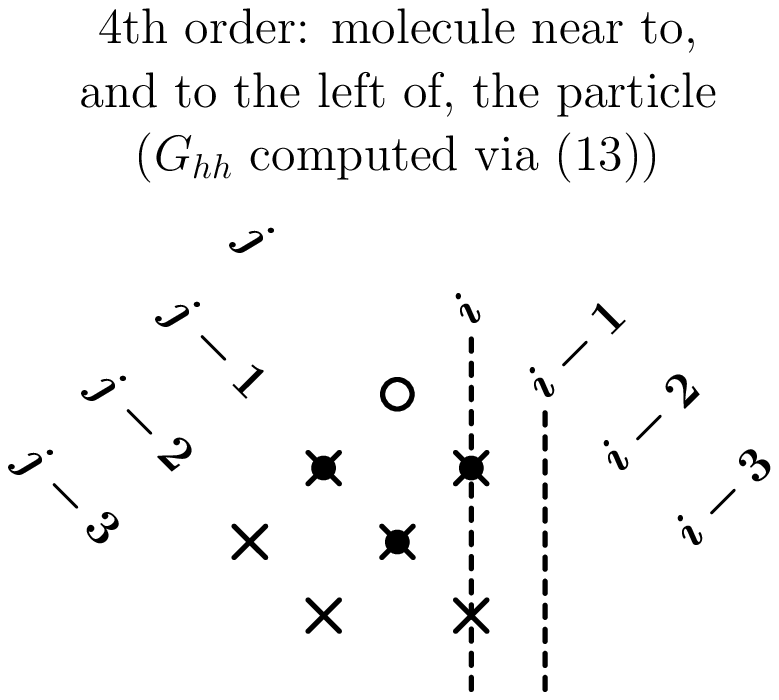}}
\put(66,50){\includegraphics[scale=0.50]{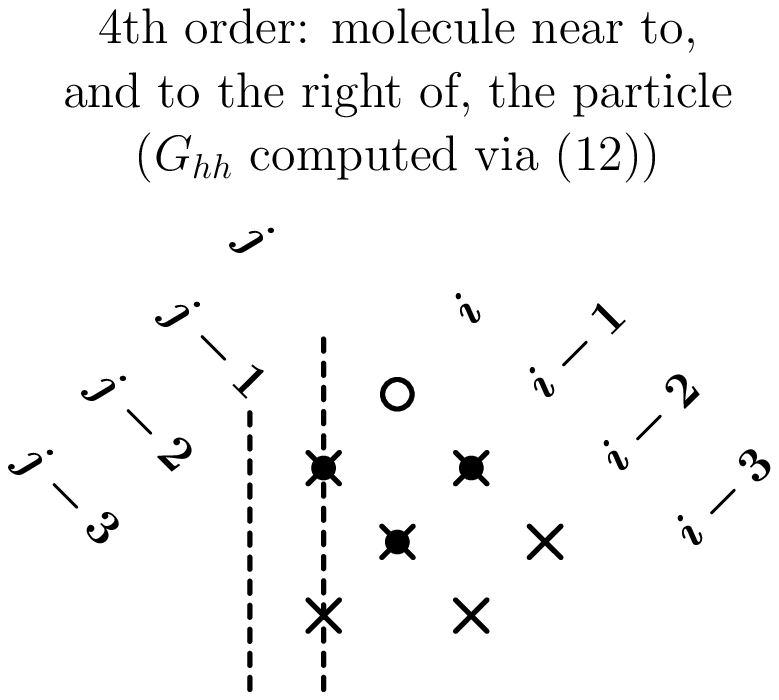}}
\put(0,0){\includegraphics[scale=0.50]{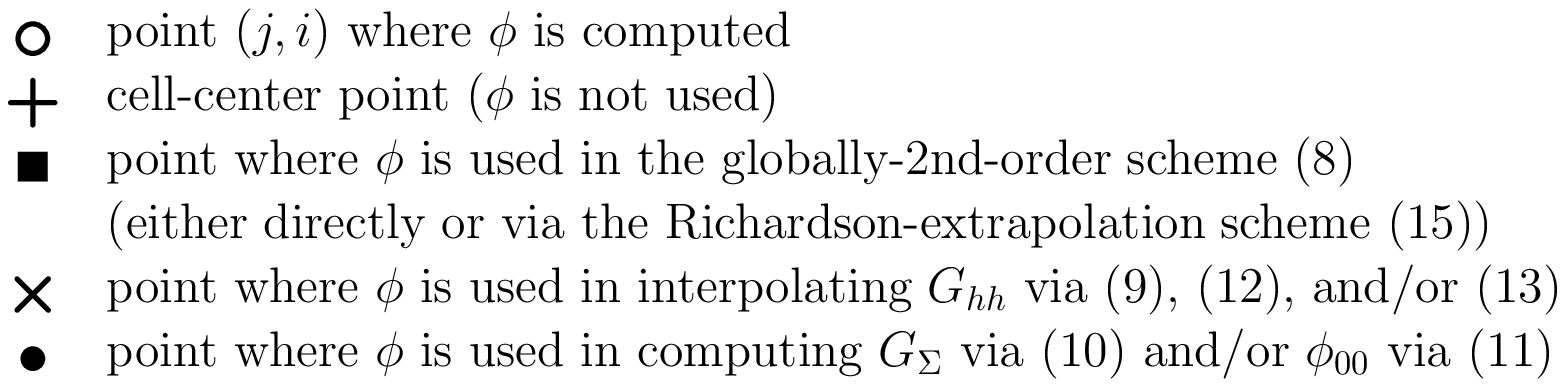}}
\put(66,0){\includegraphics[scale=0.50]{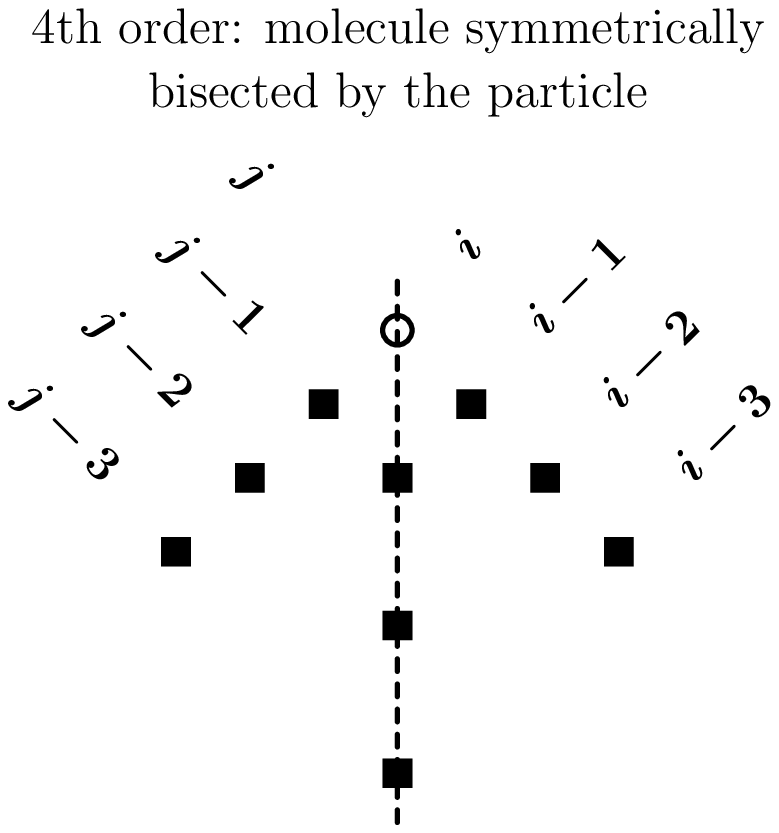}}
\end{picture}
\end{center}
\caption[Unigrid Finite Differencing Schemes]
	{
	This figure shows the unigrid finite differencing molecules.
	For clarity, the 2nd~order molecule is enlarged relative to
	the 4th~order molecules.  In the lower 3~subfigures the
	vertical dashed lines show possible positions for the
	particle worldline.
	}
\label{fig-FD-molecules-details}
\end{figure}


\subsubsection{Initial Data}
\label{app-FD-details/4th-order/initial-data}

As shown in figure~\ref{fig-FD-molecules-details}, the
globally--4th-order finite difference molecules for computing
$\phi_{j,i}$ generally include points on the $j{-}1$, $j{-}2$, and
$j{-}3$~slices, and at the $i{-}1$, $i{-}2$, and $i{-}3$ $u$~positions.
This means that starting the integration on any grid requires
``extended initial data'' on 3~slices and on 3~points on each succeeding
slice.  This poses a problem for setting up the base grid at the
very beginning of the integration,
\footnote{
	 There's no problem in starting the integration of
	 other grids, since the extended initial data for
	 any $\grid{\ell}$ with $\ell > 0$ is interpolated
	 from its parent grid $\grid{\ell{-}1}$
(line~20 of figure~\ref{fig-Berger-Oliger-algorithm/regrid}).
	 }
{} since physical boundary data is only specified along the southwest
and southeast faces of the problem domain, i.e., on a single initial
slice and at a single initial grid point of each succeeding slice.

There are several possible ways of obtaining the extended initial data:


\paragraph{Replication of the Physical Boundary Data}

For the self-force problem (\citet{Barack-Ori-2002}) the precise
choice of boundary data doesn't matter, and it's acceptable to approximate
the PDEs to a lower order of accuracy near the boundary.  Thus for
this use, the physical boundary data (in this case $\phi=0$) can
simply be replicated throughout the extended--initial-data region.


\paragraph{Taylor-series Approximation}

For more general purposes, it's usually desirable to approximate
the PDE to full accuracy everywhere in the problem domain, i.e.,
to compute the extended initial data to $\O(\Delta^4)$ accuracy.
\citet[section~4.1]{Lousto-2005} describes a Taylor-series
approximation scheme to do this.


\paragraph{Two-Level Subsampling}
\label{app-FD-details/4th-order/initial-data/two-level-subsampling}

Another scheme for computing extended initial data is to define an
auxiliary ``subsampling'' grid with spacing $\Delta_\ss \ll \Delta$,
which only covers the extended--initial-data region.  Choose
$\Delta_\ss$ such that it integrally divides~$\Delta$ and such that
$\Delta_\ss \propto \Delta^2$ for sufficiently high resolution.
The extended initial data can now be computed by integrating the
auxiliary grid using the globally--2nd-order scheme, then
subsampling data from the auxiliary grid to the main grid.  Because
$\Delta_\ss \propto \Delta^2$ at sufficiently high resolution,
$\O({\Delta_\ss}^2) = \O(\Delta^4)$, i.e., the global accuracy
remains 4th~order with respect to the main-grid resolution~$\Delta$.

The auxiliary grid only needs to cover the first 3~slices of the
main grid, together with the first 3~grid points of each later
main-grid slice.  However, this corresponds to $\O(1/\Delta)$
auxiliary-grid spacings, so integrating all the auxiliary-grid
points requires $\O(1/\Delta^3)$~CPU time and $\O(1/\Delta^2)$~memory.
This is much more expensive than the main-grid computation, which
only requires $\O(1/\Delta^2)$~CPU time and $\O(1/\Delta)$~memory.


\paragraph{Recursive Doubling}

\begin{figure}[!bp]
\begin{flushleft}
\includegraphics[scale=0.75]{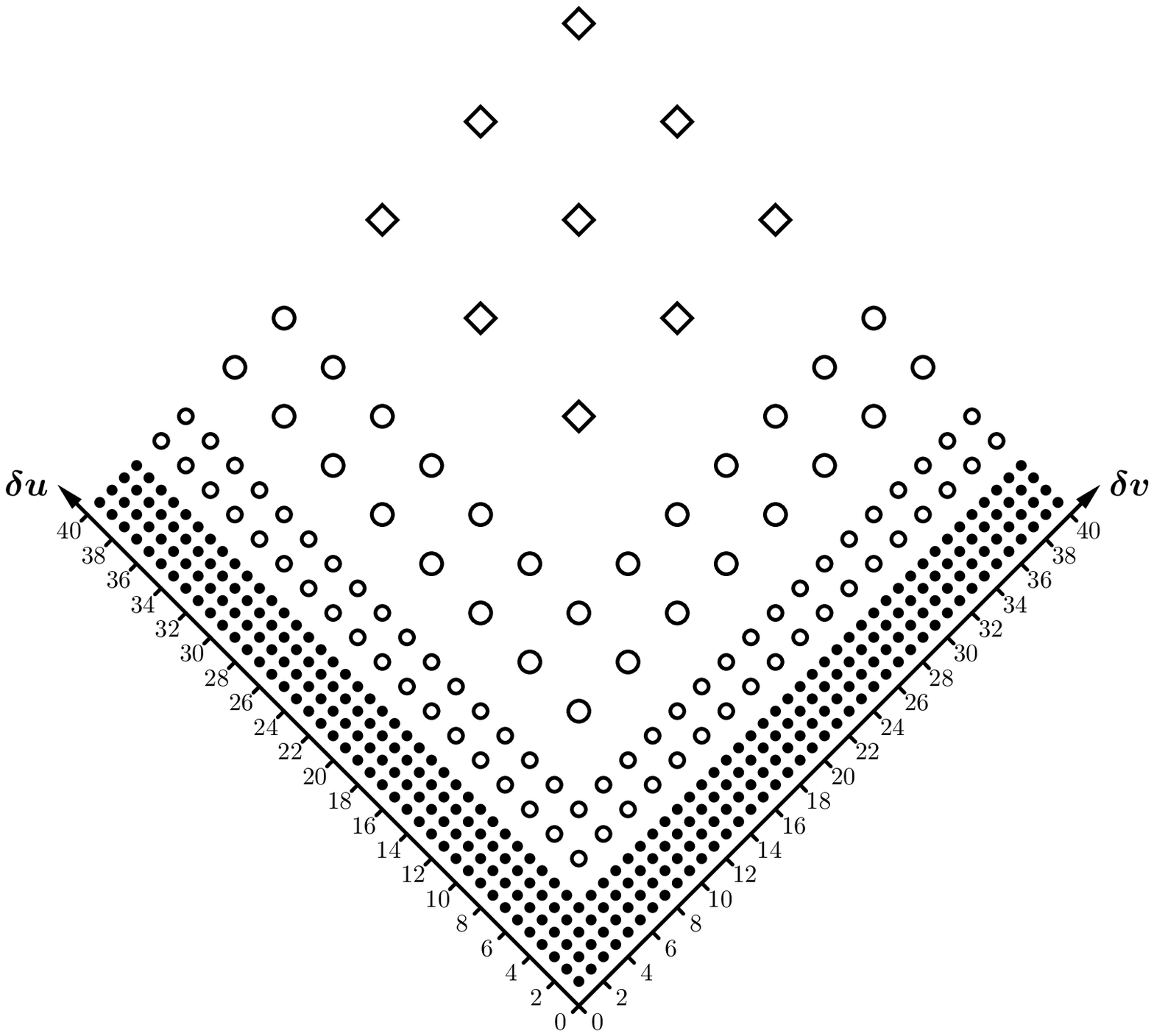}\\
\vspace{5mm}
\includegraphics[scale=0.75]{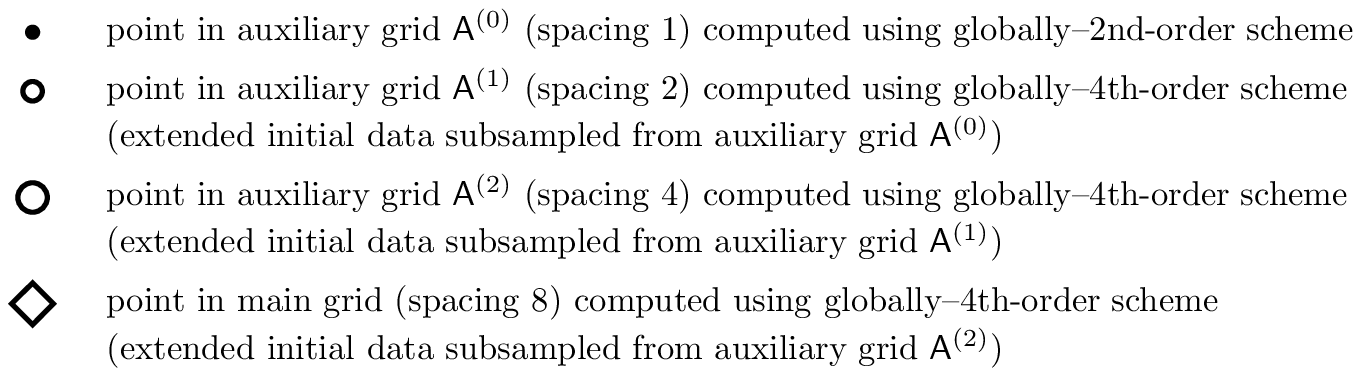}
\end{flushleft}
\caption[Example of Recursive-Doubling Extended Initial Data]
	{
	This figure shows an example of the recursive-doubling
	technique for constructing extended initial data for the
	globally--4th-order initial data scheme.  The grid axes
	are plotted in terms of $\delta u = u - u_{\min}$ and
	$\delta v = v - v_{\min}$, and the grid spacings are
	in units of finest ($\Delta_\rd^{(0)}$) auxiliary grid
	spacing.  In this example $q=3$~auxiliary grids are used.
	}
\label{fig-recursive-doubling-example}
\end{figure}

A more efficient approach is to use a recursive-doubling scheme.
Here we use a sequence of $q$~auxiliary grids $\auxgrid{0}$,
$\auxgrid{1}$, $\auxgrid{2}$, \dots, $\auxgrid{q{-}1}$, with
$\auxgrid{k}$ having spacing $2^{k{-}q} \Delta$.  We choose $q$
such that $2^{-q} \Delta \le \Delta_* < 2^{1{-}q} \Delta$ for some
$\Delta_* \propto  \Delta^2$ for sufficiently high resolution.
The extended initial data for the main grid can now be computed as
follows (see figure~\ref{fig-recursive-doubling-example} for an example):
First integrate $\auxgrid{0}$ using the globally--2nd-order scheme
for 4~slices, and for 4~points on each succeeding slice.
Then for each $k = 1$, $2$, $3$, \dots, $q{-}1$, subsample from
$\auxgrid{k{-}1}$ to obtain the extended initial data to integrate
$\auxgrid{k}$ using the globally--4th-order scheme for 2~slices,
and for 2~points on each succeeding slice.
Finally, subsample from $\auxgrid{q{-}1}$ to obtain the extended
initial data to integrate the main grid using the globally--4th-order
scheme.

Each auxiliary grid only has to cover 2~or 4~slices, and 2~or 4~points
on each succeeding slice, so this scheme is much more efficient than
the two-level subsampling scheme: in the high-resolution limit the
total cost of all the auxiliary-grid integrations is $\O(1/\Delta^2)$~CPU
time, the same order as the main-grid computation.  Unfortunately,
even with careful memory management (discarding grid points as soon
as they're no longer needed) the auxiliary grids still require
$\O(1/\Delta^2)$ memory, much more than the main-grid computation's
$\O(1/\Delta)$~memory requirement.

\begin{figure}[bp!]
\begin{flushleft}
\includegraphics[scale=0.75]{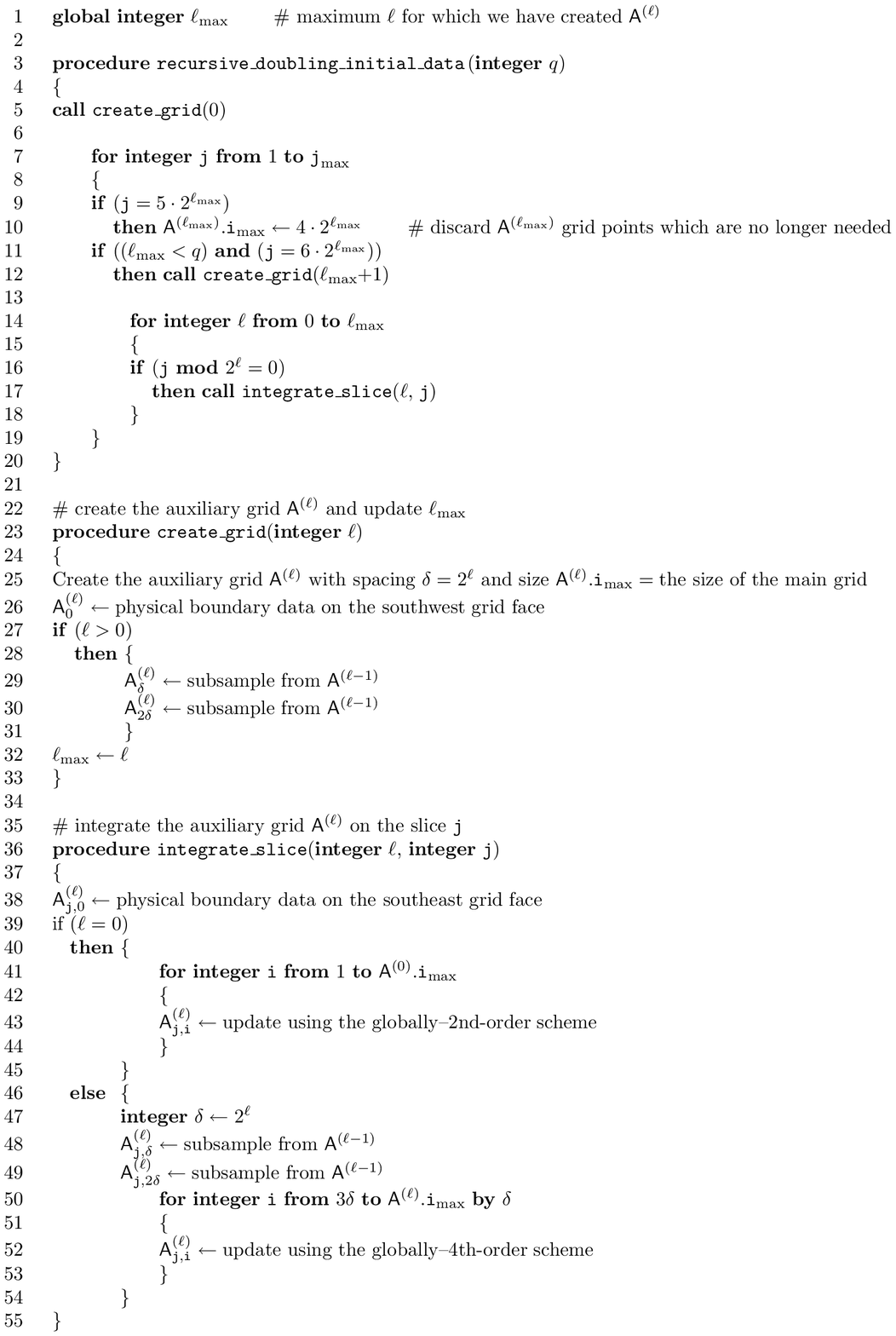}
\end{flushleft}
\caption[Outline of Recursive-Doubling Algorithm for Extended Initial Data]
	{
	This figure gives an outline of the recursive-doubling
	algorithm for constructing extended initial data for the
	globally--4th-order initial data scheme.  Coordinates for
	\emph{all} grids are measured in units of the finest
	($\auxgrid{0}$) auxiliary grid spacing, relative to the
	south corner of the problem domain.  Thus, for example,
	the grid $\auxgrid{3}$ has spacing $2^3 = 8$ and uses
	grid-point indices $\{0,8,16,24,32,40,\dots\}$.
	}
\label{fig-recursive-doubling-algorithm}
\end{figure}

It's relatively easy to implement the recursive-doubling scheme in
a code using the type of fine-grained linked-list data structures
described by \citet{Pretorius:2003wc}, where the integration can
``flow'' in either the $v$~direction or the $u$~direction at any
stage in the computation.  However, for the slice-recursion algorithm
we typically require that the integration proceed sequentially in the
$v$~direction and that any data reuse or subsampling take place within
the small (typically 4--7) number of slices kept in memory at each
refinement level.  The recursive-doubling initial-data scheme thus
requires interleaving the integration of the different auxiliary grids
along the southeast face of the grid.
Figure~\ref{fig-recursive-doubling-algorithm} gives a pseudocode
outline of an algorithm to generate the appropriate sequence of
integrations, subsamplings, and other grid operations for the
recursive-doubling scheme.

\subsection{The Coarse-Grid Instability}
\label{app-FD-details/coarse-grid-instability}

\begin{figure}[bp!]
\begin{center}
\includegraphics[scale=0.60]{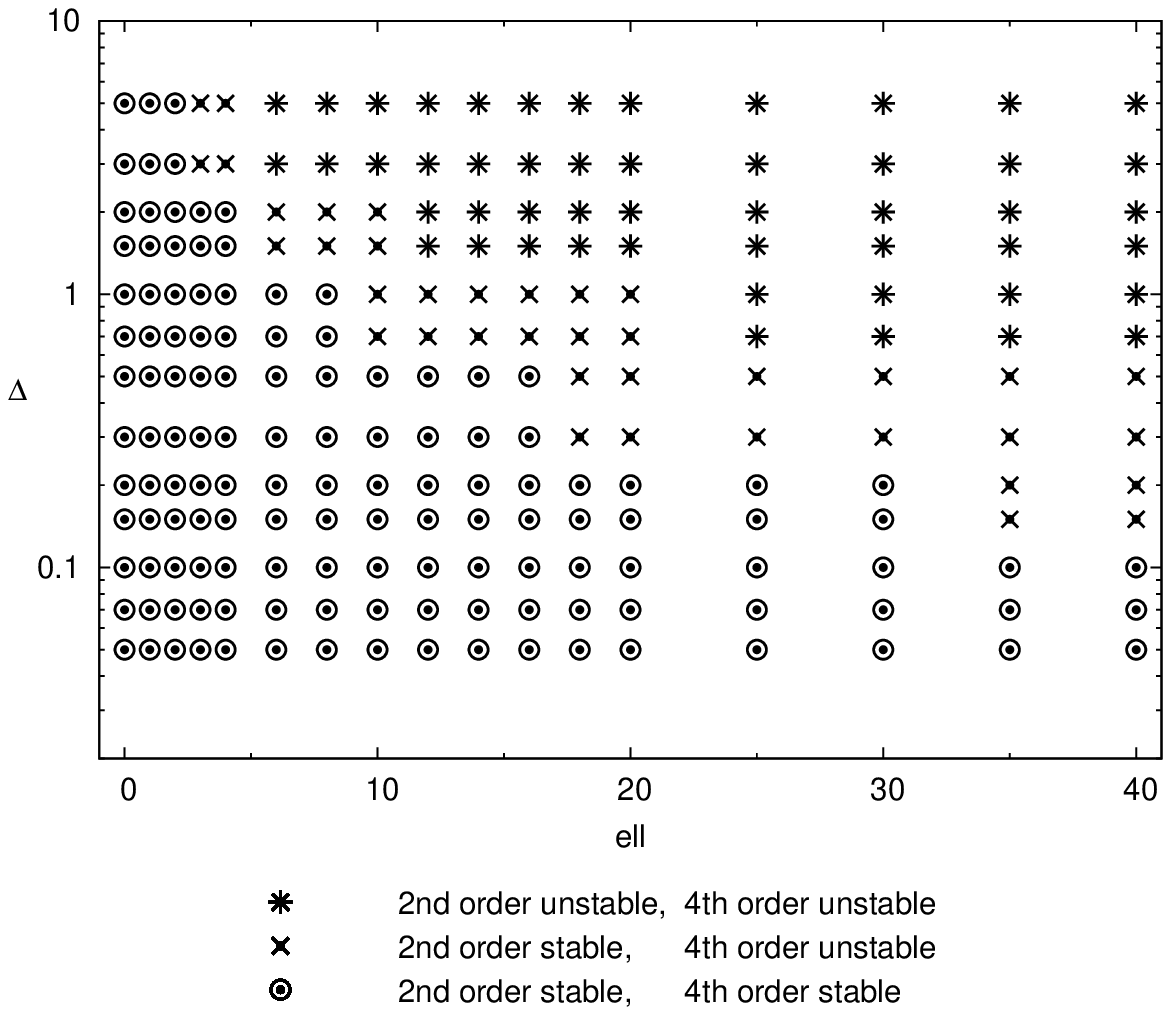}
\vspace{-2.5mm}
\end{center}
\caption[$(\ell,\Delta)$ for which the Coarse-Grid Instability is Present]
	{
	This figure shows the stability behavior of unigrid
	evolutions with varying $\ell$ and $\Delta$, using
	Gaussian initial data, no source term, and a problem domain
	size $D = 100M$.  Notice that for each $\ell$, the evolutions
	are always stable for $\Delta$ less than some threshold value.
	}
\label{fig-coarse-grid-instability}
\end{figure}

The finite differencing schemes discussed here become unstable at
very low resolutions, in a manner somewhat resembling the classic
Courant-Friedrichs-Lewy instability of Cauchy finite differencing.
I have not mathematically analyzed this instability,
\footnote{
\citet{Gomez-Winicour-1992:sssf-2+2-evolution-and-asymptotics,
Gomez-Winicour-Isaacson-1992:sssf-2+2-numerical-methods}
	 discuss the stability of diamond-cell integration
	 schemes in spherical symmetry;
	 \citet{Welling-PhD}; \citet[section~3.3]{Winicour-2009:living-review}
	 discuss subtleties in applying the CFL condition
	 to a more general null-cone evolution algorithm
	 in axisymmetry.
	 }
{} but empirically it only occurs for very low resolutions
(large $\Delta$), with the instability threshold (the smallest~$\Delta$
for which the instability appears) depending on $\ell$, but not on $m$.
Figure~\ref{fig-coarse-grid-instability} shows the $\ell$ and $\Delta$
for which the instability occurs.  Notice that the instability threshold
decreases gradually with $\ell$, and is somewhat smaller for the
globally-4th-order scheme than for the globally-2nd-order scheme.

In practice, it's rare for this ``coarse-grid instability'' to be a
significant problem because reasonable accuracy requirements normally
force much higher resolutions than those where the instability would
occur.  The one exception to this is the base grid, which might
otherwise be made very coarse (allowing the AMR to refine it as
needed); the coarse-grid instability prevents this by requiring the
base grid to be finer than the instability threshold.


\section{Implementation Details}
\label{app-implementation}


\subsection{Computing $r(r_*)$}
\label{app-implementation/computing-r(r_*)}

The finite differencing schemes discussed here use finite-difference
grids which are locally uniform in $v$ and $u$, so it's trivial to
compute the $r_*$~coordinate of any grid point.  However, the coefficients
in the wave equation~\eqref{eqn-wave} are all given as explicit
functions of $r$, so the code needs to know the $r$~coordinate of
each grid point (and, for the globally--4th-order scheme, also of
the center of each grid zone).  My code computes this as follows:
Define
\begin{subequations}
\begin{eqnarray}
y	& \eqdef &	\ln \left( \frac{r}{2M} - 1 \right)		\\
x_*	& \eqdef &	\frac{r_*}{2M}
								      \,\text{,}
\end{eqnarray}
\end{subequations}
so that $r = 2M \left( 1 + e^y \right)$ and the
definition~\eqref{eqn-rstar-defn} becomes $x_* = 1 + y + e^y$.
Then $y(x_*)$ (and hence $r(r_*)$) can be found by using Newton's
method to find a zero of the function
\begin{equation}
h(y) = 1 + y + e^y - x_*
						     \label{eqn-f(y)-for-r(r_*)}
\end{equation}
An initial guess for Newton's method can be obtained by neglecting
either $y$ or $e^y$ in~\eqref{eqn-f(y)-for-r(r_*)}, giving
\begin{equation}
y_\text{initial guess}
	= \left\{
	  \begin{array}{ll}
	  \log(x_* - 1)	& \text{if $x_* >   1$ ($y \gtsim -0.577$)}	\\
	  x_* - 1	& \text{if $x_* \le 1$ ($y \ltsim -0.577$)}	
	  \end{array}
	  \right.
\end{equation}

The Newton's-method solution is moderately expensive for a computation
which (logically) is needed at each grid point: it typically requires
$3$--$10$ iterations, with each iteration needing an \var{exp()}
computation and several other floating-point arithmetic operations.
Fortunately, within any single grid $r$ depends only on $j-i$, so in
a unigrid code it's easy to precompute~$r$ for all possible values of
$j{-}i$ (of which there are only~$\O(N)$ for an $N {\times} N$~grid)
when the grid is first set up.  For the slice-recursion algorithm a
somewhat more dynamic ``radius cache'' of $r$~coordinates is needed,
with updates each time regridding grows, shrinks, or relocates a grid.
However, the set of $j{-}i$ involved is still always a contiguous
interval, so the cache bookkeeping overhead (over and above the
storage arrays for the $r$~coordinates themselves) is only~$\O(1)$
per grid.


\subsection{Local Coordinates for each Refinement Level}
\label{app-implementation/local-coords}

Consider a single slice, and a pair of adjacent grid points in it
at some refinement level~$\ell$, say $\grid{\ell}_{\var{j},\var{i}}$
and $\grid{\ell}_{\var{j},\var{i}{+}1}$, viewed as events in spacetime.
Since the grid spacing of $\grid{\ell{+}k}$ is $2^k$~times finer
than that of $\grid{\ell}$, these same two events are necessarily
$2^k$~grid points apart in $\grid{\ell{+}k}$.  This means that it's
impossible to define integer grid-point coordinates which simultaneously
(a) have adjacent grid points separated by~$1$ in the integer coordinates
at each refinement level, and (b) assign a given event the same
integer coordinates at each refinement level.

In my AMR code I keep property~(a), but discard property~(b): each
mesh-refinement level has its own local integer coordinate system for
indexing grid points, and the code maintains explicit fine-to-coarse
and coarse-to-fine coordinate transformations between each pair
$(\ell, \ell{+}1)$~of adjacent refinement levels.  These transformations
are used when interpolating data from coarse to fine grids
(discussed in detail in section~\ref{app-implementation/interp-ops}),
when injecting fine-grid results back to coarse grids
(lines~36-41 in figure~\ref{fig-Berger-Oliger-algorithm/main}),
in setting up the ``tail'' re-integration
(lines~43--48 of figure~\ref{fig-Berger-Oliger-algorithm/main}),
and in checking the proper-nesting condition in regridding
(lines~31--48 of figure~\ref{fig-Berger-Oliger-algorithm/regrid}).
This scheme has worked very well, and I recommend its use to others
implementing Berger-Oliger mesh-refinement codes.

However, for the extended--initial-data algorithm of
figure~\ref{fig-recursive-doubling-algorithm}, it's convenient
to use integer coordinates which keep property~(b), but discard
property~(a).  The \program{Carpet} code
(\citet{Schnetter-etal-03b,Schnetter-2001:Carpet-code})
also uses integer coordinates of this latter type.


\subsection{Interpolation Operators}
\label{app-implementation/interp-ops}

As discussed in sections~\ref{sect-AMR/Berger-Oliger-algorithm}
and~\ref{sect-AMR/slice-recursion-algorithm}, the slice-recursion
algorithm needs to interpolate data from coarse to fine grids in
several situations:
\begin{itemize}
\item	When creating a new grid~$\grid{\ell}$, the first few slices
	(1~[3]~slices for globally 2nd~[4th]~order finite differencing)
	of the newly-created grid are initialized by interpolating
	from the next coarser grid~$\grid{\ell{-}1}$
	(line~20 of figure~\ref{fig-Berger-Oliger-algorithm/regrid};
	figure~\ref{fig-slice-recursion-example}c).
\item	When time-integrating any grid $\grid{\ell}$ finer than the
	base grid, the extended initial data on each new $\grid{\ell}$~slice
	(i.e., the first 1~[3] points on the slice for globally
	2nd~[4th]~order finite differencing), must be interpolated
	from the next coarser grid~$\grid{\ell{-}1}$
	(line~9 of figure~\ref{fig-Berger-Oliger-algorithm/main};
	figure~\ref{fig-slice-recursion-example}c,d).
	before the main integration of the slice can be started.
\item	When moving an existing grid~$\grid{\ell}$ to a new position
	in the current $v = \text{constant}$ slice, newly-created
	points are initialized by interpolating from the next
	coarser grid~$\grid{\ell{-}1}$
	(lines~25--26 of figure~\ref{fig-Berger-Oliger-algorithm/regrid}).
\end{itemize}

\begin{figure}[!bp]
\begin{center}
\includegraphics[scale=0.50]{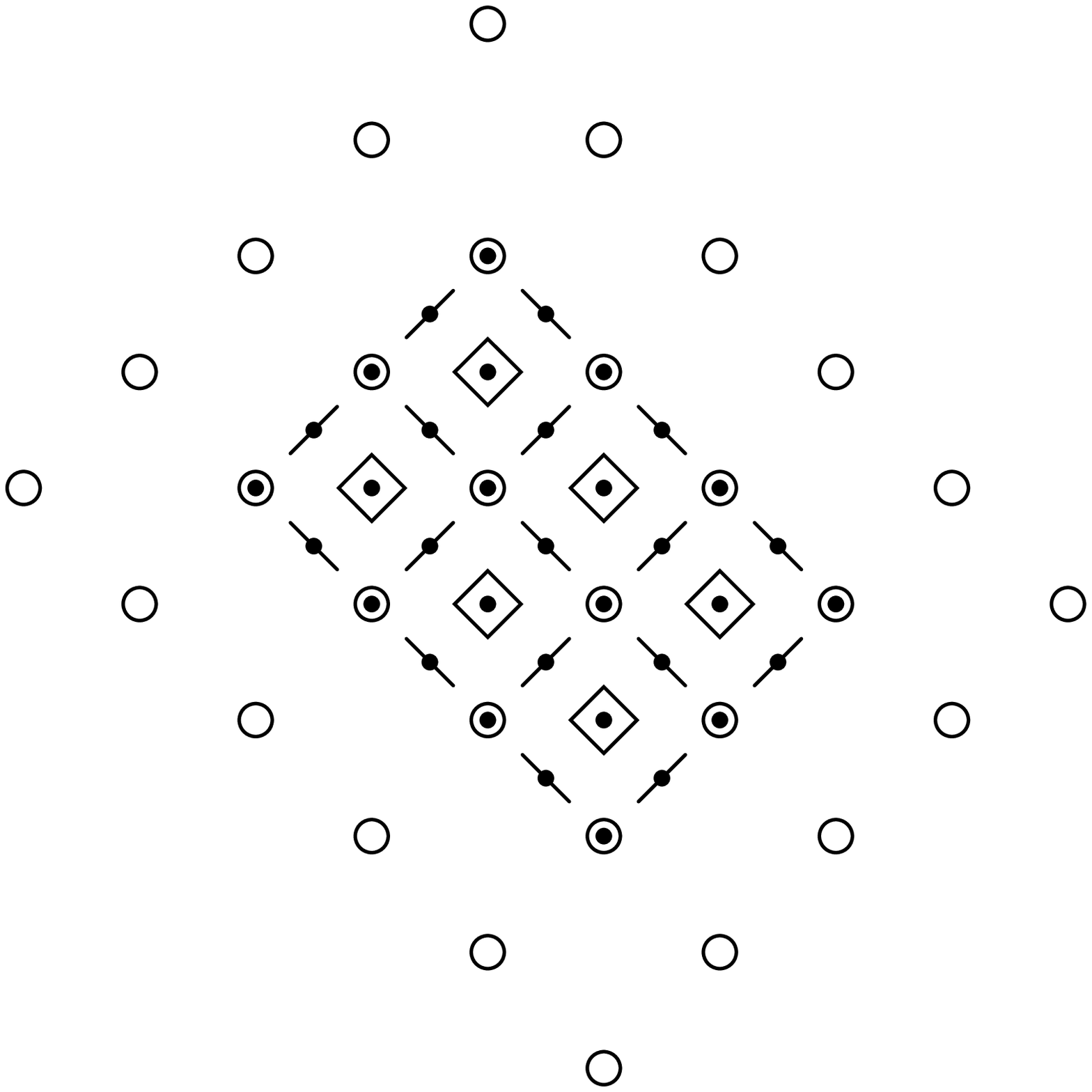}\\
\vspace{5mm}
\includegraphics[scale=0.50]{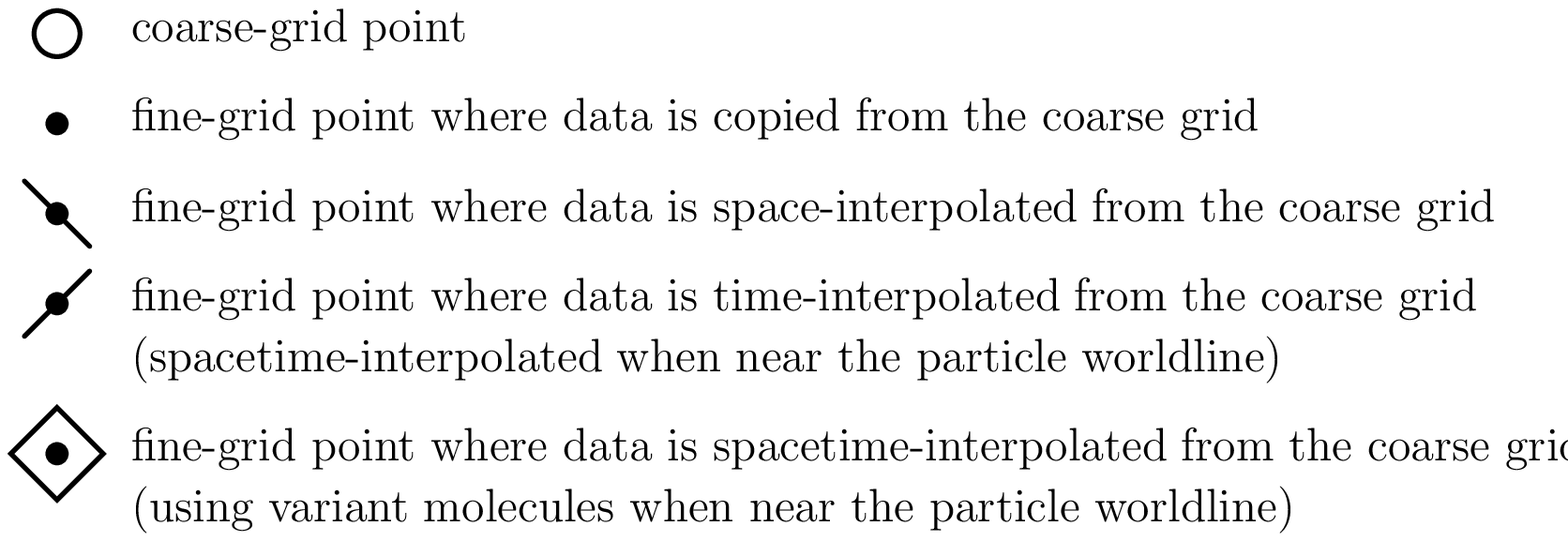}
\end{center}
\caption[Summary of Interpolation Operators]
	{
	This figure shows the type of interpolation operator used
	for each possible relative position of a fine-grid point
	with respect to the next coarser grid.  The space-interpolation
	and time-interpolation operators are described in the text.
	The spacetime-interpolation molecules are shown in
	figure~\ref{fig-interp-2D}.
	}
\label{fig-interp-vtoc}
\end{figure}

As shown in figure~\ref{fig-interp-vtoc}, the precise choice
of interpolation operator (which is made independently at each
$\grid{\ell}$~grid point) depends on the relative position of the
$\grid{\ell}$~interpolation point with respect to the next coarser
grid~$\grid{\ell{-}1}$.  All the interpolation operators considered
here are Lagrange polynomial interpolants, which assume smoothness,
so if the interpolation position is within a few grid points of the
particle worldline (where $\phi$ is only $C^1$), then a different
interpolation operator needs to be chosen so as to avoid crossing
the particle worldline:
\begin{itemize}
\item	If the interpolation point coincides with a coarse-grid
	($\grid{\ell{-}1}$) point, then the ``interpolation'' is
	just a copy of the data.
\item	Otherwise, if the interpolation point's time~($v$) coordinate
	coincides with that of a coarse-grid ($\grid{\ell{-}1}$)
	$v = \text{constant}$ slice, then the interpolation is
	a 1-dimensional Lagrange polynomial interpolation in space~($u$)
	within this slice, using 4~[6]~points for globally 2nd~[4th]~order
	finite differencing.  The interpolation is constrained not
	to cross the particle worldline and not to use data from
	outside the spatial~($u$) extent of the coarse slice.  The
	interpolation is centered if this is possible within these
	constraints, otherwise it's as minimally off-centered as
	is necessary to satisfy them.
\item	Otherwise, if the interpolation point's spatial~($u$)
	coordinate coincides with that of a coarse-grid
	($\grid{\ell{-}1}$) $u = \text{constant}$ line of grid
	points, then depending on the relative position of the
	interpolation point and the particle worldline, there are
	two cases:
	\begin{enumerate}
	\item
	\label{case-time-interp-far-from-particle}
		If the interpolation point is not close to the
		particle worldline, then the interpolation is a
		1-dimensional Lagrange polynomial interpolation
		in time~($v$) within the $u = \text{constant}$
		line of coarse-grid ($\grid{\ell{-}1}$) points,
		again using 4~[6]~points for globally 2nd~[4th]~order
		finite differencing.  The set of input points
		for this interpolation is always the most recent
		4(6)~slices of the coarse grid ($\grid{\ell{-}1}$).
	\item	Alternatively, if the interpolation point is
		too close to the particle worldline (i.e., if
		the 1-dimensional Lagrange polynomial interpolation
		molecule of case~\ref{case-time-interp-far-from-particle}
		would cross the particle worldline), then the
		interpolation is a 2-dimensional Lagrange
		polynomial interpolation in spacetime, chosen
		so as to not cross the particle worldline.  This
		case is described further below.
	\end{enumerate}
\item	Otherwise (i.e., if the interpolation point lies in the
	center of a coarse-grid ($\grid{\ell{-}1}$) cell),
	the interpolation is a 2-dimensional Lagrange polynomial
	interpolation in spacetime, again chosen so as to not
	cross the particle worldline.  This case is described
	further below.
\end{itemize}

While it is straightforward to construct Lagrange polynomial
interpolation operators in 1~dimension, doing so in 2~(or more)
dimensions is more difficult.  The basic concept is the same --
an interpolating polynomial is matched to the known grid function
values at some set of molecule points, then evaluated at the
interpolation point -- but there are several complications.

In 1~dimension the choice of interpolating polynomial is obvious,
but in multiple dimensions different choices are possible.  That is,
let $(v_*,u_*)$ be a fixed reference point somewhere near
the interpolation point $(v,u)$, and define the relative
coordinates $x \eqdef v_* - v$ and~$y \eqdef u_* - u$.
Then an $n$th~degree interpolating polynomial in $x$ and~$y$
might reasonably be defined as either
\begin{equation}
f(x,y) = \sum_{\substack{0 \le p+q \le n \\ p \ge 0, q \ge 0}} a_{pq} x^p y^q
					  \label{eqn-2D-interp/generic-triangle}
\end{equation}
or as
\begin{equation}
f(x,y) = \sum_{\substack{0 \le p \le n \\ 0 \le q \le n}} a_{pq} x^p y^q
								      \,\text{.}
\end{equation}
In my code I (somewhat arbitrarily) always use interpolating polynomials
of the form~\eqref{eqn-2D-interp/generic-triangle}.  I use $n = 3$~[$5$]
for 4th~[6th] order LTE (corresponding to 2nd~[4th]~order GTE).

\begin{figure}[!bp]
\begin{center}
\begin{picture}(114,165)
\put(7.5,126.25){\includegraphics[scale=0.50]{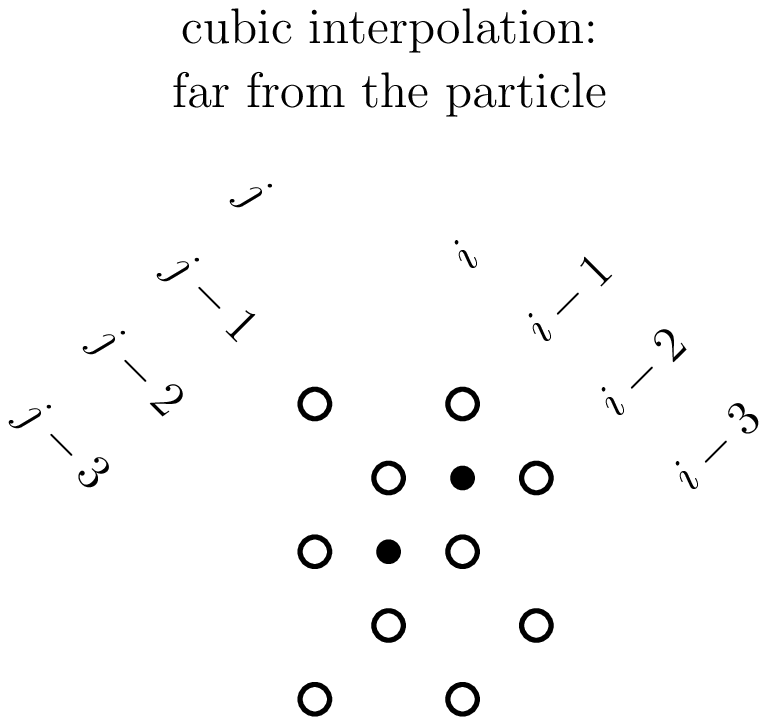}}
\put(60,115){\includegraphics[scale=0.50]{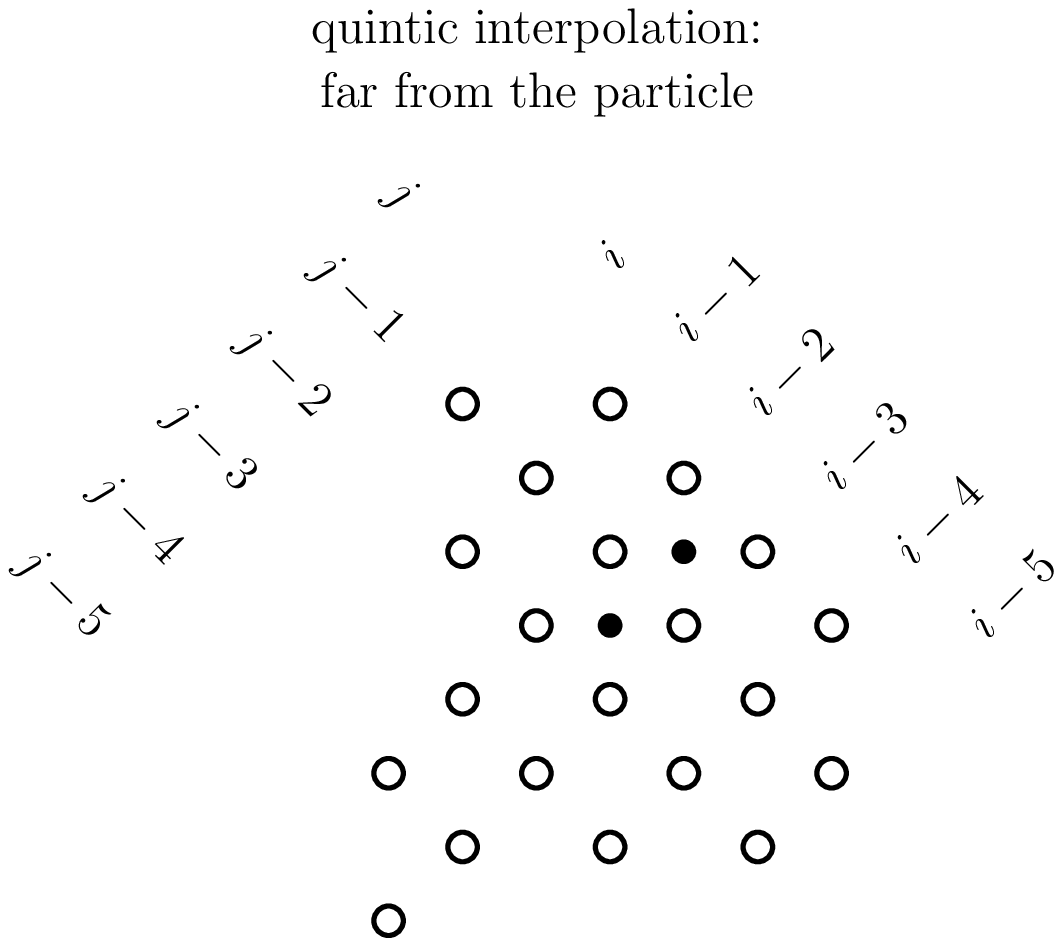}}
\put(7.5,65){\includegraphics[scale=0.50]{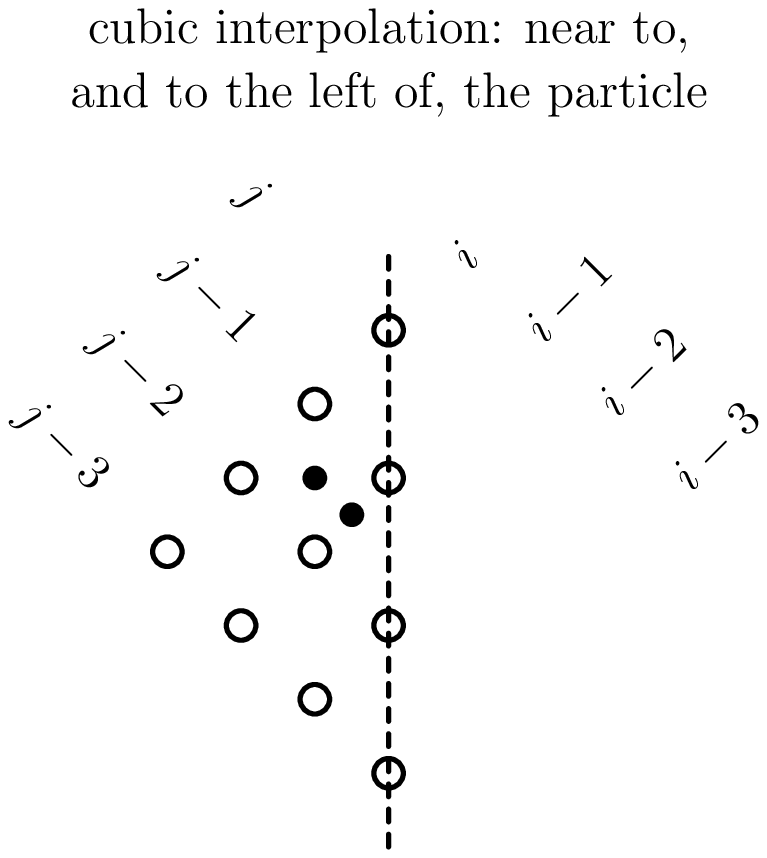}}
\put(67.5,65){\includegraphics[scale=0.50]{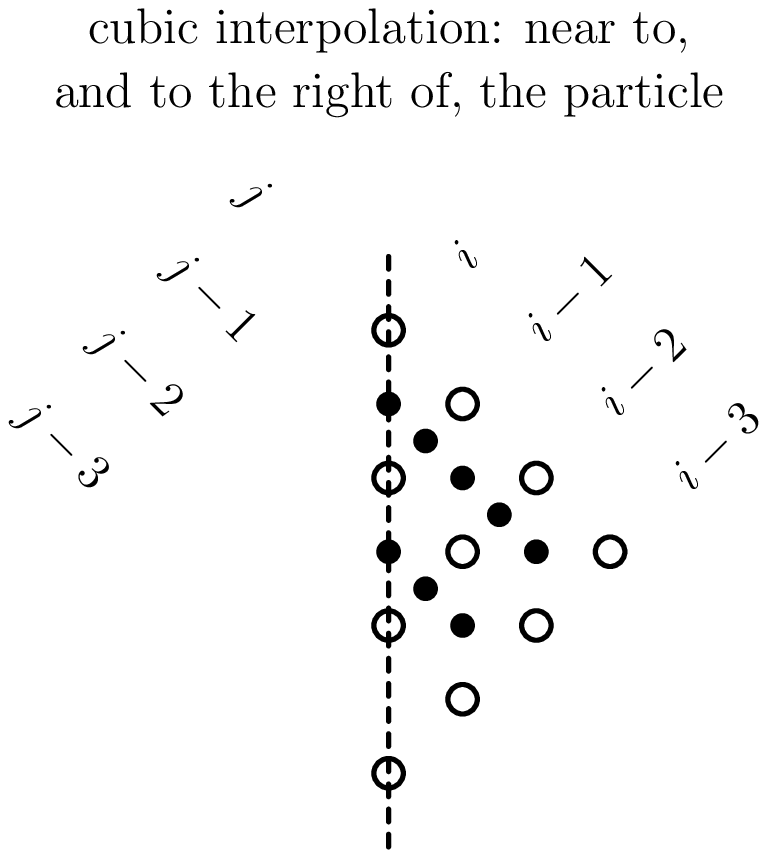}}
\put(0,0){\includegraphics[scale=0.50]{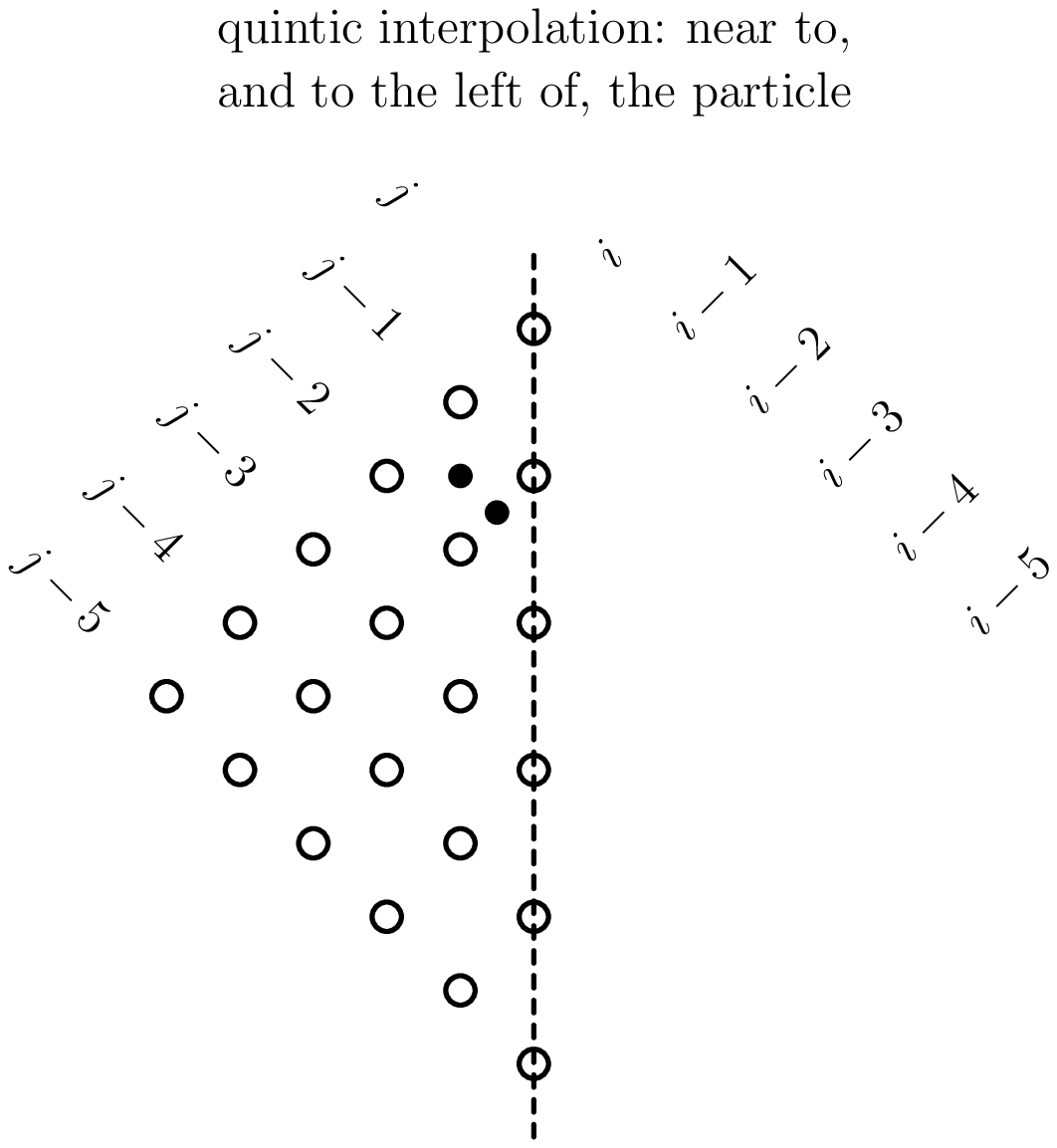}}
\put(60,0){\includegraphics[scale=0.50]{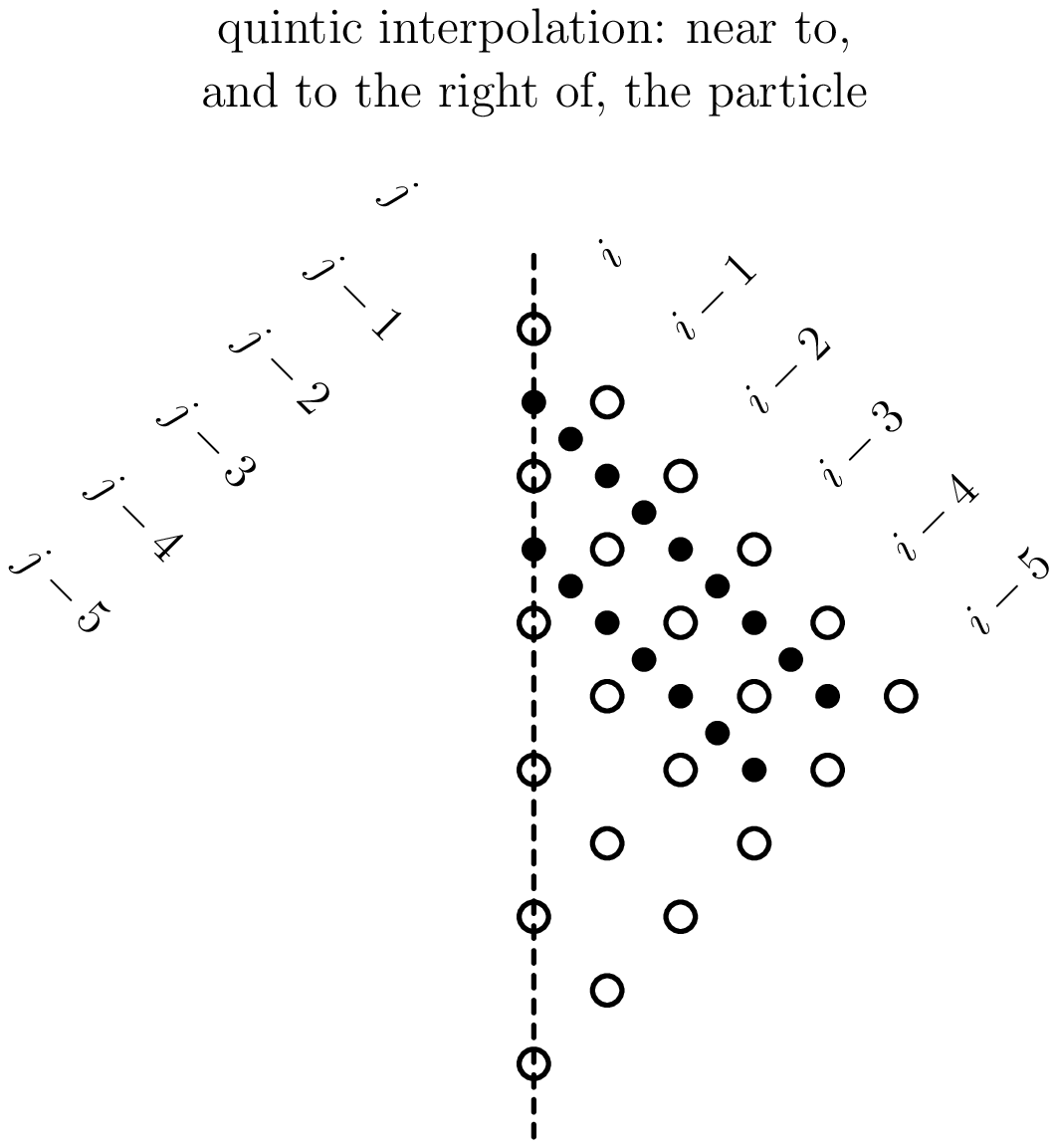}}
\end{picture}
\end{center}
\caption[Summary of Spacetime-Interpolation Molecules]
	{
	This figure shows the spacetime-interpolation molecules.
	In each subfigure, the open circles show the input points
	of the interpolation molecule (some of which may have zero
	weight in any given molecule), the solid circles show the
	various interpolation points, and in the lower 4~subfigures,
	the dashed line shows the particle worldline.
	}
\label{fig-interp-2D}
\end{figure}

Given the choice of an interpolating polynomial, there are still
many different molecules possible, even given the requirement that
the values of the interpolating polynomial at the molecule points
uniquely determine the polynomial coefficients.  I have used the
Maple symbolic algebra system
(\citet[version~11, \url{http://www.maplesoft.com/}]{Char-etal-1983:Maple-design})
to experiment with different interpolation molecules and to compute
their coefficients.  Figure~\ref{fig-interp-2D} summarizes the set
of spacetime-interpolation molecules used in my code.  The actual
coefficients may be obtained from the Maple output files in the
\texttt{sfevol/coeff/} directory of the source code included in the
electronic supplementary materials accompanying this article
(online resource~2).

\subsection{Data Structures}
\label{app-implementation/data-structures}

As noted earlier in this paper, the largest practical obstacle to the
use of Berger-Oliger mesh refinement algorithms is the complexity
of programming, debugging, and testing them.  To help reduce this
complexity for other researchers, here I briefly outline the main
data structures and debugging/testing strategies I have found useful
in implementing the slice-recursion algorithm.

A \program{slice} object represents a single $\grid{\ell}_j$ slice
at a single refinement level, i.e., it stores all the grid functions
needed to represent the solution of the PDEs on that slice.  In my code,
\program{slice} is a \Cplusplus{} template with the template parameter
selecting the PDE system (e.g., real or complex scalar field) to be
supported.

A \program{chunk} object stores enough adjacent slices at a single
refinement level to be able to take time steps, i.e., it stores
4~[7]~adjacent \program{slice} objects for the 2nd~[4th]~order GTE
finite differencing schemes described in this paper.  \program{chunk}
also maintains the radius cache discussed in
appendix~\ref{app-implementation/computing-r(r_*)}.  In my code,
\program{chunk} is a \Cplusplus{} template with 2~template parameters,
one selecting the PDE system and the other selecting the finite
differencing scheme (2nd versus 4th~order GTE) and thus implicitly
the number of adjacent slices to be stored.  To avoid unnecessary
data copying, at each time step \program{chunk} circularly rotates
pointers to a fixed set of \program{slice} objects.  When debugging
the code, \program{chunk} (and \program{slice}) can be thoroughly
tested using unigrid evolutions.

A \program{mesh} object represents an entire grid hierarchy as described
in section~\ref{sect-AMR/Berger-Oliger-algorithm}.  That is, a \program{mesh}
object stores a stack of \program{chunk} objects, one for each refinement
level, together with the necessary bookkeeping information to compute
the fine-to-coarse and coarse-to-fine coordinate transformations
described in appendix~\ref{app-implementation/local-coords}.  When
debugging the code, \program{mesh} can be tested by manually creating
a grid hierarchy and testing that the expected results are obtained
for various operations on it such as adding a new refinement level,
dropping a refinement level, moving the chunk at some refinement level
to a new location, interpolating or copying data from one refinement
level to another, or transforming the per--refinement-level coordinates
from one refinement level to another.

Finally, the actual slice-recursion Berger-Oliger and regridding
algorithms are implemented in terms of the various \program{mesh}
operations.  I found it difficult to thoroughly test the Berger-Oliger
and regridding logic, but this comprises a relatively small body of
code -- most of the overall complexity of the software lies in
\program{mesh} and lower-level code, which is relatively straightforward
to test.


\end{document}